\documentclass[fleqn,usenatbib]{config/mnras}

\usepackage{newtxtext,newtxmath}
\usepackage[LGR,T1]{fontenc}
\usepackage{anyfontsize}
\usepackage{ulem}

\usepackage{pdflscape}
\DeclareRobustCommand{\VAN}[3]{#2}
\let\VANthebibliography\thebibliography
\def\thebibliography{\DeclareRobustCommand{\VAN}[3]{##3}\VANthebibliography}

\usepackage{hyperref}
\hypersetup{
    colorlinks=true,
    linkcolor=Cerulean,
    citecolor=NavyBlue,
    filecolor=magenta,
    urlcolor=Violet,
    pdftitle={Francesco D'Eugenio - BlackJADES - JWST/NIRSpec},
    }
\usepackage{graphicx}	% Including figure files
\usepackage{amsmath}	% Advanced maths commands
\usepackage{makecell}	% Advanced maths commands
\usepackage{subcaption}
\usepackage{tabularx}
\usepackage{textgreek}
\usepackage{xargs}
\usepackage[dvipsnames]{xcolor}
\usepackage{xspace}

\usepackage{etoolbox}
\makeatletter
\newcommand\sendemail[4]{%                %\newcommand\tpj@compose@mailto[3]{%
\edef\@tempa{mailto:#1?subject=#2&body=#3 }%
\edef\@tempb{\expandafter\html@spaces\@tempa\@empty}%
\href{\@tempb}{#4}}

\catcode\%=11
\def\html@spaces#1 #2{#1%20\ifx#2\@empty\else\expandafter\html@spaces\fi#2}
\catcode\%=14
\makeatother
\usepackage{adjustbox}
\newcommand{\todo}[1]{\textcolor{magenta}{[#1]}}
\newcommand{\orcid}[2]{\href{http://orcid.org/#2}{#1}}
\newcommand{\orcidsymb}[2]{\href{http://orcid.org/#2}{#1\adjustbox{trim={-.15\width} {0\height} {-.15\width} {0\height},clip}{\includegraphics[height=10pt]{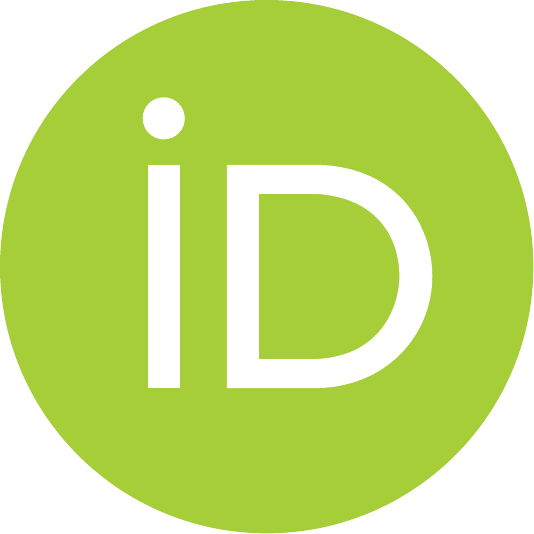}}}}

\newcommand{\citationneeded}{\textcolor{ForestGreen}{$^{\rm citation\;needed}$}}
\let\oldtextsigma\textsigma
\renewcommand{\textsigma}{\oldtextsigma\xspace}
\let\oldAA\AA
\renewcommand{\AA}{\text{\oldAA}\xspace}
\let\oldtextdegree\textdegree
\renewcommand{\textdegree}{\oldtextdegree\xspace}

\newcommand{\kms}{\ensuremath{\mathrm{km\,s^{-1}}}\xspace}
\newcommand{\Msun}{\ensuremath{{\rm M}_\odot}\xspace}
\newcommand{\Zsun}{\ensuremath{{\rm Z}_\odot}\xspace}
\newcommand{\yr}{\ensuremath{{\rm yr}}\xspace}
\newcommand{\Myr}{\ensuremath{{\rm Myr}}\xspace}
\newcommand{\Gyr}{\ensuremath{{\rm Gyr}}\xspace}
\newcommand{\peryr}{\ensuremath{{\rm yr^{-1}}}\xspace}
\newcommand{\Lsun}{\hbox{\,${\rm L}_\odot$}}
\newcommand{\mum}{\text{\textmu m}\xspace}
\newcommand{\kpc}{\text{kpc}\xspace}
\newcommand{\ZH}{\text{[Z/H]}\xspace}
\newcommandx{\pcm}[1][1=3]{\ensuremath{\mathrm{cm}^{-#1}}\xspace}	% per cm-squared

\newcommandx{\lambdar}[2][1=R,2=]{\ensuremath{\lambda_{\rm {#1}}{#2}}\xspace}
\newcommand{\eps}{\ensuremath{\epsilon}\xspace}
\newcommand{\mstar}{\ensuremath{M_\star}\xspace}
\newcommand{\mdyn}{\ensuremath{M_\mathrm{dyn}}\xspace}
\newcommand{\re}{\ensuremath{R_\mathrm{e}}\xspace}
\newcommand{\vstar}{\ensuremath{v_\star}\xspace}
\newcommand{\vnai}{\ensuremath{v_{\NaI}}\xspace}
\newcommand{\sigmastar}{\ensuremath{\sigma_\star}\xspace}
\newcommand{\sigmaestar}{\ensuremath{\sigma_{\star,\mathrm{e}}}\xspace}
\newcommand{\vperc}[1]{\ensuremath{v_{#1}}\xspace}

\newcommand{\vesc}{\ensuremath{v_\mathrm{esc}}\xspace}
\newcommand{\nelec}{\ensuremath{n_\mathrm{e}}\xspace}
\newcommand{\Telec}{\ensuremath{T_\mathrm{e}}\xspace}
\newcommand{\Rout}{\ensuremath{R_\mathrm{out}}\xspace}
\newcommand{\vout}{\ensuremath{v_\mathrm{out}}\xspace}
\newcommandx{\Mout}[2][1=,2=]{\ensuremath{M_{\mathrm{out}{#2}}^{#1}}\xspace}
\newcommandx{\Mdotout}[2][1=,2=]{\ensuremath{\dot{M}_{\mathrm{out}{#2}}^{#1}}\xspace}

\newcommandx{\fluxdcgs}[1][1=-20]{\ensuremath{\times 10^{#1}~\mathrm{erg\,s^{-1}\,cm^{-2}\,\AA^{-1}}}\xspace}
\newcommandx{\fluxcgs}[2][1=-20,2=\times]{\ensuremath{{#2}10^{#1}~\mathrm{erg\,s^{-1}\,cm^{-2}}}\xspace}
\newcommandx{\powercgs}[1][1=44]{$\times 10^{#1}$~erg~s$^{-1}$\xspace}
\newcommand{\Av}{\ensuremath{A_V}\xspace}

\newcommand{\jwst}{\textit{JWST}\xspace}
\newcommand{\hst}{\textit{HST}\xspace}
\newcommand{\ppxf}{{\sc ppxf}\xspace}
\newcommand{\prospector}{{\sc prospector}\xspace}
\newcommand{\emcee}{{\sc emcee}\xspace}
\newcommand{\cloudy}{{\sc cloudy}\xspace}
\newcommand{\pyneb}{{\sc pyneb}\xspace}
\newcommandx{\mappings}[1][1=]{{\sc mappings{#1}}\xspace}
\newcommand{\galfit}{{\sc galfit}\xspace}
\newcommand{\qubespec}{{\sc qubespec}\xspace}
\newcommand{\pysersic}{{\sc pysersic}\xspace}

\newcommand{\blackthunder}{BlackTHUNDER\xspace}
\newcommand{\Mdynvalue}{$\Mdyn = 2.0\pm0.5 \times 10^{11}$~\MSun}

\defcitealias{juodzbalis+2024b}{J24}
\defcitealias{gordon+2003}{G03}

\newcommand{\Lyalpha}{\text{Ly\,\textalpha}\xspace}
\newcommand{\Halpha}{\text{H\,\textalpha}\xspace}
\newcommand{\Hbeta}{\text{H\,\textbeta}\xspace}
\newcommand{\Hgamma}{\text{H\,\textgamma}\xspace}
\newcommand{\Hdelta}{\text{H\,\textdelta}\xspace}
\newcommand{\Paalpha}{\text{Pa\,\textalpha}\xspace}
\newcommand{\Pabeta}{\text{Pa\,\textbeta}\xspace}
\newcommand{\Hepsilon}{\text{H\,\textepsilon}\xspace}

\newcommandx{\permittedEL}[6][1=O,2=III,3=,4=,5=,6=]{\text{{#1}\,{\sc {#2}}{#3}{#4}{#5}{#6}}\xspace}
\newcommandx{\semiforbiddenEL}[6][1=O,2=III,3=,4=,5=,6=]{\text{{#1}\,{\sc{#2}}]{#3}{#4}{#5}{#6}}\xspace}
\newcommandx{\forbiddenEL}[6][1=O,2=III,3=,4=,5=,6=]{\text{[{#1}\,{\sc{#2}}]{#3}{#4}{#5}{#6}}\xspace}

\newcommand{\EW}[1]{\text{EW(#1)}\xspace}

\newcommand{\HI}{\permittedEL[H][i]}
\newcommand{\HII}{\permittedEL[H][ii]}

\newcommand{\NV}{\permittedEL[N][v]}
\newcommandx{\NVL}[1][1=1243]{\permittedEL[N][v][\textlambda][#1]}
\newcommandx{\NVall}{\permittedEL[N][v][\textlambda][\textlambda][1239,][1243]}

\newcommandx{\CIIL}[1][1=232x]{\semiforbiddenEL[C][ii][\textlambda][#1]}
\newcommandx{\CIIall}{\semiforbiddenEL[C][ii][\textlambda][\textlambda][2324--][2329]}

\newcommand{\NIV}{\semiforbiddenEL[N][iv]}
\newcommandx{\NIVL}[1][1=1486]{\semiforbiddenEL[N][iv][\textlambda][#1]}

\newcommand{\CIV}{\permittedEL[C][iv]}
\newcommandx{\CIVL}[1][1=1550]{\permittedEL[C][iv][\textlambda][#1]}
\newcommand{\CIVall}{\permittedEL[C][iv][\textlambda][\textlambda][1548,][1551]}

\newcommand{\HeII}{\permittedEL[He][ii]}
\newcommandx{\HeIIL}[1][1=1640]{\permittedEL[He][ii][\textlambda][#1]}

\newcommand{\semiOIII}{\semiforbiddenEL[O][iii]}
\newcommandx{\semiOIIIL}[1][1=1666]{\semiforbiddenEL[O][iii][\textlambda][#1]}
\newcommand{\semiOIIIall}{\semiforbiddenEL[O][iii][\textlambda][\textlambda][1661,][1666]}

\newcommand{\NIII}{\semiforbiddenEL[N][iii]}
\newcommandx{\NIIIL}[1][1=1750]{\semiforbiddenEL[N][iii][\textlambda][#1]}
\newcommand{\NIIIall}{\semiforbiddenEL[N][iii][\textlambda][\textlambda][1747--][1754]}

\newcommandx{\CIII}{\semiforbiddenEL[C][iii]}
\newcommandx{\CIIIL}[1][1=1909]{\semiforbiddenEL[C][iii][\textlambda][#1]}
\newcommand{\CIIIall}{\semiforbiddenEL[C][iii][\textlambda][\textlambda][1907,][1909]}

\newcommand{\NeIV}{\forbiddenEL[Ne][iv]}
\newcommandx{\NeIVL}[1][1=2424]{\forbiddenEL[Ne][iv][\textlambda][#1]}
\newcommand{\NeIVall}{\forbiddenEL[Ne][iv][\textlambda][\textlambda][2422,][2424]}

\newcommand{\MgII}{\permittedEL[Mg][ii]}
\newcommandx{\MgIIL}[1][1=2803]{\permittedEL[Mg][ii][\textlambda][#1]}
\newcommand{\MgIIall}{\permittedEL[Mg][ii][\textlambda][\textlambda][2796,][2803]}

\newcommand{\NeV}{\forbiddenEL[Ne][v]}
\newcommandx{\NeVL}[1][1=3426]{\forbiddenEL[Ne][v][\textlambda][#1]}
\newcommand{\NeVall}{\forbiddenEL[Ne][v][\textlambda][\textlambda][3346,][3426]}

\newcommand{\OII}{\forbiddenEL[O][ii]}
\newcommandx{\OIIL}[1][1=3726]{\forbiddenEL[O][ii][\textlambda][#1]}
\newcommand{\OIIall}{\forbiddenEL[O][ii][\textlambda][\textlambda][3726,][3729]}

\newcommand{\NeIII}{\forbiddenEL[Ne][iii]}
\newcommandx{\NeIIIL}[1][1=3869]{\forbiddenEL[Ne][iii][\textlambda][#1]}
\newcommand{\NeIIIall}{\forbiddenEL[Ne][iii][\textlambda][\textlambda][3869,][3967]}

\newcommand{\OIII}{\forbiddenEL[O][iii]}
\newcommandx{\OIIIL}[1][1=5007]{\forbiddenEL[O][iii][\textlambda][#1]}
\newcommand{\OIIIall}{\forbiddenEL[O][iii][\textlambda][\textlambda][4959,][5007]}

\newcommandx{\NIL}[1][1=5200]{\forbiddenEL[N][i][\textlambda][#1]}
\newcommand{\NIall}{\forbiddenEL[N][i][\textlambda][\textlambda][5198,][5200]}

\newcommand{\OI}{\forbiddenEL[O][i]}
\newcommandx{\OIL}[1][1=6300]{\forbiddenEL[O][i][\textlambda][#1]}
\newcommand{\OIall}{\forbiddenEL[O][i][\textlambda][\textlambda][6300,][6364]}

\newcommand{\HeI}{\permittedEL[He][i]}
\newcommandx{\HeIL}[1][1=5875]{\permittedEL[He][i][\textlambda][#1]}

\newcommand{\OIres}{\permittedEL[O][i]}
\newcommandx{\OIresL}[1][1=8446]{\permittedEL[O][i][\textlambda][#1]}

\newcommand{\NII}{\forbiddenEL[N][ii]}
\newcommandx{\NIIL}[1][1=6583]{\forbiddenEL[N][ii][\textlambda][#1]}
\newcommand{\NIIall}{\forbiddenEL[N][ii][\textlambda][\textlambda][6548,][6583]}

\newcommand{\SII}{\forbiddenEL[S][ii]}
\newcommandx{\SIIL}[1][1=6716]{\forbiddenEL[S][ii][\textlambda][#1]}
\newcommand{\SIIall}{\forbiddenEL[S][ii][\textlambda][\textlambda][6716,][6731]}

\newcommandx{\OIIAuL}[1][1=7325]{\forbiddenEL[O][ii][\textlambda][#1]}
\newcommand{\OIIAuall}{\forbiddenEL[O][ii][\textlambda][\textlambda][7319--][7331]}

\newcommandx{\CIIFIRL}{\forbiddenEL[C][ii][\textlambda][158\,\mum]}

\newcommand{\hda}{\ensuremath{\mathrm{H\text{\textdelta}_A}}\xspace}
\newcommand{\hga}{\ensuremath{\mathrm{H\text{\textgamma}_A}}\xspace}

\newcommand{\target}{159717\xspace}
\newcommand{\qsoone}{Abell2744-QSO1\xspace}
\newcommand{\sattiny}{159717C\xspace}
\newcommand{\satsmall}{159717B\xspace}
\newcommand{\satlarge}{159716\xspace}
\newcommand{\interlop}{159715\xspace}
\newcommand{\jadesgs}[1]{JADES-GS-{#1}\xspace}
\newcommand{\vabs}{\ensuremath{v_\mathrm{abs}}\xspace}
\newcommand{\sigabs}{\ensuremath{\sigma_\mathrm{abs}}\xspace}
\newcommand{\Avhatn}{\ensuremath{A_{V}}\xspace}
\newcommand{\Avhatb}{\ensuremath{A_{{V,{\rm b}}}}\xspace}
\newcommand{\fwhm}{\ensuremath{FWHM}\xspace}
\newcommand\sbullet[1][.5]{\mathbin{\vcenter{\hbox{\scalebox{#1}{$\bullet$}}}}}
\newcommand{\mbh}{\ensuremath{M_{\sbullet[0.85]}}\xspace}
\newcommand{\Lbol}{\ensuremath{L_\mathrm{bol}}\xspace}
\newcommand{\dBIC}{\text{\textDelta BIC}\xspace}
\newcommand{\ledd}{\ensuremath{\lambda_\mathrm{E}}\xspace}
\newcommand{\ergs}{\text{erg\,s\ensuremath{^{-1}}}\xspace}
\newcommand{\mulens}{\ensuremath{\mu_\mathrm{lens}}\xspace}

\title[A rest-frame gas absorber in an AGN at z=5]{JADES and BlackTHUNDER: rest-frame Balmer-line absorption and the local environment in a Little Red Dot at $z=5$}

\author[\sendemail{francesco.deugenio@gmail.com}{Questions about your JADES paper about the rest-frame absorber.}{!Hola Francesco!\%0A\%0Ahow are you doing? I have a question about this paper, do you mind? (By the way, the paper is a bit too long and even tedious, at times).\%0ANow back to the question, ...\%0A\%0ARegards,\%0A}{F. D'Eugenio}~et al.]{\parbox{\textwidth}{
\orcidsymb{Francesco D'Eugenio}{0000-0003-2388-8172}$^{\hyperlink{aff1}{1},\hyperlink{aff2}{2}}$\thanks{E-mail: francesco.deugenio@gmail.com},
\orcidsymb{Ignas Juod{\v z}balis}{0009-0003-7423-8660}$^{\hyperlink{aff1}{1},\hyperlink{aff2}{2}}$,
\orcidsymb{Xihan Ji}{0000-0002-1660-9502}$^{\hyperlink{aff1}{1},\hyperlink{aff2}{2}}$,
\orcidsymb{Jan Scholtz}{0000-0001-6010-6809}$^{\hyperlink{aff1}{1},\hyperlink{aff2}{2}}$,
\orcidsymb{Roberto Maiolino}{0000-0002-4985-3819}$^{\hyperlink{aff1}{1},\hyperlink{aff2}{2},\hyperlink{aff3}{3}}$,
\orcidsymb{Stefano Carniani}{0000-0002-6719-380X}$^{\hyperlink{aff4}{4}}$,
\orcidsymb{Michele Perna}{0000-0002-0362-5941}$^{\hyperlink{aff5}{5}}$,
\orcidsymb{Giovanni Mazzolari}{0009-0005-7383-6655}$^{\hyperlink{aff6}{6},\hyperlink{aff7}{7},\hyperlink{aff8}{8}}$,
\orcidsymb{Hannah \"Ubler}{0000-0003-4891-0794}$^{\hyperlink{aff6}{6}}$,
\orcidsymb{Santiago Arribas}{0000-0001-7997-1640}$^{\hyperlink{aff5}{5}}$,
\orcidsymb{Rachana Bhatawdekar}{0000-0003-0883-2226}$^{\hyperlink{aff9}{9}}$,
\orcidsymb{Andrew J. Bunker}{0000-0002-8651-9879}$^{\hyperlink{aff10}{10}}$,
\orcidsymb{Giovanni Cresci}{0000-0002-5281-1417}$^{\hyperlink{aff11}{11}}$,
\orcidsymb{Emma Curtis-Lake}{0000-0002-9551-0534}$^{\hyperlink{aff12}{12}}$,
\orcidsymb{Kevin Hainline}{ 0000-0003-4565-8239}$^{\hyperlink{aff13}{13}}$,
\orcidsymb{Kohei Inayoshi}{0000-0001-9840-4959}$^{\hyperlink{aff14}{14}}$,
\orcidsymb{Yuki Isobe}{0000-0001-7730-8634}$^{\hyperlink{aff1}{1},\hyperlink{aff2}{2}}$,
\orcidsymb{Zhiyuan Ji}{0000-0001-7673-2257}$^{\hyperlink{aff13}{13}}$,
\orcidsymb{Benjamin D.~Johnson}{0000-0002-9280-7594}$^{\hyperlink{aff15}{15}}$,
\orcidsymb{Gareth C.~Jones}{0000-0002-0267-9024}$^{\hyperlink{aff1}{1},\hyperlink{aff2}{2}}$,
\orcidsymb{Tobias J.~Looser}{0000-0002-3642-2446}$^{\hyperlink{aff15}{15}}$,
\orcidsymb{Erica J. Nelson}{0000-0002-7524-374X}$^{\hyperlink{aff16}{16}}$,
\orcidsymb{Eleonora Parlanti}{0000-0002-7392-7814}$^{\hyperlink{aff4}{4},\hyperlink{aff6}{6}}$,
\orcidsymb{D\'avid Pusk\'as}{0000-0001-8630-2031}$^{\hyperlink{aff1}{1},\hyperlink{aff2}{2}}$,
\orcidsymb{Pierluigi Rinaldi}{0000-0002-5104-8245}$^{\hyperlink{aff13}{13}}$,
\orcidsymb{Brant Robertson}{0000-0002-4271-0364}$^{\hyperlink{aff17}{17}}$,
\orcidsymb{Bruno Rodr\'iguez Del~Pino}{0000-0001-5171-3930}$^{\hyperlink{aff5}{5}}$,
\orcidsymb{Irene Shivaei}{0000-0003-4702-7561}$^{\hyperlink{aff5}{5}}$,
\orcidsymb{Fengwu Sun}{0000-0002-4622-6617}$^{\hyperlink{aff15}{15}}$,
\orcidsymb{Sandro Tacchella}{0000-0002-8224-4505}$^{\hyperlink{aff1}{1},\hyperlink{aff2}{2}}$,
\orcidsymb{Giacomo Venturi}{0000-0001-8349-3055}$^{\hyperlink{aff4}{4}}$,
\orcidsymb{Marta Volonteri}{0000-0002-3216-1322}$^{\hyperlink{aff18}{18}}$,
\orcidsymb{Christina C.\ Williams}{0000-0003-2919-7495}$^{\hyperlink{aff19}{19}}$,
\orcidsymb{Christopher N.~A. Willmer}{0000-0001-9262-9997}$^{\hyperlink{aff13}{13}}$,
\orcidsymb{Chris~Willott}{0000-0002-4201-7367}$^{\hyperlink{aff20}{20}}$
and 
\orcidsymb{Joris Witstok}{0000-0002-7595-121X}$^{\hyperlink{aff21}{21},\hyperlink{aff22}{22}}$
}\vspace{0.4cm}
\\
\parbox{\textwidth}{
\hypertarget{aff1}{$^{1}$}Kavli Institute for Cosmology, University of Cambridge, Madingley Road, Cambridge, CB3 0HA, United Kingdom\\
\hypertarget{aff2}{$^{2}$}Cavendish Laboratory - Astrophysics Group, University of Cambridge, 19 JJ Thomson Avenue, Cambridge, CB3 0HE, United Kingdom\\
\hypertarget{aff3}{$^{3}$}Department of Physics and Astronomy, University College London, Gower Street, London WC1E 6BT, UK\\
\hypertarget{aff4}{$^{4}$}Scuola Normale Superiore, Piazza dei Cavalieri 7, I-56126 Pisa, Italy\\
\hypertarget{aff5}{$^{5}$}Centro de Astrobiolog\'ia (CAB), CSIC–INTA, Cra. de Ajalvir Km.~4, 28850 - Torrej\'on de Ardoz, Madrid, Spain\\
\hypertarget{aff6}{$^{6}$}Max-Planck-Institut f\"ur extraterrestrische Physik (MPE), Gie{\ss}enbachstra{\ss}e 1, 85748 Garching, Germany\\
\hypertarget{aff7}{$^{7}$}Dipartimento di Fisica e Astronomia, Universit\`a di Bologna, Via Gobetti 93/2, I-40129 Bologna, Italy\\
\hypertarget{aff8}{$^{8}$}INAF – Osservatorio di Astrofisica e Scienza dello Spazio di Bologna, Via Gobetti 93/3, I-40129 Bologna, Italy\\
\hypertarget{aff9}{$^{9}$}European Space Agency (ESA), European Space Astronomy Centre (ESAC), Camino Bajo del Castillo s/n, 28692 Villanueva de la Cañada, Madrid, Spain\\
\hypertarget{aff10}{$^{10}$}Department of Physics, University of Oxford, Denys Wilkinson Building, Keble Road, Oxford OX1 3RH, UK\\
\textit{\normalsize Remaining affiliations are listed at the end of the paper}
}
}

\date{Accepted 17 Nov. 2025. Received 24 Nov. 2025; in original form 19 Jun. 2025}

\pubyear{\the\year{}}

\begin{document}
\label{firstpage}
\pagerange{\pageref{firstpage}--\pageref{lastpage}}
\maketitle

\begin{abstract}
We present a broad-line active galactic nucleus (AGN)
at $z=5.077$, observed with both NIRSpec/MSA and NIRSpec/IFU by the JADES and \blackthunder
surveys. The target exhibits all the hallmark features of a `Little Red Dot' (LRD) AGN. The combination of spatially resolved and high-resolution spectroscopy offers deeper insight into its nature. The \Halpha line has multiple components,
including two broad Gaussians, yielding a black-hole mass of
$\log (\mbh/\Msun) = 7.65$, while the narrow \OIIIL gives a galaxy dynamical mass
of $\log (\mdyn/\Msun) = 9.1$, suggesting a dynamically overmassive
black hole relative to the host galaxy. The target is immersed in a 7-kpc wide pool of ionized gas and has three neighbours: a satellite galaxy, a possible satellite/gas cloud, and a tentatively detected spatially detached outflow. \Halpha shows strong absorption, deeper than the continuum, thus ruling out
a stellar origin, and with velocity and velocity dispersion of $\vabs = -13~\kms$ and
$\sigabs = 120~\kms$. There is tentative evidence (2.6~\textsigma) of temporal variability
in the EW of the \Halpha absorber over two
rest-frame months. If confirmed, this would suggest a highly dynamic
environment. Notably, while the \Halpha absorber is clearly visible and even dominant in the
high-resolution G395H observations, it is not detected in the medium-resolution G395M data of the same epoch. This implies that the current incidence rate of absorbers in LRDs -- and especially of rest-frame absorbers --
may be severely underestimated, because most LRDs rely on lower-resolution spectroscopy.
In this context, the high incidence rate of rest-frame absorbers
in LRDs may indicate a configuration that is either intrinsically stationary, such as a
rotating disc, or that exhibits time-averaged stability, such as an oscillatory `breathing mode' accretion of cyclic expansion and contraction of the gas around the SMBH.
\end{abstract}


%%%%%%%%%%%%%%%%%%%%%%%%%%%%%%%%%%%%%%%%%%%%%%%%%%

%%%%%%%%%%%%%%%%% BODY OF PAPER %%%%%%%%%%%%%%%%%%

\section{Introduction}

In the standard \textLambda CDM cosmology, a key yet poorly understood agent in shaping
galaxies is feedback from accreting supermassive black holes (SMBHs), which manifests
via episodes of Active Galactic Nuclei (AGNs).
In the past few years, SMBH feedback has been invoked mainly to explain the
low stellar-to-total mass fraction at dark-matter halo masses higher than 
$\sim10^{12}~\Msun$ \citep{moster+2010,behroozi+2010}. However, more recently, 
AGNs are being revisited as drivers in the low-mass range too. This paradigm shift
stems from the convergence of new observational evidence \citep{greene+2004,greene+2006,schramm+2013,satyapal+2014,greene+2020}
and of recent theoretical progress \citep{silk2017,koudmani+2022}.

Observationally, the pivot of this revolution in our view of AGNs has been the discovery
of a large population of low-luminosity AGN ($\Lbol\lesssim10^{45}~\mathrm{erg\,s^{-1}}$) 
at redshifts $z=2\text{--}9$, thanks to the superior sensitivity of \jwst.
These objects are primarily identified via their broad permitted line emission, with no 
matching counterpart in the forbidden lines \citep{kocevski+2023,ubler+2023}. In 
addition, a complementary population of narrow-line AGN is also thought to exist
\citep{scholtz+2023}, but these objects have received far less attention, because 
identifying these hidden AGNs is significantly more challenging than for their broad-line
counterparts.
Due to a series of unfavourable coincidences, the optical tracers 
of narrow-line AGNs used at $z\lesssim3$ and for more massive AGNs,
\citep[such as the BPT diagram;][]{baldwin+1981} are inapplicable. In particular,
low metallicity and lack of an extended low-ionization zone conspire to cluster both 
AGN and star-forming galaxies in the same region of the BPT diagram. As a result, 
these standard diagrams have been shown to fail to identify even
secure AGN, such as AGN with broad permitted lines \citep[e.g.,][]{kocevski+2023,ubler+2023,juodzbalis+2025}.

Nevertheless, even the broad-line population alone presents new challenges to
our view of AGN, with the discovery of their X-ray and radio weakness, relative
to the expectations derived from more luminous AGN. This weakness could be due
to high covering factors of neutral and/or ionized gas \citetext{\citealp{juodzbalis+2024b}, hereafter: \citetalias{juodzbalis+2024b};
\citealp{rusakov+2025}}, or super-Eddington accretion
rates $\lambda_\mathrm{E}>1$ \citep{pacucci+2024,lambrides+2024}. Modelling the spectral energy 
distribution (SED) of these AGN has led to conflicting claims of `over-massive' 
SMBHs \cite[relative to the host stellar mass \mstar and to local scaling 
relations;][]{maiolino+2024,harikane+2023,juodzbalis+2024a}, or of normal SMBH--\mstar scaling 
\citep{sun+2024}, of possibly extremely massive host galaxies 
\citep{wang+2024a,wang+2024b}, or even of no SMBHs at all
\citep{baggen+2024,kokubo+2024}. Part of this confusion may be due to the use of
the same terminology for different sample selections, so a clarification of
our subjective language choice may be helpful.

Several low-luminosity, spectroscopically identified broad-line AGN appear red
and compact in NIRCam imaging \citep[`Little Red Dots', LRDs;][]{matthee+2024}. Many of these broad-line AGN have `v'-shaped SEDs \citetext{\citealp{furtak+2024,wang+2024a,wang+2024b}; \citetalias{juodzbalis+2024b}}, which has led to searching for similar objects
in photometry \citep{greene+2024,kocevski+2024}. However, many of these 
photometrically-identified LRDs appear consistent with stellar-dominated SEDs 
\citep{williams+2024,perez-gonzalez+2024}. Conversely, out of the population of spectroscopically confirmed broad-line AGN, the fraction of
LRDs is fairly low \citetext{20--30~percent; \citet{hainline+2024,taylor+2024}}.
In light of the above, in this work we aim to study and discuss only broad-line
AGN, with no bearing on photometrically identified samples.

Among low-luminosity, broad-line AGN, there is a high detection rate of 
Balmer-series absorption \citep{matthee+2024}, with velocities ranging 
from -340 to +50~\kms. The equivalent width (EW) of this absorption is 
generally too high to be explained by stellar atmospheric absorption
\citep{matthee+2024,juodzbalis+2024b,deugenio+2025c}, implying the 
existence of dense absorbing gas clouds near the broad-line region 
\citetext{\citealp{matthee+2024,wang+2024a}; \citetalias{juodzbalis+2024b}}.
The high detection rate of these absorbing clouds provides essential 
information
for deciphering the structure of this new population of AGN. In particular,
a high covering factor could explain the X-ray weakness of these sources  
\citetext{\citealp{wang+2024a}; \citetalias{juodzbalis+2024b}}. Moreover, the
physical conditions, kinematics and chemical composition of these clouds may
help us understand the balance between SMBH fuelling and outflows in this 
unexplored regime.

In this work, we present \jwst/NIRCam and NIRSpec observations of a
broad-line AGN at $z=5$ with rest-frame Balmer-line absorption (Section~\ref{s.data}). This galaxy was initially discovered by JADES 
\citep{eisenstein+2023a} and then re-observed by \blackthunder one year 
later. We present a full analysis of the images and integrated spectroscopy
(Section~\ref{s.an}), and present the resulting physical properties in
Section~\ref{s.phys}; among these, we identify a rest-frame \Halpha absorber, with tentative evidence of EW variation.
In Section~\ref{s.sats} we present the surrounding environment. We conclude
with a discussion of our findings (Section~\ref{s.disc}) and with a brief
summary (Section~\ref{s.conc}).

Throughout this work, we assume a flat \textLambda CDM cosmology with $H_0 = 67.4$~\kms~Mpc$^{-1}$ and $\Omega_\mathrm{m}=0.315$ \citep{planck+2020}, giving a physical scale of 6.37~kpc~arcsec$^{-1}$ at redshift $z=5.08$ (all physical scales are given as proper quantities). Stellar masses are total stellar mass formed, assuming a \citet{chabrier2003} initial mass function, integrated between 0.1 and
120~\Msun. All magnitudes are in the AB system \citep{oke+gunn1983} and all EWs are in the rest frame, with negative EW corresponding to line emission.

\section{Data}\label{s.data}

\subsection{Target galaxy \target}\label{s.data.ss.target}

Galaxy JADES-GS+033223.41-275404.5 at $z=5.07781\pm0.00003$ (hereafter, \target)
is located in the GOODS-South cosmological deep field \citep{giavalisco+2004}.
It was first reported as a $V$-band drop-out, GSWV~2234354045, with a photometric
redshift $z_\mathrm{phot}=5.3$ \citep{bouwens+2015}.
It was selected for follow-up spectroscopy as part of the \jwst Advanced Deep 
Extragalactic Survey \citep[JADES;][]{bunker+2020,rieke+2020,eisenstein+2023a},
as a relatively bright  NIRCam source (F444W$<27$~mag) with photometric
redshift $4.5<z_\mathrm{phot}<5.7$ \citetext{\citealp{hainline+2024}; priority
7.5 in table~4 of \citealp{deugenio+2025a}}. JADES spectroscopy identified this
galaxy as a \Lyalpha emitter \citep{jones+2024}. Subsequently, the source was
also targeted by \blackthunder (Section~\ref{s.data.ss.blackthunder}).
As a result, this galaxy is one of few such systems with deep, multi-epoch
\jwst/NIRSpec spectroscopy to date \citep[in addition to
Abell~2744-QSO1;][]{ji+2025,furtak+2025}.

\subsection{JADES Data}\label{s.data.ss.jades}

\begin{figure*}
  \includegraphics[width=\textwidth]{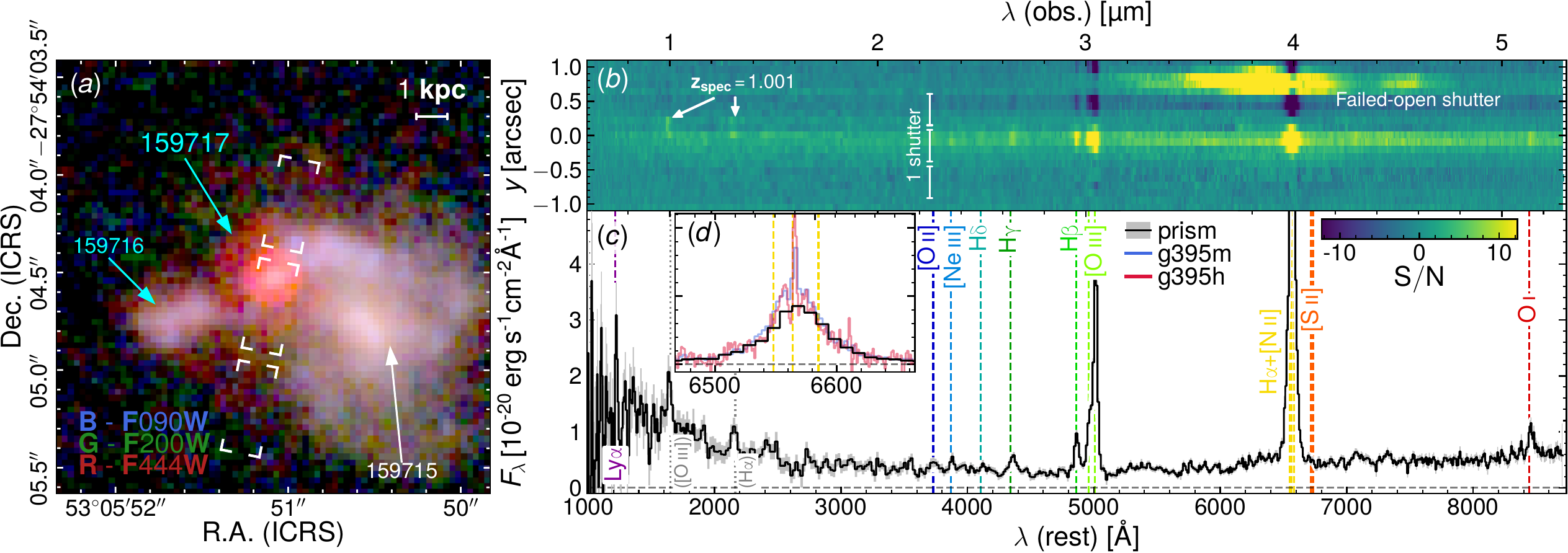}
  {\phantomsubcaption\label{f.data.a}
   \phantomsubcaption\label{f.data.b}
   \phantomsubcaption\label{f.data.c}
   \phantomsubcaption\label{f.data.d}}
  \caption{Summary of the JADES observations. Panel~\subref{f.data.a}; RGB NIRCam
  image, highlighting the target AGN at $z=5.077$ (red hues), a satellite to the
  East, and a foreground spiral galaxy. The white corners trace the vertices of
  the NIRSpec/MSA microshutters. Panel~\subref{f.data.b}; 2-d signal-to-noise
  map from NIRSpec/MSA prism. Balmer and high-ionization lines from the AGN are
  clearly visible; in the rest-frame UV, there are several faint signatures of
  the blended complexes \Hbeta and \OIIIL, and \Halpha and \NIIL from the
  foreground star-forming regions. A relatively strong line at $\lambda=5.14$~\mum
  is identified as the resonant \OIresL. Panel~\subref{f.data.c}; 3-pixel
  (0.3-arcsec) extraction prism spectrum centred on the AGN. Foreground emission-line
  contamination in the rest-frame UV is marked by vertical dotted lines. Note the red
  optical spectrum (i.e., positive slope), and the large Balmer decrement
  \Halpha/\Hbeta, typical of LRDs.
  The inset Panel~\subref{f.data.d} shows a zoom in around \Halpha, with the G395M
  and G395H spectra in blue and pink, respectively.}\label{f.data}
\end{figure*}

We use \jwst/NIRCam imaging from JADES programme IDs PID~1180, 1210 and~1286
\citep{rieke+2023,eisenstein+2023a} and from FRESCO \citep[the First Reionization
Epoch Spectroscopic COmplete Survey, PID~1895;][]{oesch+2023}, in combination
with legacy \hst/ACS and WFC3 imaging 
from the Hubble Legacy Field data \citep{whitaker+2019}. The NIRCam data
reduction from JADES has been presented in the public data release (DR) articles
\citep{rieke+2023,eisenstein+2023b,deugenio+2025a}.
We also use NIRSpec \citep{jakobsen+2022} Micro-Shutter Assembly spectroscopy
\citep[MSA;][]{ferruit+2022}, obtained as part of PID~1286. These data were 
selected and observed in the `medium jwst gs' tier of JADES 
\citep{eisenstein+2023a}. An RGB false-colour image of the target, with the
MSA shutters overlaid is shown in Fig.~\ref{f.data.a}, with a bright interloper
in the foreground. The MSA observations
consist of 8009 s with prism, G140M and G395H, and 9322 s with G235M and G395M.
These integration times were split into two times three nodded exposures, using the
NRSIRS2 readout mode \citep{rauscher+2012,rauscher+2017}. We used the nodded
exposures to subtract the background, leading to possible foreground-source subtraction
due to the contaminant. For the prism, we use only half the nodded pairs, to
avoid subtracting the light from a `disobedient' open shutter, which contains a bright feature (possibly PAH emission; visible in the top 0.5 arcsec of Fig.~\ref{f.data.b}). Apart from this
tailored setting, the data reduction process is the same as described in the JADES DR1
and~DR3 articles \citep{bunker+2024,deugenio+2025a}. We used point-source path-loss
corrections appropriate for an unresolved source.
Due to the presence of multiple nearby sources, it is essential to separate
physically associated systems from foreground and background systems.
Where spectra are not available, we use photometric redshifts $z_\mathrm{phot}$ 
from \textsc{eazy} \citep{brammer+2008}, using photometry from JADES~DR2 
\citep{eisenstein+2023b} and the methods outlined
in \citet{hainline+2024}.
The JADES DR1 and DR3 NIRSpec data reduction has a 15~percent systematic offset
between the gratings and prism fluxes \citep{bunker+2024,deugenio+2025a}. Since the
prism flux calibration is in agreement with NIRCam \citep{bunker+2023}, we downscale
all the grating spectra from JADES by a factor of 0.85. This brings the \Halpha
flux measured from the gratings in agreement with both the prism value and with
the value from the \blackthunder G395H spectrum, which uses a more recent
calibration (Section~\ref{s.data.ss.blackthunder}).

NIRCam shows a crowded field (Fig.~\ref{f.data.a}; see also Fig.~\ref{f.ifsimg}), with a satellite
JADES-GS+033223.44-275404.7 (hereafter: ID~159716) at photometric redshift
$z_\mathrm{phot}=5.22\pm 0.08$
($z_\mathrm{spec}=5.07622\pm0.00002$; Section~\ref{s.sats.ss.apspec}),
and behind the foreground spiral galaxy J033223.38-275404.6 (CANDELS~ID GS-1826;
JADES~ID \jadesgs{\interlop}; hereafter: \interlop) at
$z_\mathrm{phot}=1.18\pm 0.06$ ($z_\mathrm{spec}=1.00115\pm0.00001$; Appendix~\ref{a.interl}).
The presence of this interloper complicates significantly our analysis; most 
relevant to this work, we estimate that the additional foreground dust 
attenuation correction for the AGN is a secondary effect 
(Section~\ref{s.an.ss.dust}) and there is no significant magnification from
gravitational lensing (Appendix~\ref{s.lens}).

The prism spectrum shows the characteristic `v'-shaped inflection of some LRDs, 
combining a blue UV slope $\beta_\mathrm{UV}=-2$ with a red rest-frame optical 
continuum, with the inflection point near the Balmer limit \citep{setton+2024}.
We lack MIRI coverage to explore if the SED of our target displays
the rest-frame 1.6-\mum turnover typical of stellar-dominated spectra
\citep{williams+2024,perez-gonzalez+2024}.

The rest-UV spectrum shows tentative evidence of \Lyalpha emission \citep[][but our spatially resolved data suggests this may be \OIIall emission at $z=1$ from \interlop; Appendix~\ref{a.interl}]{jones+2024}, and 
a possible 2-\textgamma\ or damped \Lyalpha absorption continuum. These features 
have been associated with AGN in the past \citetext{\citealp{tacchella+2024,wu+2024,li+2024}, but see \citealp{cameron+2024,terp+2024} for different interpretations}.
The rest-frame UV spectrum also shows evidence for two lines at 
$\lambda = 1.05$ and 1.36~\mum, present in both the prism and medium-resolution G140M data.
The 1.05-\mum line is spectrally resolved into a doublet; these are therefore the \OIIIall
doublet and \Halpha from the northern spiral arm of \interlop at $z_\mathrm{spec}=1.00115$.
At this redshift, and due to the low resolution of the prism, \Lyalpha could also correspond
to \OIIall in the foreground. Similarly, a tentative line
at 1.16~\mum matches very well both \CIIIL in the AGN but also \HeIL[5875] at
$z=1$.

The rest-frame optical spectrum shows both Balmer and metal-line emission. The latter include \OIIall, \NeIIIL, \OIIIall, and the fluorescent \OIresL. The Balmer emission is broad, as highlighted also in Fig.~\ref{f.data.d}, where we show \Halpha as observed in the medium and high-resolution gratings. The width of \Halpha exceeds a full-width half-maximum of $FWHM = 1,500~\kms$; coupled with the narrow emission from the forbidden \OIIIall lines (dispersion $\sigma \sim 50~\kms$), this is strong evidence for an AGN broad-line region. This explanation resonates with the detection of the \OIresL line, a high-energy resonant line which is very faint in star-forming galaxies \citep{strom+2023}, but is common in AGN hosts \citep{rudy+1989,juodzbalis+2024b}.
The Balmer emission is dominated by the broad component, but a narrow component is clearly visible. The high-resolution spectrum shows clear evidence for \Halpha absorption too (Section~\ref{s.an}).

\subsection{\blackthunder Data}\label{s.data.ss.blackthunder}

\blackthunder (Black holes in THe early Universe aNd their DensE
surRoundings) is a \jwst/NIRSpec programme (PID~5015; PIs H.~\"Ubler and
R.~Maiolino) that targets 20 low-luminosity broad-line AGN using NIRSpec
integral-field spectroscopy \citep[IFS;][]{boker+2022}.
The sample consists of broad-line AGN at $z>5$, so \target was included for
follow-up spectroscopy after being first identified in JADES. The observing
setup has two visits, using the prism and the G395H grating,
consisting of 14 dithered integrations each.
For the prism, we used the NRSIRS2RAPID readout mode, with 37 groups per
integration and a single integration per dither, totalling 7,761 seconds. For the
grating, we used NRSIRS2 and 23 groups per integration, totalling 23,692
seconds. The observations were strongly affected by
anomalous proton flux, resulting in a higher rate of artefacts.

We processed the raw files using the \jwst Science Calibration pipeline
\textsc{jwst} \citep{alvesdeoliveira+2018}, version 1.15.0, with the calibration files specified by the
Calibration Reference Data System (CRDS) context file number 1281. Additional processing steps were performed, following the procedure developed
by \citet{perna+2023}. Residual pink noise was corrected using a polynomial fit.
We manually masked regions affected by open shutters from the NIRSpec
micro-shutter assembly \citep{ferruit+2022}, and by strong cosmic rays.
Remaining outliers were flagged in individual exposures using the Laplacian
edge detection algorithm \citep{vandokkum2001}, as implemented by
\citet{deugenio+2024a}. From the reduced 2-d frames, we created the rectified
datacubes using the `drizzle' algorithm and an output grid with 0.05-arcsec
spaxels.

The background was subtracted by creating a white image and an emission-line
image, obtained respectively by taking the median across all wavelengths,
and across a narrow wavelength window centred on \OIIIL. We ran
\textsc{sextractor} \citep{bertin+arnouts1996} to create a segmentation
map for each of the two images, we then padded these segmentation maps by two
spaxels, and finally defined the source mask as the union of the two
padded segmentation maps.
We estimate the background for each wavelength pixel using the background
algorithm from the \textsc{astropy} package \citep{astropy+2013,astropy+2018},
with a 5$\times$5 filtering window, and with spatial interpolation across the 
windows and across the source mask. The resulting background datacube was
smoothed in wavelength using median filtering and a window of 25 pixels.
The quality of the resulting background-subtracted cube was assessed by
taking random apertures (outside of the source mask) and verifying that
their flux was consistent with zero.

\subsection{Field characterisation and redshift determination}\label{s.data.ss.redshifts}

\begin{figure}
    \centering
    \includegraphics[width=\columnwidth]{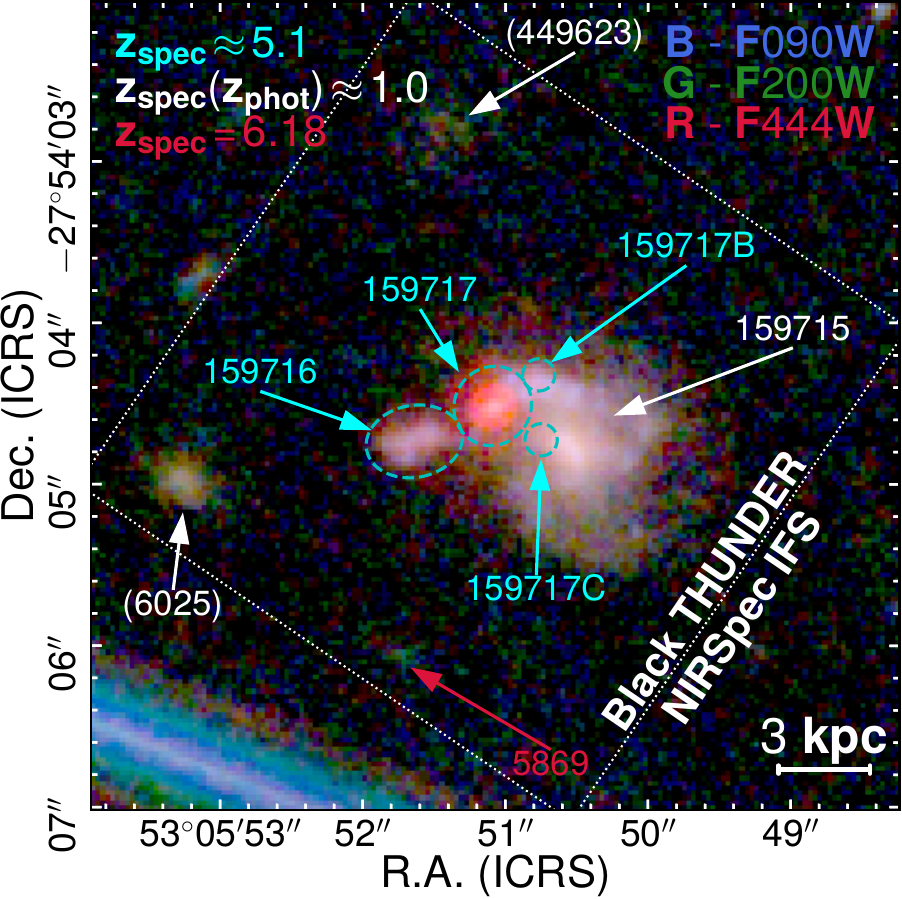}
    \caption{The NIRSpec/IFS field of view overlaid on false-colour NIRCam image.
    Dashed ellipses and circles are the apertures used to extract the spectra of
    \target and physically associated galaxies, \satlarge and two more
    emission-line detections (\satsmall and \sattiny) which are not visible in this image.
    Other clearly detected sources in the field of view are indicated. Names in
    cyan are physically associated with \target; names in white are in the foreground,
    (with names in parentheses indicating sources without a spectroscopic redshift). The
    galaxy in red is in the background. Galaxy names that do not end in a letter are
    from the JADES catalogue v1.0.
    }\label{f.ifsimg}
\end{figure}

The IFS field of view is illustrated in Fig.~\ref{f.ifsimg}, where we also indicate foreground and background sources detected from photometry.
The known interloper \interlop has already been discussed (Section~\ref{s.data.ss.jades}),
but \blackthunder high-resolution spectroscopy detects \Paalpha in the northern spiral
arm, enabling a precise redshift measurement (Appendix~\ref{a.interl}).
The two galaxies \jadesgs{449623} and \jadesgs{6025} are identified in the JADES catalogue v1.0 (JADES Collaboration, in prep.);
they have photometric redshift $\approx1$ \citep{hainline+2024}, but we are unable to provide
spectroscopic confirmation, because no emission lines are detected by \blackthunder.
We confirm spectroscopically \jadesgs{5869}, via detection of \OIIIall and \Halpha in the
prism spectrum, yielding $z = 6.190\pm0.001$. We also
indicate the locations of \satsmall, a much fainter emission-line source at the same redshift as \target, and
yet another emission-line source (\sattiny); these two
objects are not visible in
Fig.~\ref{f.ifsimg}, but are discussed in
Section~\ref{s.sats}.

\subsection{\blackthunder aperture spectra}\label{s.data.ss.apspec}

To study the main target \target, we define two aperture spectra: a `total' aperture of semi-major axis $R_\mathrm{ap}=0.25$~arcsec (1.6~kpc; to capture the total flux of the galaxy) and a smaller aperture with
$R_\mathrm{ap}=0.125$~arcsec (0.8~kpc), to maximise the signal-to-noise ratio (SNR) at the expense of
inaccurate aperture losses, due to the resolved nature of the target in \OIIIL.
To define these apertures, we use a curve-of-growth approach. We create an image 
of the broad-line \Halpha by co-adding the cube along the wavelength slices in the wings 
of the broad line, at a wavelength of 4~\mum (see Section~\ref{s.an.ss.g395h.bt}). We then model the resulting image as a
Gaussian light profile plus linear background, using \galfit \citep{peng+2002,peng+2010}, with the 
assumption that the source is unresolved in broad \Halpha, and that this emission line traces the instrument point spread
function (PSF).
The best-fit Gaussian model has axis ratio 0.9 and position angle -25.4\textdegree. The 
PSF is elongated in the direction along the IFU slices, in agreement with \citet{deugenio+2024a},
but here we find a considerably larger full-width at half maximum than in
\citet{deugenio+2024a}, with $FWHM=0.175$~arcsec along the
slicers. This larger PSF is in agreement with independent measurements \citetext{the bright quasar J0224-4711, Perna M., in~prep.; \citealp{jones+2025c,zamora+2024}}.
The disagreement is due to \citet{deugenio+2024a} using a less accurate method
than here, based on forward modelling an extended target starting from the NIRCam images.

We create a set of aperture spectra by summing the light  inside elliptical apertures with the same shape and centre as the best-fit Gaussian model, with increasing semi-major axis in steps of one 0.05-arcsec spaxel, starting from one and reaching ten spaxels.
For each wavelength pixel of the aperture spectrum, the flux is determined by adding
all spaxels from the corresponding wavelength slice, weighting each spaxel only by the fractional area inside the current elliptical aperture. This approach does not benefit from the increased precision of inverse variance, but avoids biasing the result. To remove outliers, we use 
4-\textsigma clipping for each wavelength slice: we first divide the current slice 
by the PSF (point spread function), then we calculate the median and define \textsigma as half the 84\textsuperscript{th}-16\textsuperscript{th} inter-percentile range of the data. Finally,
we assign weight zero to all voxels deviating more than 4 \textsigma from the median. The uncertainties are estimated by repeating the same procedure on the variance datacube. The resulting error spectrum is upscaled to match the empirical noise observed in the aperture spectrum. To do so, we estimate the effective noise on the aperture spectrum by calculating the standard deviation about the median-filtered spectrum inside a moving window, neglecting emission-line regions. We then upscale the original error spectrum to match the median value of the empirical error spectrum.
This preserves wavelength-specific noise features like photon noise around bright emission lines and under-exposed pixels.

\begin{figure}
  \includegraphics[width=\columnwidth]{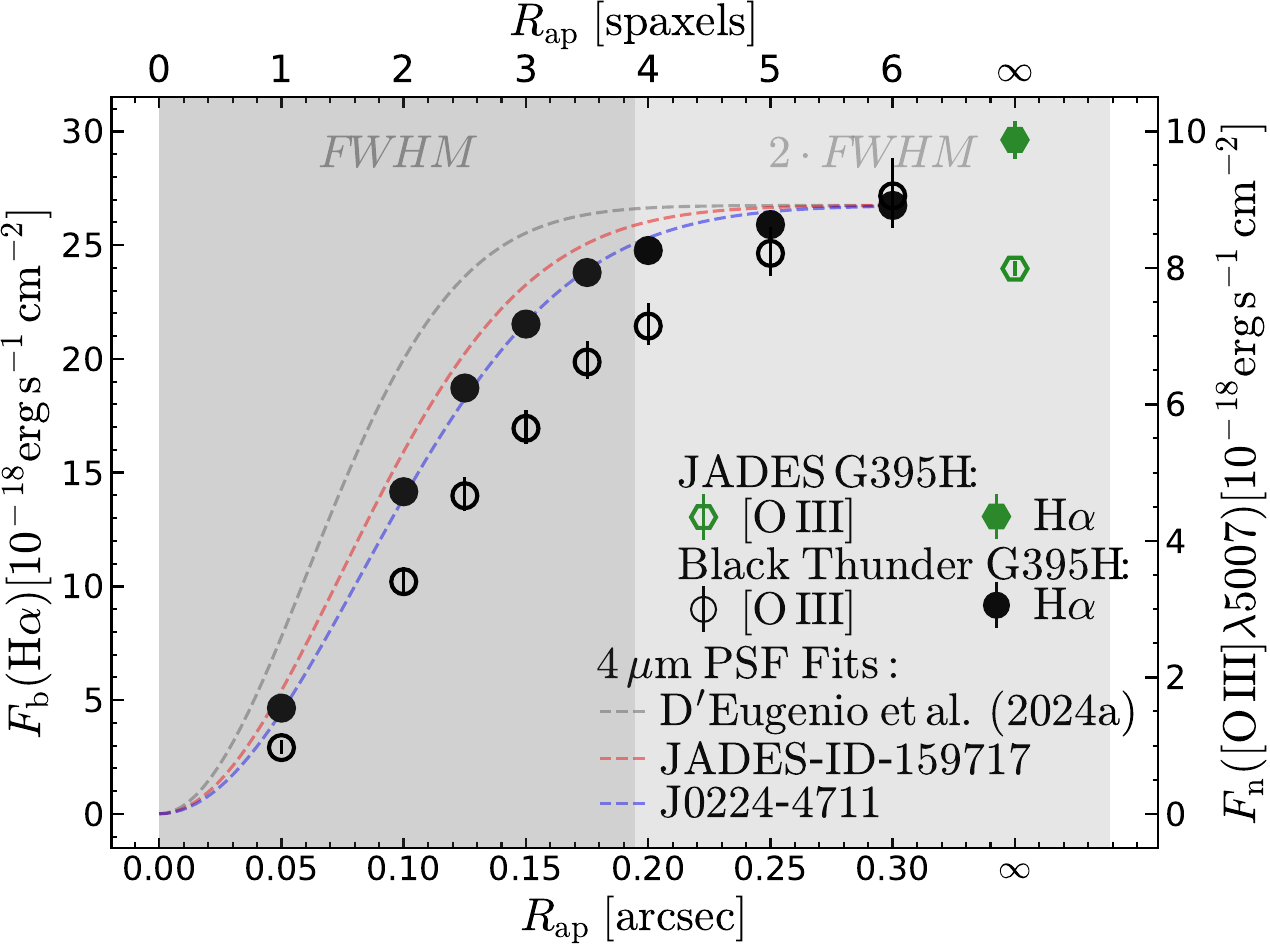}
  \caption{Curve of growth of emission lines from \blackthunder (black circles),
  compared to the JADES aperture-corrected values (green). The empty symbols
  trace narrow \OIIIL flux (right axis), the filled symbols are broad
  \Halpha (left axis). The \Halpha fluxes are corrected for $n=2$ hydrogen absorption
  (Sections~\ref{s.an.ss.fitting}--\ref{s.an.ss.g395h.jades}). The dashed lines are
  curves of growth of Gaussians of varying FWHM. Broad \Halpha closely follows
  a Gaussian profile, while \OIIIL displays a less steep slope which implies a
  spatially resolved nature. Our fiducial aperture has $R_\mathrm{ap} = 0.25$ arcsec,
  so it does not require large aperture corrections. We also use a smaller aperture
  to maximise the SNR, but this smaller aperture, with point-source aperture corrections,
  may underestimate the flux of \OIIIL, which is not point-like, unlike broad \Halpha.
  The grey curve is
  estimated
  from an extended source; the blue curve is estimated
  from the broad \Halpha emission line of the QSO J0224-4711.
  }\label{f.psf}
\end{figure}

Our fiducial aperture to study \target has semi-major axis of 0.25~arcsec, which encloses 
$\sim90$~percent of the PSF flux at the wavelength of \Halpha
(cyan dashed circle in Fig.~\ref{f.ifsimg}; see also \citealp{zamora+2024}). We also consider a smaller aperture of
radius 0.125-arcsec; this second aperture maximises the SNR of the spectrum, but we apply an
aperture-correction factor of 1.22 to capture the total \Halpha flux.
This accuracy of aperture choice is validated in Fig.~\ref{f.psf}, where we show the
emission-line fluxes from narrow \OIIIL and from broad \Halpha as a function of aperture
radius, without applying any aperture-loss 
correction. The circles with uncertainties are
measurements from \blackthunder (sections \ref{s.an.ss.fitting}--\ref{s.an.ss.g395h.bt}), while the green hexagons are measurements from
JADES (Section~\ref{s.an.ss.g395h.jades}). The
empty/filled symbols refer to the fluxes of
\OIIIL and of broad \Halpha. The dashed lines
compare the curve of growth of three PSFs: the model from \citet[][grey line]{deugenio+2024a},
which clearly under-estimates the FWHM; a simple Gaussian fit to the broad-\Halpha image
(red), and a Gaussian fit to the broad \Halpha wing of the J0224-4711 QSO.

The different curves of growth between the filled and empty symbols highlight the
spatially extended nature of \OIIIL, while the broad \Halpha follows closely a Gaussian
profile as expected from a point source. The
neighbours contribution to \OIIIL begins around $R>R_\mathrm{ap}\sim0.2$~arcsec.

Having located the neighbours, we create aperture spectra to measure their systemic
redshift and emission-line ratios. For \satsmall and \sattiny, which are
not spatially resolved (Section~\ref{s.an.ss.size}), we use circular apertures of
radius 0.1 arcsec, which ensure maximum possible SNR. For \satlarge, which is
extended, we use an elliptical aperture that encloses the full NIRCam-detected
flux. These three apertures are outlined in Fig.~\ref{f.ifsimg}.

As noted by \citet{ubler+2023}, aperture spectra that assume uncorrelated noise
result in severely under-estimated uncertainties. We therefore upscale the
uncertainties spectrum by a factor of four, estimated by comparing the robust
standard deviation of the fit residuals to the nominal uncertainties.

\section{Analysis}\label{s.an}

\subsection{Size measurement}\label{s.an.ss.size}

LRDs are known to have compact sizes in the rest-frame optical \citep{furtak+2023,killi+2024},
consistent with the broad lines and optical continuum being dominated by AGN \citep{ji+2025,naidu+2025,degraaff+2025}.
Where sufficiently deep NIRCam imaging is available, the rest-frame UV sizes are often extended 
\citetext{\citealp{killi+2024}; \citetalias{juodzbalis+2024b}}, often displaying
complex UV morphologies \citep{rinaldi+2024}. However, the case in hand
is complicated by the presence of the foreground contaminant. We adopt two complementary
approaches. The first method leverages the well understood PSF and noise properties of NIRCam, but
suffers from considerable contamination by \interlop and self-contamination from
the red LRD continuum. The second method uses the \OIIIL line map from the NIRSpec/IFS
G395H observations, which suffer from no continuum contamination but have a larger PSF
and poorer noise performance than NIRCam, due to correlated noise in the aperture
(Section~\ref{s.data.ss.apspec}).

\begin{figure}
  \includegraphics[width=0.97\columnwidth]{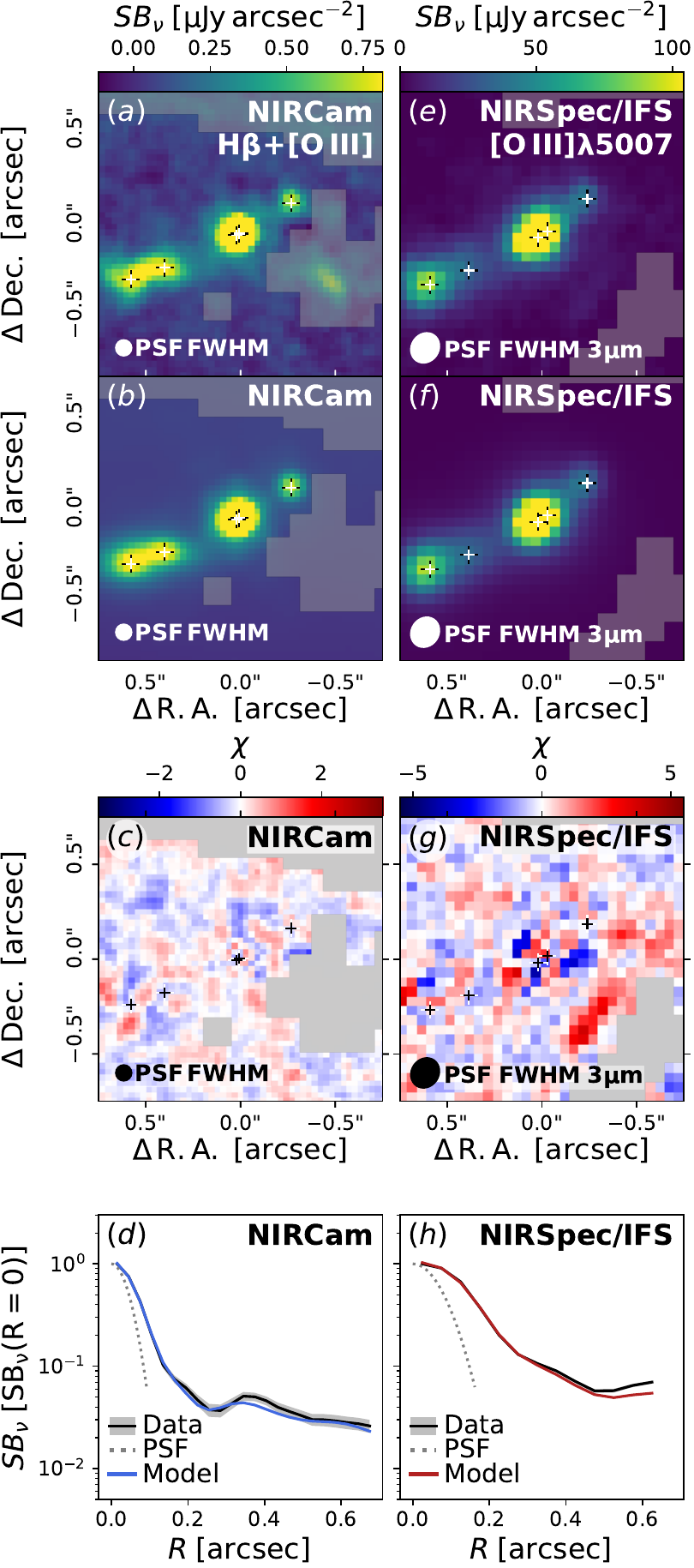}
  {\phantomsubcaption\label{f.size.a}
   \phantomsubcaption\label{f.size.b}
   \phantomsubcaption\label{f.size.c}
   \phantomsubcaption\label{f.size.d}
   \phantomsubcaption\label{f.size.e}
   \phantomsubcaption\label{f.size.f}
   \phantomsubcaption\label{f.size.g}
   \phantomsubcaption\label{f.size.h}}
  \caption{S\'ersic models of the NIRCam F277W-F200W image
  (capturing \Hbeta and \OIIIall at $z=5$;
  panels~\subref{f.size.a}--\subref{f.size.d}) and of the NIRSpec/IFS \OIIIL
  map (panels~\subref{f.size.e}--\subref{f.size.h}). We model the scene using
  five components (centred on the black/white crosses).
  The four rows are the data, MAP model, residuals, and the surface brightness
  profile. The main galaxy \target (centre of cutouts) is modelled as a superposition
  of a point-source and a S\'ersic component, with a flux ratio of 0.6--0.7.
  The galaxy is spatially resolved, with
  the S\'ersic component having intrinsic half-light semi-major axis $\re = 200$~pc and 
  900~pc
  in the NIRCam and NIRSpec/IFS models, respectively. Given the red F277W-F200W
  continuum colour of LRDs, the NIRCam image in panel~\subref{f.size.a} also
  includes continuum flux, which is absent in the NIRSpec/IFS image (we used a
  wavelength band of 150~\kms).
  }\label{f.size}
\end{figure}

For the first method, we create an emission-line map of the target by subtracting the
PSF-matched F200W image (which does not contain strong emission lines) from the F277W image
(which contains \Hbeta and \OIIIL). The resulting image is shown in Fig.~\ref{f.size.a}.
Due to the complex nature of the field, we adopt generous masking around the bulge of
\interlop and, to the north, a diffraction spike from a bright star.
We model five sources, based on fitting and inspecting the residuals. These consist of
a point source and a S\'ersic profile at the location of \target, two S\'ersic profiles
to describe the `cigar-shaped' \satlarge, and a point source to describe \satsmall.
The centre of each source is marked by a cross marker in Fig.~\ref{f.size.a}-\subref{f.size.c}.
We infer the model parameters using \pysersic \citep{pasha+miller2023}, adopting the
empirical F277W PSF from \citet{ji+2024b}. The fitting setup is the same as
\citet{deugenio+2025d}.

The marginalized posterior probabilities on key model
parameters are reported in Table~\ref{t.size}, and the fiducial (maximum a-posteriori; MAP)
model and the data-model $\chi$ residuals are illustrated in
Fig.~\ref{f.size.b}--\subref{f.size.c}. The $\chi$ map highlights the presence of
significant residual sub-structure, but we deem this a satisfactory model, given the
complexity of the field and the ongoing interaction between the galaxies.
For \target, we find a flux ratio between the point-source and S\'ersic component
of $0.7\pm0.2$; the S\'ersic
component has a very large index $n=4.5$, albeit the uncertainties are also large.
The semi-major axis half-light radius is extremely compact, $\re=0.21\pm0.06$~kpc; for
reference, a galaxy with stellar mass $\mstar = 10^8\text{--}10^9~\Msun$ has
$\re = 0.5\text{--}0.8$~kpc at redshift $5<z<6$ \citep{miller+2024}.
For \target, we tested forcing an exponential disc profile using a Gaussian prior probability
with mean $n=1$ and standard deviation 0.1; this results in a twice larger \re, but
also in two times lower flux, with the point-source component becoming the brightest
of the two.
The satellite galaxy \satlarge has larger size than \target, with the two (spatially offset)
components having $\re = 0.8$ and 0.9~kpc, respectively. These values are close to
the expectations for a galaxy as massive as \satlarge (Section~\ref{s.sats.ss.mstar}).

The second approach is equivalent, aside from the data and mask used, and from the
larger and non-circular PSF of NIRSpec/IFS (Section~\ref{s.data.ss.apspec}). The emission-line map has been created by
taking the median of the datacube in a 150-\kms window centred on \OIIIL
(Fig.~\ref{f.size.e}). We find no evidence of contamination from the foreground galaxy,
but there are low-intensity artefacts that we mask. A relatively bright, elongated
feature is not masked; this could be an artefact, but is very close to the location
of \sattiny. We use the same five components to model the system and the same
inference method as for NIRCam.

The inference results are displayed in Fig.~\ref{f.size.e}--\subref{f.size.h} and
are reported in the bottom five rows of Table~\ref{t.size}.
For \target, we find $\re = 0.92\pm0.02$~kpc, 4.5~times larger than for NIRCam.
A likely explanation is that the NIRCam emission-line map suffers from substantial
contamination from the object continuum, which may arise from a compact or even
unresolved component \citep{naidu+2025,degraaff+2025}. This is confirmed by the
prism observations, which indeed detect a compact continuum source (Appendix~\ref{a.compcont}).
In contrast, NIRSpec/IFS can
accurately identify \OIIIL emission and eliminate any underlying continuum.
The S\'ersic index is significantly lower than for NIRCam $n=1.9\pm0.2$ (3-\textsigma
difference) and the shape is considerably rounder.
The origin of this discrepancy is unclear, but a possibility is that the NIRCam
emission-line map F277W-F200W is heavily biased by a point-source continuum, thus yielding a more
compact solution.
However, the flux ratio between the point-source and S\'ersic components is
$0.57\pm0.04$, consistent with the NIRCam value, so overall the difference may be
dominated by systematics, e.g. in the PSF determination. For \satlarge, the sizes are in
good agreement with NIRCam.

In Section~\ref{s.sats},
we will show that the kinematic properties of the \OIIIL-emitting gas surrounding \target
suggest a merger scenario. In this context, it is reasonable to expect the gas to be more extended than the continuum -- particularly because we find no evidence of gaps in the gas distribution or in its velocity field (see again Section~\ref{s.sats}). This suggests that
the half-light \OIIIL size measured from
NIRSpec/IFS should not be used for calculating the dynamical mass of the
system. For this reason, hereafter we adopt the NIRCam \re value of $0.21\pm0.06$~kpc
as the fiducial size measurement for \target. However, we also present the corresponding NIRSpec measurements for comparison. Note that these \re values already remove the
contribution of the unresolved component. Had we modelled \target with a single S\'ersic
profile without separating the point-source contribution, the resulting \re would be even smaller, giving even lower constraints on the dynamical mass.

\begin{table*}
    \setlength{\tabcolsep}{2.5pt}
    \caption{Morphological parameters of the sources at $z=5.077$, derived from \pysersic modelling. Each row corresponds
    to a source, as indicated in Column 1. The first five and bottom five rows report the parameters inferred from
    the NIRCam and NIRSpec/IFS data (Column 2). \target and \satlarge
    are modelled with two components, \satsmall with a single point-source component, and \sattiny is not modelled.
    The columns report the marginalized posterior distributions on the model parameters, with the median and
    16\textsuperscript{th}--84\textsuperscript{th} percentile range.
    }\label{t.size}
    \begin{tabularx}{\textwidth}{lXXXXXXXX}
Source ID    & Data        & Profile      &  $F_\nu^a$       &      $n$    &     $q$       &     P.A.      &     \re      &     \re       \\
             &             &              &  \textmu Jy      &      ---    &     ---       &    rad        & pixels$^b$   &     kpc       \\
\hline
\target      &    NIRCam   & S\'ersic     &  0.063$\pm$0.011 & 4.5$\pm$1.6 & 0.44$\pm$0.17 & 2.0$\pm$0.2   & 1.1$\pm$0.3  & 0.21$\pm$0.06$^c$ \\
\target      &      "      & Point Source &  0.042$\pm$0.011 &    ---      &       ---     &     ---       &      ---     &      ---      \\
\satlarge    &      "      & S\'ersic     &  0.053$\pm$0.008 & 1.8$\pm$0.7 & 0.40$\pm$0.09 & 1.7$\pm$0.1   & 4.6$\pm$0.9  & 0.9$\pm$0.2   \\
\satlarge    &      "      & S\'ersic     &  0.036$\pm$0.006 & 2.5$\pm$1.1 & 0.16$\pm$0.06 & 1.6$\pm$0.07  & 4.0$\pm$0.8  & 0.8$\pm$0.2   \\
\satsmall    &      "      & Point Source &  0.015$\pm$0.001 &    ---      &       ---     &     ---       &      ---     &      ---      \\
\hline
\target      & NIRSpec/IFS & S\'ersic     &  8.2$\pm$0.4     & 1.9$\pm$0.2 & 0.98$\pm$0.02 & 1.7 $\pm$0.8  & 2.9$\pm$0.06 & 0.92$\pm$0.02$^c$ \\
\target      &      "      & Point Source &  4.7$\pm$0.2     &    ---      &       ---     &     ---       &      ---     &      ---      \\
\satlarge    &      "      & S\'ersic     &  6.5$\pm$0.5     & 6.0$\pm$1.0 & 0.95$\pm$0.05 & 1.5 $\pm$0.9  & 2.5$\pm$0.3  & 0.8$\pm$0.1   \\
\satlarge    &      "      & S\'ersic     &  2.9$\pm$0.3     & 0.8$\pm$0.1 & 0.53$\pm$0.05 & 1.86$\pm$0.07 & 3.4$\pm$0.3  & 1.1$\pm$0.1   \\
\satsmall    &      "      & Point Source &  1.0$\pm$0.1     &    ---      &       ---     &     ---       &      ---     &      ---      \\
\hline
    \end{tabularx}
\raggedright{$^a$ For NIRCam, $F_\nu$ is the difference in flux density between the F200W and F277W wide-band filters. For NIRSpec/IFS,
$F_\nu$ is the average flux density inside a narrow wavelength window of 150~\kms, centred on \OIIIL (hence the much higher $F_\nu$
in NIRSpec than in NIRCam).
$^b$ We use 0.03-arcsec pixels for NIRCam and 0.05-arcsec spaxels for NIRSpec/IFS.
$^c$ The large difference between NIRCam and NIRSpec could be due to NIRSpec being more sensitive to the emission-line morphology, while the NIRCam emission-line map may suffer from contamination from the point-source continuum.}
\end{table*}

\subsection{Emission-line measurement methods}\label{s.an.ss.fitting}

To measure the emission lines, we use a Bayesian approach, using different models as specified in the following sections. To integrate the posterior distribution, we use the Markov Chain Monte Carlo method, with the software \textsc{emcee} \citep{foreman-mackey+2013}. Before comparing to the data,
all models are convolved with the wavelength-dependent line-spread function (LSF) of NIRSpec  \citep{jakobsen+2022}. For JADES only, we use the LSF corrected for slit underfill in NIRSpec/MSA 
\citep{degraaff+2024}. After this step, the model
is integrated over each spectral pixel.
To initialize the chains, we first identify the minimum-$\chi^2$ solution 
using the specified model and ordinary least-squares minimization. We mask any spectral pixel deviating
more than 3~\textsigma from the best-fit model; this bad-pixel mask is saved and used later for the Bayesian estimate step.
We estimate the uncertainties on this solution using the resulting Jacobian matrix. 
We use 140 chains initialized from truncated Gaussians, with mean equal to 
the minimum-$\chi^2$ solution, and dispersion equal to twice the least-squares uncertainty. The truncation 
uses very generous bounds; we inspected
the marginalised posterior probabilities after each fit, and increased the
truncation bounds whenever a bound was within 3~\textsigma away from the 
median. The exception being when a
physically motivated solution is being enforced (e.g, non-negative flux for forbidden emission lines, covering factors between 0 and 1).
We run 10,000 steps for each chain, with 50~percent burn-in steps. All chains are concatenated and visually inspected for convergence.
As fiducial parameters, we adopt the median and 16\textsuperscript{th}--84\textsuperscript{th} percentile range of the marginalized posterior. Hereafter, the fiducial model is always the MAP model.

\subsection{High-resolution \texorpdfstring{\blackthunder}{Black THUNDER} spectrum -- black-hole properties}\label{s.an.ss.g395h.bt}

For the properties of the SMBH, we rely on the high-resolution \blackthunder G395H 
data around \Hbeta--\OIIIall and \Halpha, using the `total' elliptical aperture with $R_\mathrm{ap}=0.25$~arcsec. 
We use the Bayesian approach outlined in \citetalias{juodzbalis+2024b}, but model simultaneously both line groups. The continuum is parametrized as two first-order
polynomials in two windows centred at 3 and 4~\mum (rest-frame 4935 and 6565~\AA), requiring 4 free parameters.
All narrow lines share the same redshift $z_\mathrm{n}$ and
velocity dispersion $\sigma_\mathrm{n}$ (2 free parameters); $z_\mathrm{n}$ is adopted
as the spectroscopic redshift of the galaxy 
$z_\mathrm{spec}$.
We assume Gaussian velocity distributions, which for the range of velocities considered here, can be 
approximated as Gaussians in wavelength space too, as
\begin{equation}\label{eq.gaussl}
\begin{split}
    f[v(\lambda)] = & \dfrac{\mathrm{c}}{\sqrt{2 \text{
     \textpi}}\, \sigma_\mathrm{n} (1+z_\mathrm{n})  \lambda_0(k)} \\
     & \hspace{1cm} \cdot \exp \left\{-\dfrac{1}{2} \left[ \dfrac{\mathrm{c} 
     \left( \lambda - \lambda_0(k)\cdot(1+z_\mathrm{n}) \right)}{\sigma_\mathrm{n} (1+z_\mathrm{n}) \lambda_0(k) } \right]^2 \right\},
\end{split}
\end{equation}
where $\lambda$ is the observed wavelength and $\lambda_0(k)$ is the vacuum rest-frame wavelength of a given emission line ($k=\Hbeta, \OIIIL[4959], \dots$).
The \OIIIall and \NIIall doublets require 1 extra free parameter each \citetext{in addition to the shared redshift and velocity dispersion of the narrow lines; the doublet flux ratios are fixed to 0.335, see \citealp{storey+zeippen2000}, and 0.327, see \citealp{dojcinovic+2023}, respectively}.
Narrow \Halpha and \Hbeta require 2 free parameters.
Broad \Halpha emission requires two Gaussians; their common
redshift is
$(1+z_\mathrm{n}) \cdot \exp(v_\mathrm{b}/\mathrm{c})$, parametrised by the velocity of the broad component
$v_\mathrm{b}$ relative to the redshift of the narrow component.
Their flux is parametrized by the total flux $F_\mathrm{b}(\Halpha)$ and
by the flux ratio $F_\mathrm{b,1}/F_\mathrm{b}(\Halpha)$, while their
FWHMs are free parameters, subject to $FWHM_{b,1}<FWHM_{b,2}$.
Broad \Hbeta is also modelled as a double Gaussian, but is parametrized only with the flux ratio between
\Hbeta and \Halpha, while all other parameters are tied to the broad \Halpha.
The broad components require 6 free parameters.
A simpler model using a single Gaussian is manifestly inadequate
and is statistically disfavoured \citetext{we used the Bayesian information criterion, BIC, \citealp{schwarz1978}, and we obtain $\dBIC>10$; we find consistent results using the $\chi^2$ distribution}.
As pointed out in \citet{deugenio+2025c}, a double Gaussian model for the BLR is adopted as an effective line
profile, without interpreting it as two SMBHs.
For the broad-line Gaussians only, we also model the effect 
of a foreground $n=2$ hydrogen absorber, using a geometric
covering factor $C_f$ and the optical-depth approach, with
the residual intensity at wavelengths $\lambda$ given by
\begin{equation}\label{eq.residual}
\begin{split}
    I(\lambda)/I_0(\lambda) &= 1 - C_f + C_f \cdot \exp \left(- \tau(k;\,\lambda) \right)\\
    \tau(k;\,\lambda) &= \tau_0(k) \cdot f[v(\lambda)],
\end{split}
\end{equation}
where $I_0(\lambda)$ is the spectral flux density before absorption, $\tau_0(k)$ is the optical depth at the centre of the line (with $k=\Hbeta$ or \Halpha) and $f[v(\lambda)]$ is the velocity distribution of the absorbing atoms. The latter is approximated again using Eq.~\ref{eq.gaussl},
with $\lambda_0(k) = \lambda_0(\Halpha)$, and replacing $z_\mathrm{n}$ and
$\sigma_\mathrm{n}$ with $z_\mathrm{abs}(\vabs)$ and \sigabs, where \vabs and
\sigabs describe the Gaussian velocity distribution of the absorbing gas.
\vabs is related to the absorber redshift $z_\mathrm{abs}$ by $(1+z_\mathrm{abs}) \equiv (1 + z_\mathrm{n}) \cdot \exp( \vabs/\mathrm{c})$, thus the
velocity of the absorber is measured relative to the redshift of the narrow lines.
The 5 free parameters for the absorption are $C_f$, $\tau_0(\Hbeta)$ and $\tau_0(\Halpha)$ from 
Eq.~\ref{eq.residual}, and \vabs and \sigabs from the equivalent of 
Eq.~\ref{eq.gaussl}, giving a total of 21 free parameters in the model. We also
model a broader component in \OIIIall only, which adds three free parameters, the doublet velocity and
velocity dispersion and the broad-\OIIIL flux.
In addition to the standard flat priors (Section~\ref{s.an.ss.fitting}), we also use three
erfc priors. We penalize the narrow \NIIL/\Halpha ratio against values higher than 1 \citep[based
on the very low detection rate of \NIIall at $z>5$;][]{cameron+2023}. We also penalize
the ratios $\sigma_\mathrm{n}/\sigma_\mathrm{out}$ and $F_\mathrm{out}(\OIIIL)/F_\mathrm{n}(\OIIIL)$
against values higher than 1, to avoid swapping the narrow-line flux with the much fainter
broader \OIIIL emission. Since none of these ratios can be negative, we normalize the erfc prior by
its integral over the non-negative numbers.

\begin{figure*}
  \includegraphics[width=\textwidth]{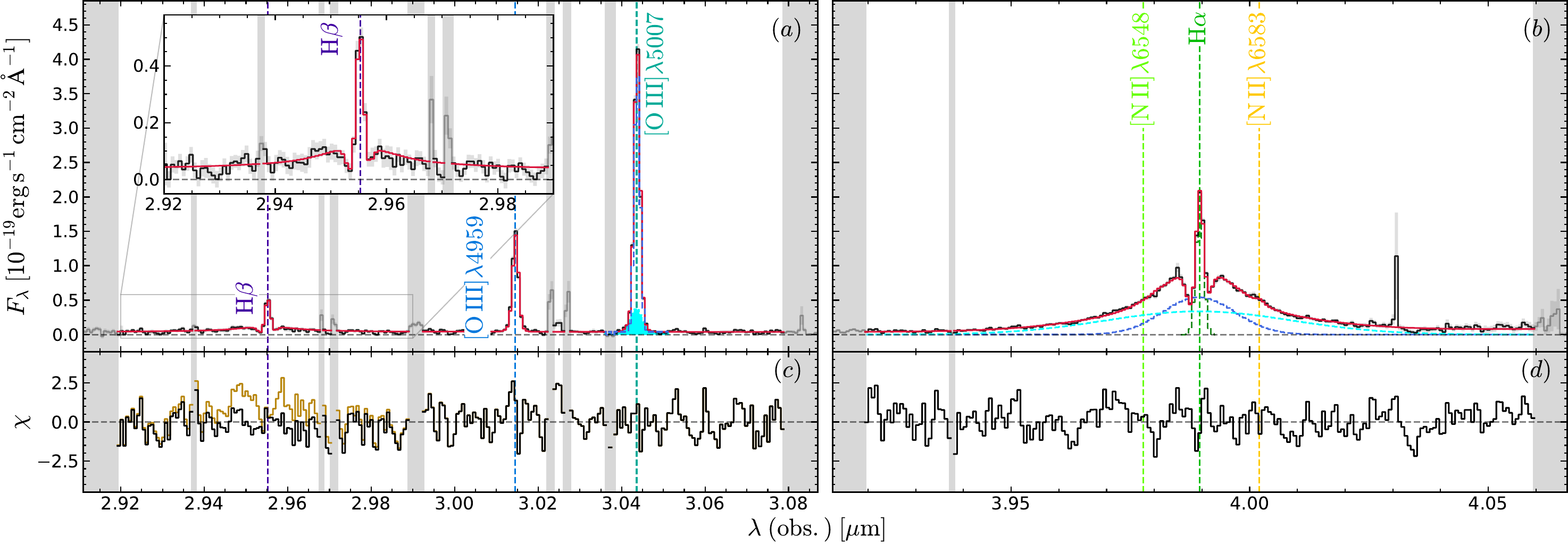}
  {\phantomsubcaption\label{f.g395h.bt.a}
   \phantomsubcaption\label{f.g395h.bt.b}
   \phantomsubcaption\label{f.g395h.bt.c}
   \phantomsubcaption\label{f.g395h.bt.d}}
  \caption{\blackthunder G395H $R_\mathrm{ap}=0.125$-arcsec aperture spectrum (black line), with the best-fit model overlaid in red. Grey vertical strips mark either spectral regions outside of the fitted range or pixels affected by masked outliers (Section~\ref{s.an.ss.fitting}).
  In the bottom panels we show the fiducial $\chi$ residuals (black line) and the residuals of the disfavoured model without a broad \Hbeta component (sand-coloured line). The blue line and filled cyan curve in
  panel~\subref{f.g395h.bt.a} show the two components (narrow and broad) of the \OIIIL emission line;
  similarly, the dashed lines in panel~\subref{f.g395h.bt.b} show the individual emission components of
  \Halpha (without absorption).
  }\label{f.g395h.bt}
\end{figure*}

The posterior parameters of the fit to the fiducial aperture spectrum are
listed in Column~1 of Table~\ref{t.pars}. In Fig.~\ref{f.g395h.bt} instead we
show the data and fiducial model from the high-SNR aperture with
$R_\mathrm{ap}=0.125$~arcsec.
Compared to the $R_\mathrm{ap}=0.25$-arcsec aperture, this
smaller aperture spectrum has biased emission-line ratios but
offers higher SNR in the broad lines. The model correctly reproduces the shape of the 
emission lines, including the complex profile around \Hbeta and \Halpha.
Broad \Hbeta emission is clearly present, with a 7~\textsigma detection ($F_\mathrm{b}(\Hbeta)=1.4\pm0.2$~\fluxcgs[-18]; Table~\ref{t.pars}, Column~2), whereas the
$R_\mathrm{ap}=0.25$-arcsec aperture yields only a 4-\textsigma result (Table~\ref{t.pars},
Column~1), consistent with the expectations for an unresolved source. Fig.~\ref{f.g395h.bt.c}
shows $\chi$, the data-minus-model residuals normalised by the noise (black line). The sand-coloured line
highlights the systematically higher residuals when omitting the broad-\Hbeta component.
Tentative \Hbeta absorption is also seen, with a slightly blue-shifted centroid (inset panel).
The broad \OIIIall component is detected too, albeit only at the 5-\textsigma level, but is discussed
more in detail in Section~\ref{s.an.ss.outflows}.

We detect a marginal blueshift of the broad \Halpha ($v_\mathrm{b} = -45\pm9~\kms$), but this is
driven by the large aperture, which may bias the narrow-line redshift estimate due to contamination from possible extended emission along the axis connecting \satsmall to \satlarge (Fig.~\ref{f.ifsimg}).
In agreement with this hypothesis, when repeating the fit for the 0.125-arcsec aperture (Table~\ref{t.pars}, Column~2), we find indeed a
lower systemic redshift (2-\textsigma significance) and, therefore, less blue-shifted broad \Halpha 
($v_\mathrm{b}=-28\pm8~\kms$).
There are some positive residuals at the spectral position of the narrow \Halpha line and at the blue 
wing of the \Halpha absorption. These residuals are not statistically significant, but they can be
removed by decoupling the narrow \Halpha from the bluer narrow lines, and by introducing a second
\Halpha absorber.

The \Halpha absorption has $EW_\mathrm{abs}(\Halpha) = 6.9^{+0.5}_{-0.4}~\AA$ relative to the broad-line flux
density. This value is deceptively small: the strength of the absorption can be fully appreciated when
related to the broad-\Halpha equivalent width, $EW(F_\mathrm{b}(\Halpha)) = -1100$~\AA, or when
relating the absorption to the continuum ($EW_\mathrm{abs}(\Halpha) = 190~\AA$).

A selected subset of model parameters is shown in Fig.~\ref{f.trian}, showing that the
G395H data can meaningfully constrain $C_f$ and $\tau_0(\Halpha)$, even though a  strong degeneracy 
between these two parameters persists \citetext{as expected; e.g., \citealp{davies+2024}, 
\citetalias{juodzbalis+2024b}}. The velocity dispersion $\sigabs = 120\pm10~\kms$ of the absorber is
similar to other results reported in the literature \citepalias[e.g.,][]{juodzbalis+2024b}, but 
strikingly, the absorber velocity is remarkably low, $\vabs = -26^{+3}_{-4}~\kms$, statistically consistent with the velocity of the BLR ($v_\mathrm{b}=-28\pm8~\kms$).

\begin{figure}
  \includegraphics[width=\columnwidth]{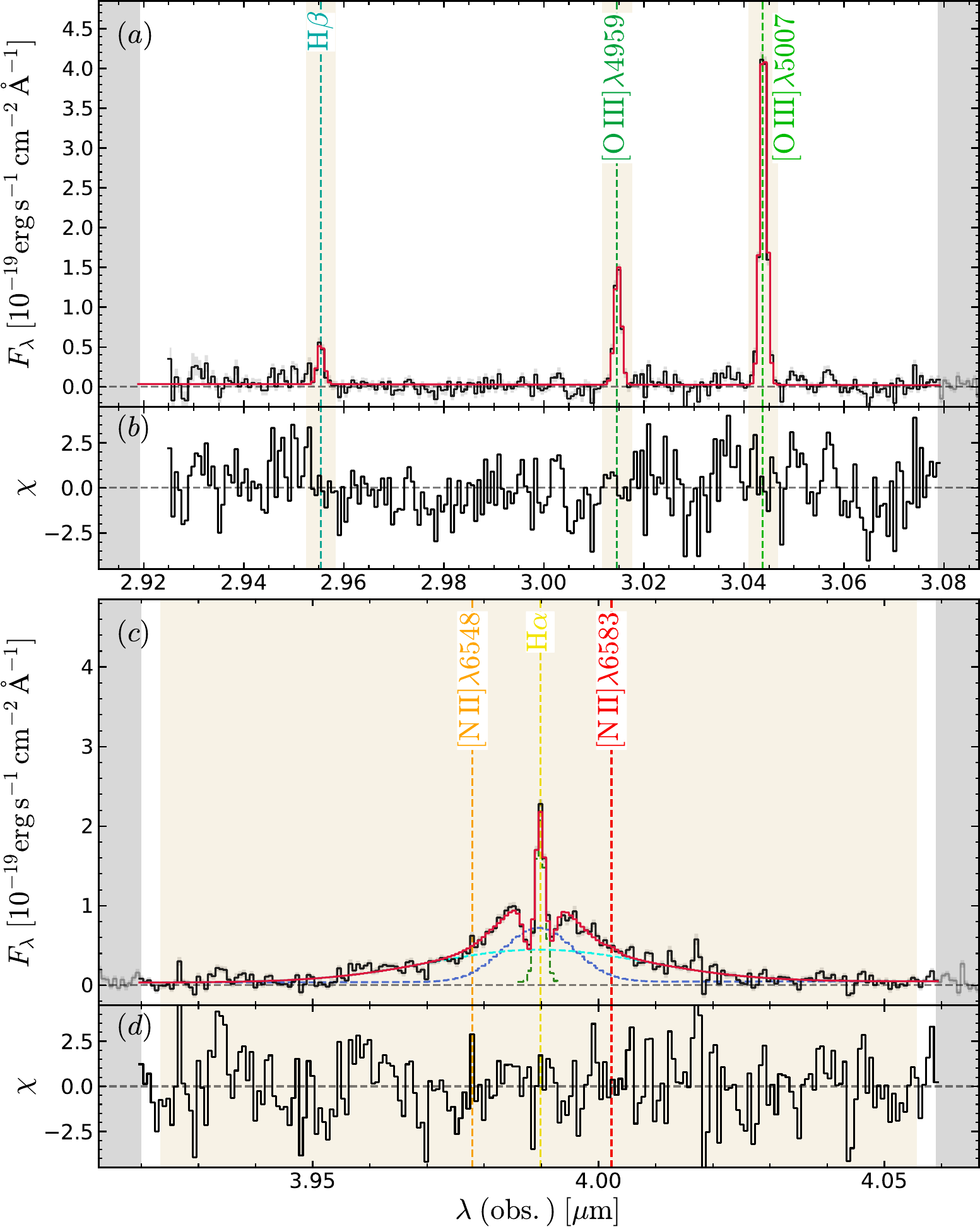}
  {\phantomsubcaption\label{f.g395h.jades.a}
   \phantomsubcaption\label{f.g395h.jades.b}
   \phantomsubcaption\label{f.g395h.jades.c}
   \phantomsubcaption\label{f.g395h.jades.d}}
  \caption{JADES G395H aperture spectrum (black line), with the best-fit model overlaid in red. Grey vertical bands are not included in the fit, while sand-coloured bands are the regions we used to calculate the BIC value and to compare different models (Section~\ref{s.an.ss.g395h.jades}). In the bottom panels we show the $\chi$ residuals.
  }\label{f.g395h.jades}
\end{figure}

\subsection{High-resolution JADES spectroscopy}\label{s.an.ss.g395h.jades}

The availability of JADES G395H spectroscopy enables us to test \target for time variability.
We fit the G395H JADES data with the same approach as for the \blackthunder G395H aperture data,
with two differences. We remove the broad \Hbeta component, the \Hbeta absorber and the broad 
\OIIIall component, since we find no supporting statistical evidence for any of these features
(five less free parameters than the model in the previous section), but we add the narrow \SIIall 
doublet, with free \SIIL flux and doublet flux ratio \citep[constrained to the physical range from][adding two 
more free parameters]{sanders+2016}.

The resulting best-fit model is shown in Fig.~\ref{f.g395h.jades}, while the percentiles of the
posterior probabilities are in Table~\ref{t.pars}, Column 3.
Several parameters have different posterior probability distributions between JADES and \blackthunder.
The lack of broad \OIII, broad \Hbeta and \Hbeta absorption is consistent with the lower SNR of the JADES observations.
The JADES data show clear detections of \SIIall, but we discard these as
unreliable (Section~\ref{s.an.ss.n2s2}).
JADES finds higher redshift $z_\mathrm{n}=5.07783\pm0.00002$ with high significance 
(11~\textsigma), but the redshift difference is only $\Delta\,z = 0.0003$, which corresponds to just one third of a spectral pixel at 4~\mum, so the discrepancy could be due to systematics in the wavelength solution, which for sources smaller than the MSA microshutter depends on the intra-shutter position -- itself known only approximately.
An error of this magnitude could be due to the bias between intra-shutter position and wavelength solution in the JADES MSA data reduction \citetext{\citealp{deugenio+2025a}; see \citealp{scoltz+2025b} for a solution to this problem}.
We tested that decoupling the redshift of the
spectral regions around \Hbeta--\OIIIall from the redshift of the lines closest to \Halpha
does not substantially improve the fit.
The velocity dispersion is marginally different from \blackthunder, with instrument-deconvolved
$\sigma_\mathrm{n} = 54\pm1$~\kms
(2-\textsigma difference). Systematic uncertainties in the LSF may be responsible for this discrepancy. In fact, 
we assumed the LSF for point sources from \citet{degraaff+2024}, but in our case this choice likely over-estimates the
actual resolution, because while \target is
compact, the narrow lines are spatially resolved (Section~\ref{s.an.ss.size}),
so the effective resolution is intermediate between the nominal value assuming
uniform slit illumination and the LSF for point-sources.
The JADES data has lower observed $F_\mathrm{n}(\Hbeta) = (0.83\pm0.08)\fluxcgs[-18]$ (2-\textsigma
significance). This small difference leads the model to infer higher dust attenuation,
$\Avhatn=1.4_{-0.2}^{+0.3}$~mag, but this latter difference is not statistically significant (1.4~\textsigma).
Lower narrow \Hbeta flux is consistent with the bias deriving from assuming point-source path-loss
corrections in the JADES data reduction, which assume a higher flux loss at redder wavelengths,
thereby potentially leading to a spurious increase in the Balmer decrement.
These non-consequential differences have plausible explanations and overall do not affect our results.

However, there are two important mismatches. The FWHM of the broad-\Halpha line is $FWHM_\mathrm{b}(\Halpha)
= 1490\pm80~\kms$ for JADES and $1790\pm60~\kms$ for \blackthunder (a 3-\textsigma discrepancy). This
20~percent difference alone propagates to an error on the SMBH mass of 45~percent
(Section~\ref{s.phys.ss.smbh}), only marginally reduced by the 14~percent larger \Halpha flux in JADES
than in \blackthunder. This flux difference (4~\textsigma) reflects the difficulty of inferring the
emitted \Halpha flux in the presence of strong absorption ($\tau_0(\Halpha) = 3.1^{+0.7}_{-0.5}$).

\subsection{Ionized outflows}\label{s.an.ss.outflows}

The broad \OIIIall component (shown separately by the filled cyan line under \OIIIL
in Fig.~\ref{f.g395h.bt.a}) is clearly much broader than the narrow lines, yet
significantly narrower than the broad Balmer lines. It is not detected in JADES ($\dBIC=6$),
while the deeper \blackthunder data favours multiple components, with an
improvement of $\dBIC=21$ for the double-component model over the single-component
model.

According to the \blackthunder data, both \OIIIall components are kinematically
narrow ($\sigma_\mathrm{out} = 140\pm20$~\kms vs. $\sigma_\mathrm{n} = 50\pm2$~\kms)
and they are consistent with the same systemic velocity (velocity offset $\delta v =
-20^{+10}_{-20}~\kms$). The narrowest component is the brightest, with an \OIIIL flux
of $8.3_{-0.4}^{+0.3}\fluxcgs[-18]$, while the broad component has flux
$F_\mathrm{out}(\OIIIL) = 1.5_{-0.3}^{+0.4}\fluxcgs[-18]$ (5~\textsigma significance).
The significance does not change with aperture size, suggesting a spatially resolved nature. However, the SNR is too low for detecting the
broader component in individual spaxels, hence we omit it from the
spatially resolved analysis (Section~\ref{s.sats}).

Repeating the deblending analysis performed on \OIIIL for \Halpha and \Hbeta is not possible due to degeneracies with the absorption profile shape and, in the case of \Hbeta, the low SNR.
Given this lack of information, it is impossible to provide a physical
interpretation of the two components. We present three simple scenarios.
The narrowest component may be tracing the host galaxy, while the broadest
component is due to tidally disrupted material following the interaction with one or more
satellites (Section~\ref{s.sats}). Alternatively, if the broadest component
traces the host galaxy, then the narrowest component could be either a star-forming
clump, or gas photo-ionized by the AGN. Faint outflows are also a possibility, with an
outflow velocity that can be estimated \citep{rupke+2005,veilleux+2005} as $|v_\mathrm{out}| + 2\cdot \sigma_\mathrm{out} = 300~\kms$, in agreement
with the values found in star-forming galaxies at $z>4$ \citep{carniani+2024}.
The possible detection of a localized outflow resonates with the finding
of the detached outflow component \sattiny (sections~\ref{s.sats.ss.apspec}
and~\ref{s.sats.ss.specres}).

\subsection{Absence of nitrogen, sulphur, and helium}\label{s.an.ss.n2s2}

Near \Halpha, we find no evidence for \NIIall and \SIIall, as expected from
the generally low detection rate of these lines at $z>5$ \citep{cameron+2023,deugenio+2025a,mascia+2024,juodzbalis+2025}.
Performing the inference while including \NIIall (with the same redshift and
intrinsic dispersion as the other narrow lines) yields a 3-\textsigma upper limit
on the \NIIL flux $<0.1\text{--}0.15$~\fluxcgs[-18] (depending on the survey and aperture considered).
For \SIIL and \SIIL[6731], the JADES G395H fit yields 6- and
5-\textsigma detections respectively. The nominal fluxes are 
$(0.36\pm0.06)\fluxcgs[-18]$ and $(0.31\pm0.05)\fluxcgs[-18]$, with a plausible doublet ratio of 1.2. The \blackthunder G395H data cannot confirm this detection, because at the position of \target, the doublet falls in the detector gap. However,
neither of these two lines are confirmed in the medium-resolution grating or in the prism (even 
though the low spectral resolution of the prism is expected to increase the SNR of the 
doublet). 
Besides, \SIIL[6731] is very near some noise features. Therefore, we dismiss
the signal observed in G395H as an artefact.
Similarly, we find no evidence for \HeII, in agreement with
similar findings in other LRDs \citep[see e.g. the objects included in the AGN sample of][]{juodzbalis+2025}.

\subsection{Other tests and model parameters}\label{s.an.ss.other}

Using G395H data from both JADES and \blackthunder, we tested the impact of wavelength-calibration issues on the recovered parameters. We repeated the Bayesian inference by decoupling the \Hbeta and~\OIIIall region of the spectrum from \Halpha. This was done by introducing a new free parameter for the line broadening of the narrow \Hbeta and \OIIIall lines ($\sigma_\mathrm{n,\,blue}$) and a velocity offset $v_\mathrm{n,\,blue}$ between this set of narrow lines and narrow \Halpha. The marginalized posterior on $\sigma_\mathrm{n,\,blue}$ is 10~percent larger than $\sigma_\mathrm{n}$, suggesting that our LSF characterization does not fully capture the instrument performance. For $v_\mathrm{n,\,blue}$ we find either no offset (for JADES, $v_\mathrm{n,\,blue}=10\pm10~\kms$), or a 4-\textsigma offset of \Hbeta and \OIIIall towards the blue (for \blackthunder, $v_\mathrm{n,\,blue}=-6.1\pm1.5~\kms$). In both cases, there is a strong correlation between $v_\mathrm{n,\,blue}$ and the velocity of the absorber, which propagates to other physical quantities like $C_f$ and $\tau_0$. This is understandable, because a velocity offset of the absorber can shift the peak of the narrow-\Halpha line. The exact effect on $C_f$ and $v_\mathrm{abs}$ is complicated by the inter-dependence of these two quantities. However, for $EW_\mathrm{abs}(\Halpha)$, using a free $v_\mathrm{n,\,blue}$ (as opposed to $v_\mathrm{n,\,blue}=0$ in the fiducial models) leads to changes in the EW measurement of 10~percent.

\subsection{Importance of high-resolution \jwst spectroscopy}\label{s.resol}

\begin{figure}
  \includegraphics[width=\columnwidth]{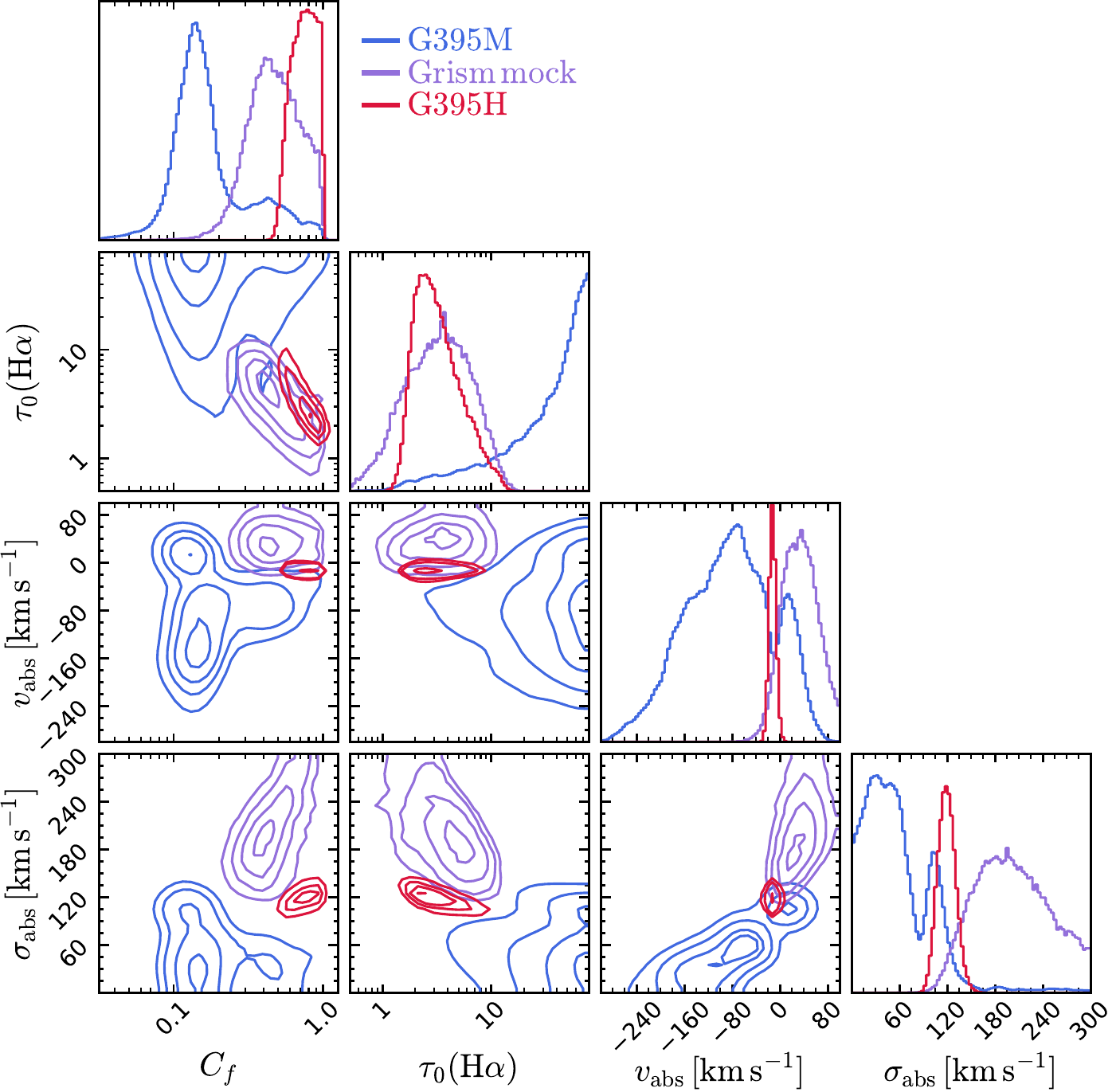}
  \caption{Posterior probability distribution of the \Halpha model, showing the different constraining power of the NIRSpec medium-resolution data (blue; G395M), of the NIRCam/WFSS grisms (purple; Grism mock), and of the high-resolution NIRSpec data (red; G395H). At a resolution of $R=1,000$, G395M is unable to constrain the absorber, and even at $R=1,600$ (corresponding to NIRCam/WFSS) the physical properties of the absorber may be biased. With NIRSpec and G395H, we can measure the velocity of the absorber with high precision, showing that it is very close to the galaxy rest-frame.}\label{f.trian}
\end{figure}

To evaluate the bias of our measurements for different dispersers, we repeat
the above modelling analysis on the medium-resolution JADES data,  with spectral
resolution $R=700\text{--}1,500$, and with a mock realisation of NIRCam/WFSS
F444W data. Besides differences in the data and LSF, the models are
identical. However, for the NIRCam mock spectrum, we exclude the model and
data near \Hbeta and \OIIIall, because these would not be normally available.
We do not correct for the generally higher noise of the NIRCam slitless
spectroscopy. The comparison between the high- and medium-resolution NIRSpec/MSA spectroscopy is shown in  Fig.~\ref{f.abs} (for NIRCam/WFSS see Appendix~\ref{a.nircamwfss}). The
contrast between panel~\subref{f.abs.a} and panels~\subref{f.abs.b}
is stark. In the medium-resolution data,
there is almost no trace and certainly no statistical evidence of \Halpha
absorption. For the NIRCam mock data, the presence of the absorber is still
clear (Appendix~\ref{a.nircamwfss}), and its physical properties $C_f$ and 
$\tau_0(\Halpha)$ are reasonably well constrained (Fig.~\ref{f.trian}),
though these constraints may stem from an unreasonably good data quality, 
which would require very long integration for NIRCam/WFSS.
However, even the best-quality NIRCam/WFSS data cannot constrain well the
kinematic properties of the absorber, with 5$\times$ larger uncertainties.
From here on, we adopt the high-resolution G395H model as the fiducial model for deriving the galaxy redshift, the narrow-line velocity dispersion, and the SMBH parameters.

\begin{figure}
  \includegraphics[width=\columnwidth]{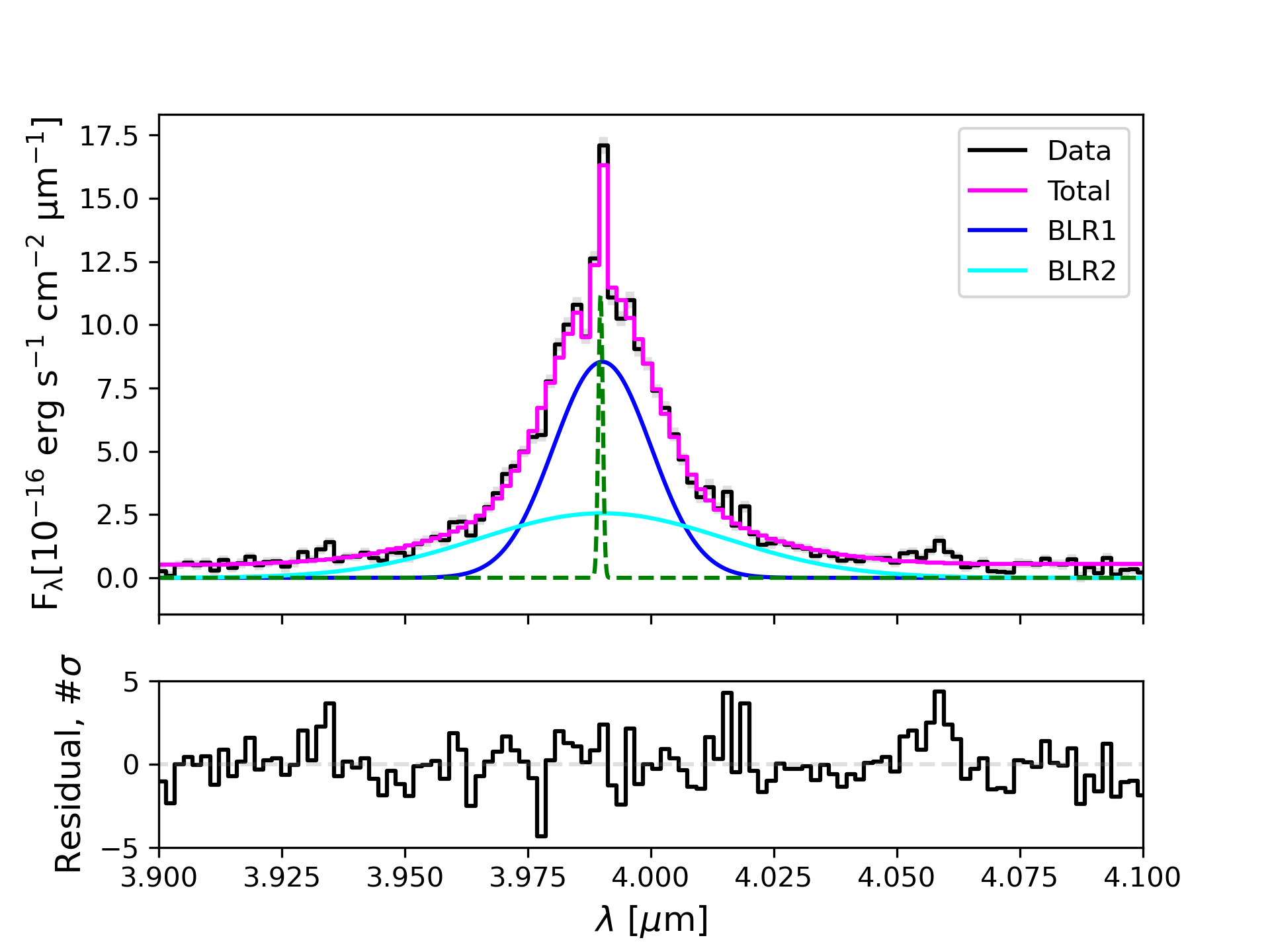}
  \includegraphics[width=\columnwidth]{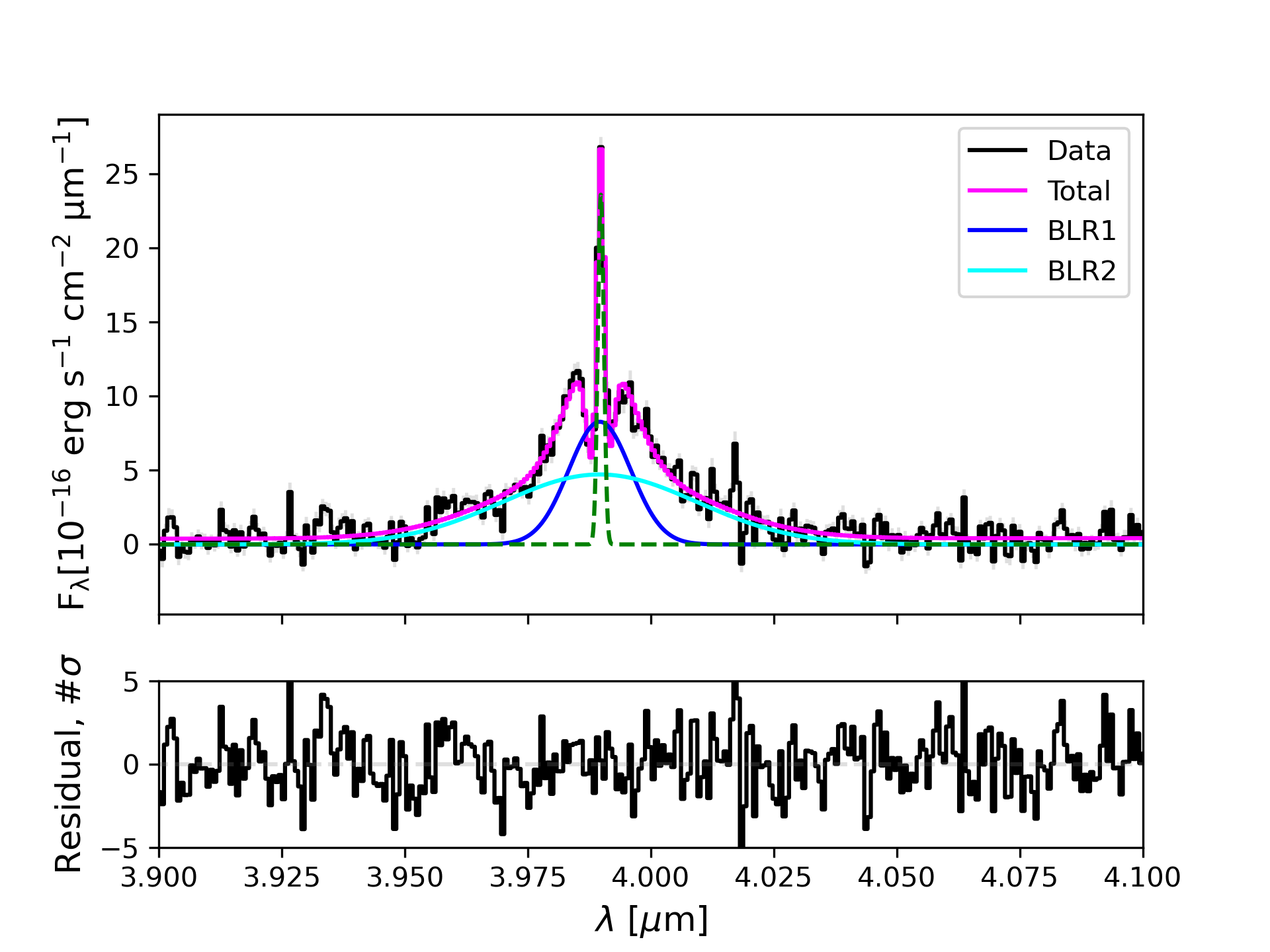}
  {\phantomsubcaption\label{f.abs.a}
   \phantomsubcaption\label{f.abs.b}}
  \caption{Medium- and high-resolution JADES spectra of \target, highlighting how
  G395M (panel~\subref{f.abs.a}) can completely miss an absorber with
  $\tau_0(\Halpha)=3$.}\label{f.abs}
\end{figure}

% g395h_fit_lsf_anna_N2_S2.py
% medium_fit_lsf_anna_N2_S2.py
% prism_fit_lsf_anna.py
\begin{table*}
    \setlength{\tabcolsep}{2.5pt}
    \caption{Posterior probabilities for key parameters of the emission-lines model. Our
    fiducial measurements use the \blackthunder 0.25-arcsec aperture, which ensures accurate line
    ratios across the wavelength range. This accuracy comes at the cost of lower SNR, as evidenced by
    the higher SNR in some lines (e.g., $F_\mathrm{b}(\Hbeta)$ goes from 4~to 7~\textsigma detection
    when restricting the aperture to 0.125~arcsec).
    Parameters in square brackets in Column~(4) are constrained using priors from the
    the posterior of the G395H model. Note the dependence of dust attenuation on the aperture size used.
    The JADES NIRSpec/MSA grating spectra have been downscaled by 0.85 to match the line fluxes of
    \blackthunder and of the JADES prism.}\label{t.pars}
    \begin{tabularx}{\textwidth}{lccXXXX}
    %\begin{tabular}{lcccccc}
  Survey                              &   &                  & \blackthunder                            &             \blackthunder                &  JADES                                   &                    JADES                      \\
  Aperture                            &   &                  &       $R_{\rm ap}\!=\!0.25$~arcsec       &     $R_{\rm ap}\!=\!0.125$~arcsec        &        MSA                               &                    MSA                        \\
  Disperser                           &   &                  &       G395H                              &   G395H                                  &   G395H                                  &              G235M \& G395M                   \\
\hline                                                           
  Parameter                           & Fit & Unit             &    (1)                                 &                 (2)                      &                (3)                       &                   (4)                         \\
\hline                                                           
$F_{\rm n}(\OIIL)$                    & Y & \fluxcgs[\!-\!18][]&      ---                                 &      ---                                 &      ---                                 & $0.32_{-0.04}^{+0.05}$  \\
$F_{\rm \!n}(\OIIL[3729])\!/\!F_{\rm\! n}(\OIIL)$ & Y & ---  &      ---                                 &      ---                                 &      ---                                 & $1.1_{-3}^{+0.2}$                             \\
$F_{\rm n}(\NeIIIL)$                  & Y & \fluxcgs[\!-\!18][]&      ---                                 &      ---                                 &      ---                                 & $0.48_{-0.07}^{+0.06}$        \\
$F_{\rm n}(\Hgamma)^a$                & Y & \fluxcgs[\!-\!18][]&      ---                                 &      ---                                 &      ---                                 & $0.35_{-0.04}^{+0.04}$   \\
$F_{\rm n}(\OIIIL[4363])$             & Y & \fluxcgs[\!-\!18][]&      ---                                 &      ---                                 &      ---                                 & $0.23_{-0.06}^{+0.06}$        \\
$F_{\rm n}(\Hbeta)^a$                 & Y & \fluxcgs[\!-\!18][]& $1.08_{-0.06}^{+0.06}$                   & $0.88_{-0.05}^{+0.05}$                  & $0.90_{-0.08}^{+0.08}$                   & $0.93_{-0.07}^{+0.07}$         \\
$F_{\rm n}(\OIIIL)$                   & Y & \fluxcgs[\!-\!18][]& $8.2_{-0.4}^{+0.3}$                      & $6.5_{-0.4}^{+0.3}$                      & $8.0_{-0.1}^{+0.1}$                      & $7.3_{-0.1}^{+0.1}$          \\
$F_{\rm n}(\Halpha)$                  & Y & \fluxcgs[\!-\!18][]& $4.5_{-0.2}^{+0.2}$                      & $3.6_{-0.2}^{+0.2}$                      & $4.3_{-0.2}^{+0.2}$                      & $4.2_{-0.2}^{+0.2}$          \\
$F_{\rm n}(\NIIL)$                    & Y & \fluxcgs[\!-\!18][]& $<0.15$                                  & $<0.1$                                   & $<0.15$                                  & $(<0.2)$                                      \\
$F_{\rm n}(\SIIL)$                    & Y & \fluxcgs[\!-\!18][]&      ---                                 &      ---                                 &      ---$^b$                             & $(<0.06)$                                     \\
\Avhatn$^a$                           & N & mag              & $1.0_{-0.2}^{+0.2}$                      & $1.0_{-0.1}^{+0.2}$                      & $1.4_{-0.2}^{+0.3}$                      &   $1.2_{-0.2}^{+0.2}$                         \\
\Avhatb                               & N & mag              & $4.4_{-0.5}^{+0.7}$                      & $4.8_{-0.4}^{+0.5}$                      &             ---                          &            ---                                \\
$\sigma_{\rm n}$                      & Y & \kms             & $49_{-2}^{+2}$                           & $47_{-2}^{+2}$                           & $54_{-1}^{+1}$                           &   $69_{-3}^{+3}$                              \\
$z_{\rm n}$                           & Y & ---              & $5.07753_{-0.00002}^{+0.00002}$          & $5.07747_{-0.00002}^{+0.00002}$          & $5.07783_{-0.00002}^{+0.00002}$          &   $5.07785_{-0.00004}^{+0.00004}$             \\
\hline                                                      
$F_{\rm out}(\OIIIL)$                 & Y & \fluxcgs[\!-\!18][]& $1.6_{-0.3}^{+0.4}$                      & $1.5_{-0.3}^{+0.4}$                      &         ---                              &              ---                              \\
$v_{\rm out}$                         & Y & \kms             &         $-20_{-20}^{+10}$                &         $-29_{-12}^{+10}$                &         ---                              &              ---                              \\
$\sigma_{\rm out}$                    & Y & \kms             &         $130_{-20}^{+20}$                &         $120_{-20}^{+20}$                &         ---                              &              ---                              \\
\hline                                                       
$v_{\rm b}$                           & Y & \kms             & $-45_{-9}^{+9}$                          & $-28_{-8}^{+7}$                          & $-20_{-10}^{+10}$                        &   $-10_{-7}^{+7}$                             \\
$F_{\rm b}(\Hbeta)/F_{\rm b}(\Halpha)$& Y & ---              & $0.06_{-0.01}^{+0.01}$                   &      $0.055_{-0.010}^{+0.009}$           &         ---                              &              ---                              \\
$F_{\rm b}(\Hbeta)^c$                 & N & \fluxcgs[\!-\!18][]&         $1.6_{-0.4}^{+0.4}$              &         $1.4_{-0.2}^{+0.2}$              &         ---                              &              ---                              \\
$F_{\rm b}(\Halpha)^c$                & Y & \fluxcgs[\!-\!18][]& $25.9_{-0.6}^{+0.6}$                     & $25.8_{-0.4}^{+0.5}$                     & $29.8_{-0.8}^{+0.8}$                     &   $27.4_{-0.3}^{+0.3}$     \\
$F_{\rm b,1}/F_{\rm b}(\Halpha)$      & Y & ---              & $0.39_{-0.07}^{+0.07}$                   & $0.37_{-0.03}^{+0.03}$                   & $0.35_{-0.02}^{+0.02}$                   &   $0.47_{-0.05}^{+0.05}$                      \\
$FWHM_{{\rm b,1}}(\Halpha)$           & Y & \kms             & $1300_{-100}^{+100}$                     & $1330_{-80}^{+80}$                       & $1100_{-70}^{+70}$                       &   $1550_{-100}^{+100}$                      \\
$FWHM_{{\rm b,2}}(\Halpha)$           & Y & \kms             & $3200_{-200}^{+300}$                     & $3200_{-100}^{+100}$                     & $3400_{-100}^{+200}$                     &   $3900_{-200}^{+200}$                        \\
$FWHM_{{\rm b}}(\Halpha)$             & N & \kms             & $1740_{-90}^{+80}$                       & $1790_{-60}^{+60}$                       & $1490_{-80}^{+80}$                       &            $1920_{-60}^{+70}$                \\
\hline                                                       
$v_{\rm abs}$                         & Y & \kms             &  $-34_{-5}^{+5}$                         &  $-26_{-4}^{+4}$                         &  $-13_{-4}^{+5}$                         &   $[-6_{-7}^{+7}]$                            \\
$\sigma_{\rm abs}$                    & Y & \kms             &      $98_{-7}^{+8}$                      &      $103_{-6}^{+6}$                     &       $108_{-8}^{+8}$                    &   $[84_{-6}^{+7}]$                            \\
$C_f$                                 & Y & ---              &      $0.94_{-0.08}^{+0.04}$              &      $0.90_{-0.12}^{+0.07}$              &      $0.95_{-0.06}^{+0.03}$              &   $[0.96_{-0.04}^{+0.03}]$                    \\
$\tau_0(\Hbeta)$                      & Y & ---              &         $20_{-8}^{+7}$                   &         $7_{-3}^{+6}$                    &         ---                              &              ---                              \\
$\tau_0(\Halpha)$                     & Y & ---              &         $2.2_{-0.4}^{+0.6}$              &         $2.0_{-0.4}^{+0.6}$              &         $3.1_{-0.5}^{+0.7}$              &   $[2.7_{-0.3}^{+0.3}]$                       \\
$EW_{\rm abs}(\Hbeta)$                & N & \AA              & $9.2_{-0.9}^{+0.7}$                      & $8.0_{-1.1}^{+1.0}$                      &         ---                              &            ---                                \\
$EW_{\rm abs}(\Halpha)$               & N & \AA              & $6.6_{-0.4}^{+0.5}$                      & $6.7_{-0.5}^{+0.7}$                      & $8.3_{-0.4}^{+0.5}$                      &            ---                                \\
\hline                                                       
$T_{\rm e}({\rm O^{2+}})$             & N & $10^4$~K         &       ---                                &       ---                                &       ---                                & $2.1_{-0.3}^{+0.4}$                           \\
$T_{\rm e}({\rm O^+})$                & N & $10^4$~K         &       ---                                &       ---                                &       ---                                & $1.8_{-0.2}^{+0.3}$                           \\
$n_{\rm e}({\rm O^+})$                & N & \pcm             &       ---                                &       ---                                &       ---                                & $400_{-300}^{+700}$                           \\
$12\!+\!\log({\rm O^{2+}/H^+})$       & N & dex              &       ---                                &       ---                                &       ---                                & $7.5_{-0.1}^{+0.2}$                           \\
$12\!+\!\log({\rm O^+/H^+})$          & N & dex              &       ---                                &       ---                                &       ---                                & $6.8_{-0.2}^{+0.2}$                           \\
$12\!+\!\log({\rm O/H})$              & N & dex              &       ---                                &       ---                                &       ---                                & $7.6_{-0.1}^{+0.2}$                           \\
\hline
%$\log(\mbh/\Msun)$                    & N & dex              &  $7.50_{-0.05}^{+0.04}$                  & $7.53_{-0.03}^{+0.03}$                   & $7.39_{-0.05}^{+0.05}\;(\pm0.3)$         &            ---                                \\
$\log(\mbh/\Msun)_{\Avhatn}$          & N & dex              &  $7.65_{-0.05}^{+0.05}$                  & $7.68_{-0.04}^{+0.04}$                   & $7.60_{-0.07}^{+0.06}\;(\pm0.3)$         &            ---                                \\
$\log(\mbh/\Msun)_{\Avhatb}$          & N & dex              &  $8.14_{-0.10}^{+0.11}$                  & $8.24_{-0.07}^{+0.08}$                   &       ---                                &            ---                                \\
  \hline
    \end{tabularx}
    
\raggedright{
\textit{Notes.} 
$^a$For the medium-gratings only, we fix the intrinsic flux ratio between the narrow Balmer lines and apply a dust attenuation. Hence, for this fit, \Avhatn is a free parameter, while the $F_\mathrm{n}(\Hgamma)$ and $F_\mathrm{n}(\Hbeta)$ are not.
$^b$\SIIL and \SIIL[6731] are formally detected at 6 and 5~\textsigma in the JADES G395H data, but not in the other configurations (and fall in the detector gap in the \blackthunder G395H data). We interpret the observed signal as spectral overlap from other sources (Section~\ref{s.an.ss.n2s2}).
$^c$For the flux of the broad-line regions we report the flux before the Balmer absorber.}
\end{table*}

\subsection{Dust attenuation}\label{s.an.ss.dust}

The dust attenuation in \target consists of an intrinsic component, 
and a foreground screen due to the interloper \interlop. In principle,
we should distinguish between the intrinsic and foreground terms. In
practice, the dust attenuation from the interloper is negligible
(Appendix~\ref{a.interl}),
due to a combination of low intrinsic attenuation in the interloper and
to the fact that the optical emission lines from \target are already
at observed-frame 1~\mum when they pass through \interlop at $z\sim1$.
For the main analysis, we assume the \citet[][hereafter: \citetalias{gordon+2003}]{gordon+2003}
attenuation law, parametrized by the $V$-band attenuation \Avhatn and
expressed as a spline interpolating the data presented in
\citetalias{gordon+2003}. To infer \Avhatn, we compare the Balmer decrement
$F(\Halpha)/F(\Hbeta)$ to the assumed intrinsic ratio.

For the galaxy ISM we use the narrow-line fluxes measured from the
fiducial \blackthunder aperture, where bias due to aperture losses is minimal.
We assume Case-B recombination, electron temperature \Telec=10,000~K, and 
electron density \nelec=500~\pcm, giving $\Halpha/\Hbeta=2.86$
\citep{osterbrock+ferland2006}.
In the fiducial aperture, we find an effective dust-attenuation value
$\Avhatn=1.0\pm0.2$~mag (Table~\ref{t.pars}, Column 1). The smaller
$R_\mathrm{ap}=0.125$-arcsec aperture instead yields $1.3\pm0.2$~mag.
While these two values are statistically consistent, the higher attenuation
in the smallest aperture suggests that the larger aperture captures
additional narrow-line emission for which our aperture correction is
inadequate (based as it is on the curve-of-growth of the broad-\Halpha line,
which is a point source). Alternatively, the innermost region may suffer from
higher dust attenuation relative to the outskirts.
Assuming a higher intrinsic \Halpha/\Hbeta ratio of 3.10 would revise \Avhatn
down by 0.21~mag.

As we have noted, \Avhatn is an equivalent dust attenuation, 
which includes reddening in \target, as well as additional 
reddening due to crossing the interloper \interlop at 
$z_\mathrm{spec}=1.001$. This latter reddening is hard to 
estimate accurately, because we do not detect \Hbeta in the 
foreground star-forming regions. We detect either the spectrally blended \Hbeta--\OIIIall complex (in the prism), or just \OIIIL 
(in the G140M grating). Flux calibration uncertainties between
the NIRSpec dispersers are of the same order of magnitude as the 
effect of moderate dust attenuation \citep{deugenio+2025a}, so
we do not attempt to infer \Hbeta by subtracting the grating 
\OIIIall flux from the \Hbeta--\OIIIall blend in the prism.
A physically motivated fit of the interloper spectrum finds $A_V = 0.4^{+0.7}_{-0.3}$, and is thus consistent with no dust attenuation (Appendix~\ref{a.interl}). We thus do not attempt to break down \Avhatn, the effective dust attenuation in \target, into a local and a foreground component.

For the BLR, we find a Balmer decrement of $15^{+5}_{-3}$. Estimating
the attenuation towards the BLR is highly uncertain, because in the
high-density BLR the Balmer lines can be powered by other processes such
as collisional excitation. Assuming an intrinsic ratio of 3.06
\citep[from observations of blue quasars;][]{dong+2008}, we obtain
$\Avhatb = 4.4^{+0.7}_{-0.5}$~mag, which would imply an intrinsic \Halpha flux 25 times brighter than what we measure. Such a high attenuation may be at odds with the
reported weakness of LRDs at MIR wavelengths \citep{wang+2024a,williams+2024,axins+20,degraaff+2025}.

On the other
hand, an intrinsic Balmer ratio of 10 \citep{ilic+2012} would
give a dust attenuation $\Avhatb = 1.2$~mag, fully consistent with the dust
attenuation inferred from the narrow-line ratios and implying no additional
dust towards the BLR.

\subsection{Medium-resolution spectrum -- host-galaxy ISM}\label{s.an.ss.g395m.jades}

\begin{figure}
  \includegraphics[width=\columnwidth]{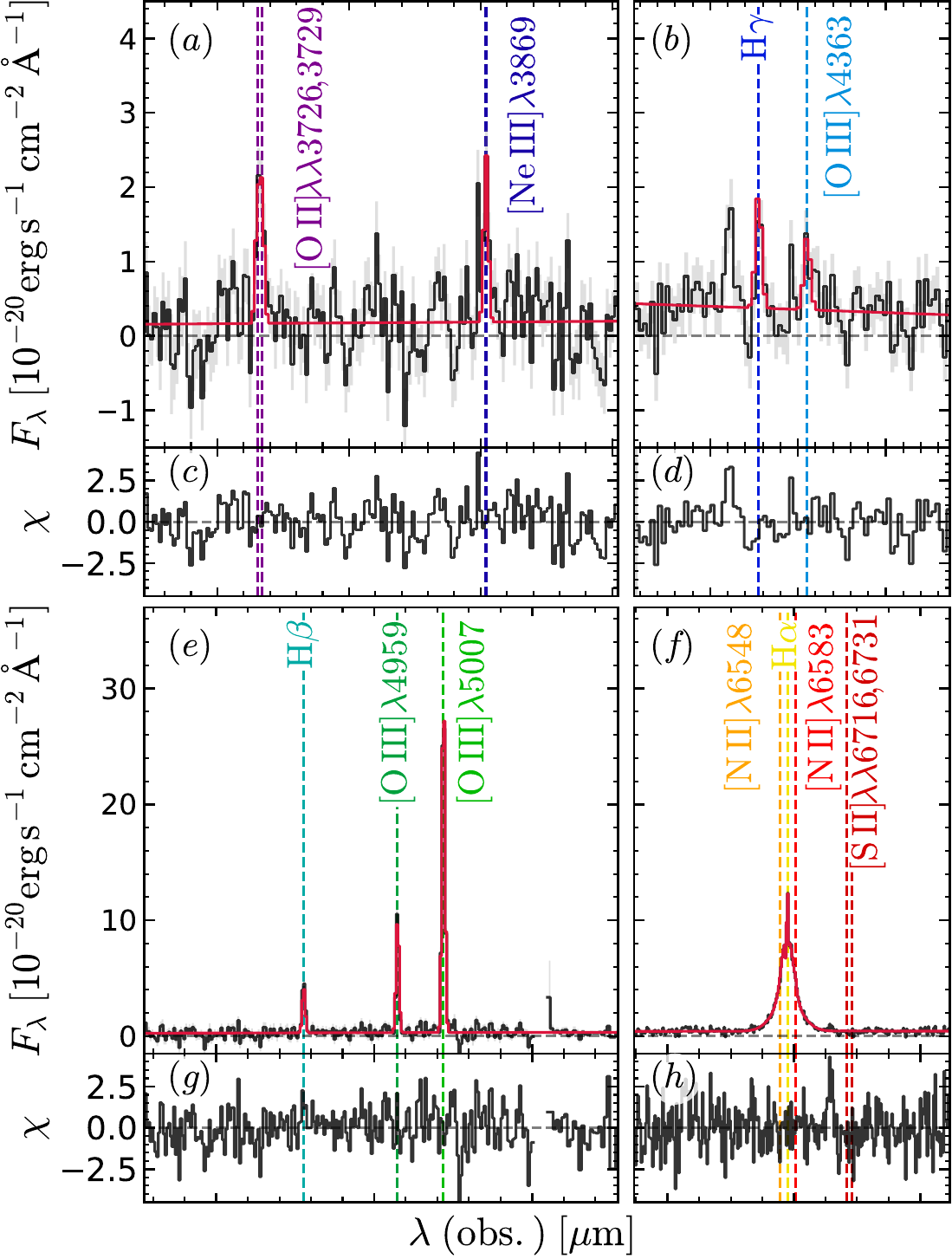}
  {\phantomsubcaption\label{f.r1000.a}
   \phantomsubcaption\label{f.r1000.b}
   \phantomsubcaption\label{f.r1000.c}
   \phantomsubcaption\label{f.r1000.d}
   \phantomsubcaption\label{f.r1000.e}
   \phantomsubcaption\label{f.r1000.f}
   \phantomsubcaption\label{f.r1000.g}
   \phantomsubcaption\label{f.r1000.h}}
  \caption{Piecewise fit of the medium-resolution spectroscopy for \target,
  showing the region around \OIIall and \NeIIIL (panel~\subref{f.r1000.a}),
  around \Hgamma and \OIIIL[4363] (panel~\subref{f.r1000.b}),
  \Hbeta and \OIIIall (panel~\subref{f.r1000.e}), and \Halpha
  (panel~\subref{f.r1000.f}). Panels~\subref{f.r1000.c}, \subref{f.r1000.d},
  \subref{f.r1000.g} and \subref{f.r1000.h} show the model residuals normalised
  by the noise.
  All narrow lines are modelled assuming the same
  redshift $z_\mathrm{n}$ and intrinsic dispersion $\sigma_\mathrm{n}$.
  The Balmer lines \Hgamma and \Hbeta are modelled assuming fixed
  ratios to \Halpha, and a \citet{gordon+2003} dust attenuation. Note that
  this model includes an unresolved \Halpha absorber near rest frame, with 
  informative priors derived from modelling the G395H data. Including this
  absorber has a strong impact ($>0.5$~mag) on the recovered dust attenuation
  $A_V$. All lines are seen in the 2-d spectrum.}\label{f.r1000}
\end{figure}

To derive the physical properties of the host galaxy, we use the medium-resolution spectra, because they cover the full range of rest-optical strong lines,
unlike high-resolution G395H spectroscopy.
We detect no line emission in the G140M spectrum (except for 
rest-frame optical lines from the interloper, Appendix~\ref{a.interl}), so we 
use only the G235M and G395M spectra. We define four spectral windows around
\OIIall and \NeIIIL, \Hgamma and \OIIIL[4363], \Hbeta and \OIIIall, and \Halpha,
and we use a joint 
model to fit the spectrum simultaneously in each window (Fig.~\ref{f.r1000}).
The background is modelled piecewise inside each window as a first-order 
polynomial (2 free parameters per window, 8 total). The narrow lines are all
modelled with the same redshift $z_\mathrm{n}$ and intrinsic velocity
dispersion $\sigma_\mathrm{n}$ (2 free parameters). 
All lines are parametrized via their dust-corrected flux, requiring one
free parameter for the \citetalias{gordon+2003} dust attenuation. This approach
models simultaneously the entire spectrum \citep[similar to ][]{greene+2024},
and ensures a Bayesian approach to estimate the dust attenuation.
\OIIall is parametrized by the \OIIL[3726] flux and the \OIIL[3729]/\OIIL[3726] ratio \citetext{2 free 
parameters; the doublet ratio is constrained to the range 0.3839--1.4558, following \citealp{sanders+2016}}. \OIIIL[4363] and \OIIIL[5007] are parametrized via their 
flux, whereas the flux of
\OIIIL[4959] is fixed to 0.335 the \OIIIL[5007] flux (2 free parameters).

The narrow-component Balmer lines \Hgamma, \Hbeta and \Halpha are parametrized
via the \Halpha flux alone (one free parameter). The fluxes of \Hbeta and \Hgamma are fixed to their Case-B recombination 
ratios with respect to \Halpha, where we assumed $\Telec=10,000$~K and
$\nelec=500$~\pcm. These values are justified as follows; for the density, we use the 
mean density of star-forming galaxies at $z\approx5$
\citep{isobe+2023}; \textit{a posteriori}, 
this value is not inconsistent with the mean value we infer from the measured \OIIL[3729]/\OII[3726] ratio (although the uncertainties are very large).
The impact of this density choice on our model is in any case small; 
for example, the mean density of narrow-line AGN in the local Universe is found
to be significantly higher, when estimated from higher-ionization species
\citep[$\nelec=10,000\text{--}30,000~\pcm$;][]{binette+2024}, but even at
these higher densities the \Halpha/\Hbeta and \Hgamma/\Hbeta ratios changes by
less than 2~percent.
\citet{mcclymont+2024} reported the discovery of galaxies at $z>4$ with observed Balmer-line ratios which are inconsistent with Case-B recombination. However, these galaxies seem associated with density-bounded nebulae, and tend to exhibit blue continuum and \Lyalpha emission,
contrary to what we observe here.
For the temperature, we reason that regions of higher 
temperature (as those producing auroral \OIIIL[4363] emission) have lower Balmer-line emissivity, as highlighted by the anti-correlation between \OIIIL/\Hbeta and
\OIIIL[4363]/\OIIIL in the narrow-line regions of type 1 AGN 
\citep{baskin+laor2005}; therefore, the overall Balmer ratio should be biased
towards lower-temperature regions. In any case, also the assumptions on \Telec do not have a 
major impact on our findings.

Four more narrow lines are included: \NIIall and \SIIall;
these are parametrized by 3 free parameters as
in Section~\ref{s.an.ss.g395h.bt}.
Based on the fit residuals, the only clearly detected broad-line 
component is that of \Halpha, which is modelled in the same way as for the
G395H data (Section~\ref{s.an.ss.g395h.bt}), using two Gaussians and
five free parameters.
Finally, we add the \Halpha absorber, with its 4 free parameters. In total,
this model has 28 free parameters, 11 for the narrow lines, 5 for the broad \Halpha, 4 for the 
absorber, and 8 for the local background.
Given that these data are unable to constrain the properties of the
absorber (Section~\ref{s.resol}), for the four free parameters 
describing the absorber we adopt informative priors based on the 1-d 
marginalised posterior probabilities from the G395H model inference from JADES (Table~\ref{t.pars}, Column 3), without 
considering the covariance in the posterior
probability.
We broaden these posteriors by a factor of two (we use a log-normal for $C_f$
and $\tau_0(\Halpha)$). The goal of this setup is to capture the
uncertainty in the narrow-line flux due to poorly constrained absorption, while 
still adopting a physically motivated range of absorber properties.

The resulting properties from the above model are reported in Table~\ref{t.pars} (Column 4).
These
measurements are the basis for our analysis of the galaxy ISM. To avoid
confusion, we do not use the posterior for the broad-\Halpha and absorber
parameters; these model components are included only to obtain an unbiased
flux of the narrow \Halpha and of the dust attenuation. For analysing 
the broad component, we used the G395H data (Section~\ref{s.an.ss.g395h.bt}; Table~\ref{t.pars}, Column 2).

From the medium gratings, we obtain a dust attenuation $\Avhatn=1.2\pm0.2$~mag, lower than the value
inferred from the G395H spectrum ($\Avhatn=1.4_{-0.2}^{+0.3}$~mag; Table~\ref{t.pars}).
The main result of our joint fit to the Balmer lines is to up-weight the
\Hgamma/\Hbeta ratio in the estimate of the dust attenuation, yielding a higher value 
compared to fitting the lines separately and then estimating \Avhatn (a
method that yields $\Avhatn = 0.9\pm0.2$~mag).
Our high attenuation values are not implausible, given the red slope of the
rest-optical continuum (e.g., Figs.~\ref{f.data.c} and~\ref{f.r1000}), and
other reported measurements of $A_V$ in the literature \citep{killi+2024,furtak+2024,akins+2024,ji+2025},
including from narrow-lines \citep{brooks+2024,deugenio+2025c}.
The uncertainties are large, correctly reflecting the uncertainties in the
properties of the absorber (which in our case are informed by the results of the G395H fit).

Including the known absorber is essential for recovering unbiased
emission-line properties.
For instance, removing the absorber altogether from the model leads to
severely under-predicting the narrow \Halpha flux; with our physically
motivated model we obtain $F_\mathrm{n}(\Hbeta)$  and $F_\mathrm{n}(\Halpha)$
of $(0.53\pm0.03)\fluxcgs[-18]$ and $(1.55\pm0.08)\fluxcgs[-18]$, respectively.
This yields a dust attenuation $A_V=0.07\pm0.03$~mag, consistent with no dust and much
lower than the true value. In fact, the \Halpha flux is so severely
underestimated that the model under-predicts \Hbeta too, since our model
enforces a minimum Balmer decrement of 2.86. Removing this constraint, we
would obtain an unphysical Balmer decrement of 1.5.

\section{Physical properties}\label{s.phys}

\subsection{The supermassive black hole}\label{s.phys.ss.smbh}

From the G395H \Halpha model of the fiducial $R_\mathrm{ap}=0.25$-arcsec
\blackthunder spectrum (Section~\ref{s.an.ss.g395h.bt}), we obtain a BLR
$\fwhm = 1740^{+80}_{-90}~\kms$ (Table~\ref{t.pars}, Column 1). The best-fit
value for the higher-SNR $R_\mathrm{ap}=0.125$-arcsec spectrum has smaller
uncertainties but is fully consistent ($\fwhm = 1790\pm60~\kms$; Column 2).
We use the larger aperture because it minimizes aperture-loss corrections,
ensuring the highest accuracy in the  inferred dust attenuation.
The absorption-corrected \Halpha luminosity is $7.3\pm0.2\times10^{42}~\ergs$,
increasing to $15\pm2\times10^{42}~\mathrm{erg\,s^{-1}}$ after correcting for
dust attenuation using $\Avhatn=1.0\pm0.2$~mag (for all derived quantities, we
calculate the uncertainties from the MCMC chains, in the same way as the
posterior probability on the other model parameters).
We assume virial equilibrium, and use the single-epoch black-hole mass estimator of \citet{reines+volonteri2015},
appropriate for SMBHs at the low-mass end of the calibration. This gives
$\log \mbh/\Msun = 7.65\pm0.05$ ($7.50\pm0.04$ without dust correction), dominated
by uncertainties in the calibration, of order 0.3~dex.

Alternatively, instead of applying a dust-attenuation correction based on the
Balmer decrement of the narrow-line region, we can apply the higher attenuation
inferred from the Balmer decrement of the BLR. We estimate a maximum value
of $\Avhatb = 4.4^{+0.7}_{-0.5}$~mag, giving a broad \Halpha luminosity of
$1.7^{\times 1.6}_{\div 1.5} \times10^{44}~\mathrm{erg\,s^{-1}}$. With this much
higher luminosity, we get $\log \mbh/\Msun = 8.1\pm0.1$, which we treat as an
upper limit on the SMBH mass, due to the large range of possible intrinsic
Balmer ratios for BLRs \citetext{\citealp{dong+2008,ilic+2012}; see
Section~\ref{s.an.ss.dust}}. Recently, joint measurements of broad \Halpha, \Hbeta and \Hgamma in a $z=6.68$ LRD have shown observed ratios that are inconsistent with standard attenuation laws \citep{deugenio+2025g,nikopoulos+2025}, which would invalidate our
maximal $A_V=4.4$~mag. This scenario may be actually quite common, because LRDs -- including \target -- tend to have remarkably large \Halpha/\Hbeta.

To obtain the bolometric luminosity, we use the calibration of
\citet{stern+laor2012} based on \Halpha. The resulting Eddington ratio is $\ledd = 0.35$ (0.24 without dust-attenuation correction).
For the extreme case of a highly obscured BLR, we would estimate $\ledd=1.4$,
driven by the fact that (with our single-epoch \mbh calibration and for fixed
FWHM), we have approximately $\ledd \propto \sqrt{ F_\mathrm{b}(\Halpha) }$.

There is also the possibility that the double-Gaussian profile is due to
two SMBHs; dual AGNs have already been reported
\citep[e.g.,][]{maiolino+2024,ubler+2024a,ubler+2025} and their
rate is more common at high redshift, at least on kpc scales \citep{perna+2025}.
However, while `double-Gaussian' profiles are widespread among LRDs \citep[e.g.,][]{juodzbalis+2025}, there is currently very little evidence for kinematic offsets between the two broad-line components.
At plausible separations of 1--100~pc, the orbital velocity of two SMBHs with mass 
comparable to \target would range between 60--600~\kms; admittedly our observations cannot 
fully constrain the low-velocity envelope of this range, due to the difficulty of measuring 
a velocity offset that is small compared to the line widths, without even considering
inclination effects or the complication of the narrow line and \Halpha absorber. 
However, while some confirmed cases do have small velocity offsets
\citep[e.g., ][$\delta\,v = 40~\kms$]{ubler+2024a}, the general paucity of larger
offsets seems to be a problem for the double-SMBH interpretation.

An alternative interpretation is that the broad Balmer lines are shaped by electron scattering, not virial motions \citep{rusakov+2025}. This scenario is discussed in
Section~\ref{s.disc.ss.overmassive}, but the relevant profile fits and calculations are
reported in Appendix~\ref{a.exponential}.

\subsection{The host galaxy}\label{s.phys.ss.host}

We do not attempt to measure the stellar mass of \target. The usual uncertainties
about the stellar mass of LRDs \citetext{\citealp{wang+2024a,wang+2024b};
\citetalias{juodzbalis+2024b}; \citealp{setton+2024,ma+2024}} are compounded by the
presence of \interlop and of its UV-bright spiral arm (Fig.~\ref{f.ifsimg}). Besides,
two other LRDs have been confirmed to have time-variable\footnote{We note that the
prevalence of time variability among LRDs is debated, with NIRCam studies of large samples of LRDs showing no time variability \citep{zhang+2025}.} rest-frame optical continuum
\citep{ji+2025,naidu+2025}, implying that the AGN dominates the optical continuum, from which
stellar masses are usually inferred.
With the interloper contaminating the UV, we have
no handle on the stellar SED.

Assuming virial equilibrium, we can estimate the galaxy dynamical mass from the galaxy size and from the narrow-line width, following the
approach outlined in \citet{ubler+2023} and \citet{maiolino+2024}.
We adopt the calibration of \citet{vanderwel+2022}, with S\'ersic index $n=4.5$,
projected axis ratio $q=0.44$, and $\re = 210$~pc (Table~\ref{t.size}).
We correct the observed narrow-line velocity dispersion $\sigma_\mathrm{n} =
49\pm2~\kms$ upward by 0.2~dex, following the calibration of \citet{bezanson+2018}
that converts gas velocity dispersions to stellar values.
With these numbers, we obtain $\log(\mdyn/\Msun) = 9.1$~dex, which we also
use as an upper limit on the stellar mass. 
The large uncertainties on $q$ and \re translate
into uncertainties on \mdyn of order 0.1~dex. For comparison, the virial 
calibration from \citet{stott+2016} gives
$\log (\mdyn/\Msun) = 8.6$, meaning that our dynamical mass is dominated by
systematics.
We remark that within a factor of 2, this \mdyn of \target is comparable
to the stellar mass of the satellite \satlarge (Section~\ref{s.sats.ss.mstar}).
Conversely, using the NIRSpec \OIIIL-inferred size we obtain
$\log(\mdyn/\Msun) = 10.1$~dex \citep[9.2~dex using][]{stott+2016}. For the \citet{vanderwel+2022} calibration, the 1-dex increase
is due to the combination of 4.5-larger \re, plus differences in $q$ and $n$. For both NIRCam and NIRSpec, significant departures from virial equilibrium would probably over-estimate \mdyn.

\subsection{ISM properties}\label{s.phys.ss.ism}

From the narrow-line parameters of the medium-resolution model (Table~\ref{t.pars}, second column), we can estimate the
physical properties and chemical abundances of the ISM of the host galaxy. In
principle, we can estimate the electron temperature
$T_\mathrm{e}(\mathrm{O^{++}}$) of the O\,\textsc{iii}-emitting gas
directly from the dust-corrected \OIIIL[4363]/\OIIIL[5007] ratio. This diagnostic is
insensitive to the electron density over a wide range of values, so we assume 
$\nelec=500~\pcm$, which is typical of the O\,\textsc{ii}-emitting
regions of star-forming galaxies at $z\sim5$ \citep{isobe+2023}, and agrees with
our (very uncertain) measurements based on the \OIIall doublet ratio. 
To infer $T_\mathrm{e}$ we use \pyneb \citep{luridiana+2015}, and we do so for 
each of the MCMC chains, obtaining $T_\mathrm{e}(\mathrm{O^{++}}) =
26,000\pm4,000$~K.
From this value, we estimate the O$^{++}$ abundance to be $12 + \log (\mathrm{O^{++}/H^+}) = 7.4\pm0.1$ using the calibration of \citet{dors+2020}.
Further using the calibration of \citet{izotov+2006}, we estimate also $T(\mathrm{O^+})=21,000\pm3,000$~K (but we note that this calibration is used outside its established validity range).
From \nelec and  $T(\mathrm{O^+})$, using again
the calibration of \citet{dors+2020}, we find that the O$^+$ abundance is small (6.6~dex).
The total metallicity is $12 + \log(\mathrm{O/H})=7.5\pm0.1$, or 10~percent solar \citep{asplund+2009}.

\begin{figure}
\includegraphics[width=\columnwidth]{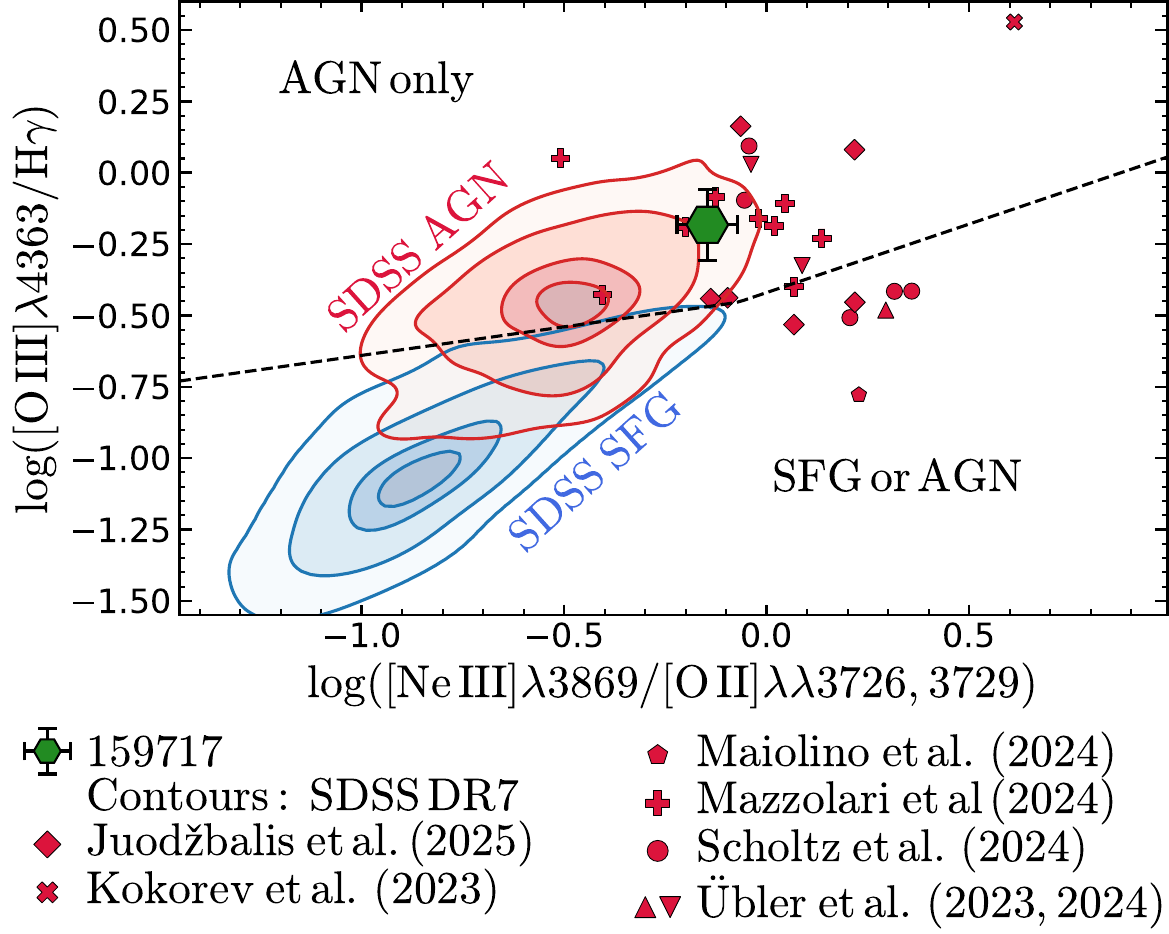}
\caption{The photo-ionization diagnostic of \citet{mazzolari+2024} places our main target \target (hexagon) firmly in the `AGN only' region, above the dashed demarcation line. For reference, we show star-forming and AGN-host galaxies from SDSS (contours), and a range
of high-redshift AGN from the literature.}\label{f.ohno}
\end{figure}

To assess the photoionization source, we use the methods of \citet[][Fig.~\ref{f.ohno}]{mazzolari+2024}.
We show the emission-line ratios \OIIIL[4363]/\Hgamma \textit{vs} \NeIIIL/\OIIall, since
these two ratios are robust against dust attenuation (unlike the version of the diagram
that replaces \NeIIIL with \OIIIL). Our main target \target is found in the AGN-only region
of the diagram, driven by the high \OIIIL[4363]/\Hgamma ratio. A number of high-redshift
AGNs are also shown, including both Type 1 AGN 
\citep{juodzbalis+2025,kokorev+2023,maiolino+2024,mazzolari+2024,ubler+2023,ubler+2024a}
and Type 2 AGN \citep{mazzolari+2024,scholtz+2023,ubler+2024a}.

\subsection{The absorbing clouds}\label{s.phys.ss.absorb}

As shown in Fig.~\ref{f.abs.b}, there is a clear absorption component near the rest-frame centroids of narrow and broad \Halpha emission lines in \target.
Such Balmer absorption is frequently seen in spectroscopically confirmed broad-line LRDs revealed by \jwst, with a detection fraction of $\sim20$ percent \citep{juodzbalis+2024b,matthee+2024}.
Clearly, the case of \target shows that the occurrence rate of Balmer absorption can be even higher because NIRSpec medium-resolution spectroscopy is not effective in selecting such narrow absorption close to the line centroids (see Fig.~\ref{f.abs}).
The absorption in LRDs is unlikely to have a stellar origin, as the absorption is usually deeper than the underlying continuum with a high EW, meaning the absorbing medium must (also) be absorbing the broad Balmer lines.
The same situation is seen in \target as shown in Fig.~\ref{f.g395h.jades}.

\begin{figure}
\includegraphics[width=\columnwidth]{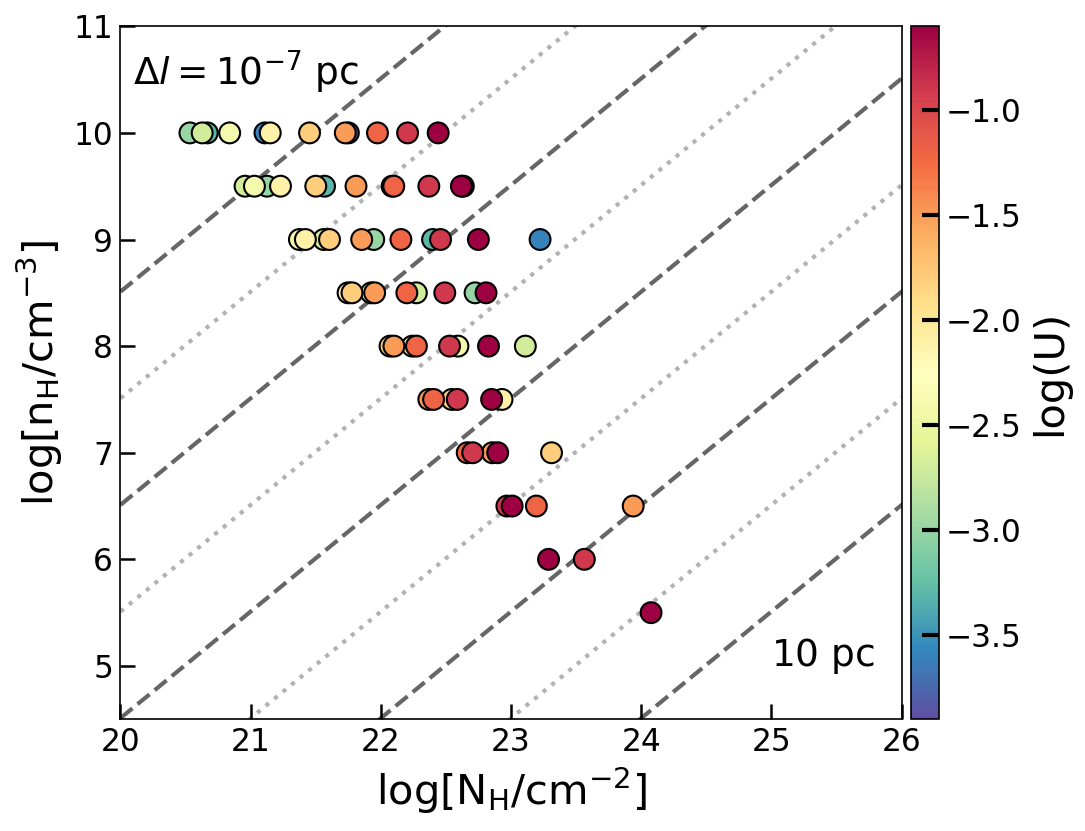}
\caption{Plausible parameter space for the physical conditions of the warm absorber in \target that produces the \Halpha absorption. Coloured circles correspond to \cloudy photoionization models for the absorber with a range of hydrogen density, column density, and ionization parameter.
Dashed lines and dotted lines represent the effective thickness, $N_{\rm H}/n_{\rm H}$, of the absorber.
With a single absorption line, $N_{\rm H}$, $n_{\rm H}$, and $U$ show significant degeneracy. Still, there is a general indication of high densities and/or high column densities for the absorber.
}\label{f.cloudy}
\end{figure}

Naively, one can model the absorption as a slab of gas obscuring our line-of-sight (LOS), as done in \citet{juodzbalis+2024b}.
To produce a strong Balmer absorption, the gas needs to be dense enough to collisionally excite hydrogen to the level $n=2$, and/or thick enough to trap \Lyalpha photons.
With the optical depths we fit for the Balmer absorption lines in Table~\ref{t.pars}, we calculate the column density of the excited hydrogen using
\begin{equation}
    N_{\mathrm{H}\,(n=2)} = \frac{m_{\rm e}c}{4\pi e^2 f_0 \lambda _0}\tau,
\end{equation}
where $m_{\rm e}$ is the electron mass, $c$ is the speed of light, $e\equiv q_{\rm e}/\sqrt{4\pi \epsilon _0}$ is the `Gaussian electron charge' ($q_{\rm e}$ is the electron charge and $\epsilon_0$ is the vacuum permittivity), $f_0$ is the oscillator strength, $\lambda _0$ is the central wavelength, and $\tau$ is the integrated optical depth with a dimension of velocity.
From the value of $N_{\rm H~(n=2)}$ we can
use \cloudy models to infer the density and column density of hydrogen, $n_\mathrm{H}$ and $N_\mathrm{H}$. Unfortunately, due to the lack of constraints on the ionization fraction, the conversion between $n=2$ and the total hydrogen column density is very uncertain, resulting in poorly constrained $N_\mathrm{H}$ (Fig.~\ref{f.cloudy}).

\begin{figure}
{\phantomsubcaption\label{f.ewvar.a}
 \phantomsubcaption\label{f.ewvar.b}}
\includegraphics[width=\columnwidth]{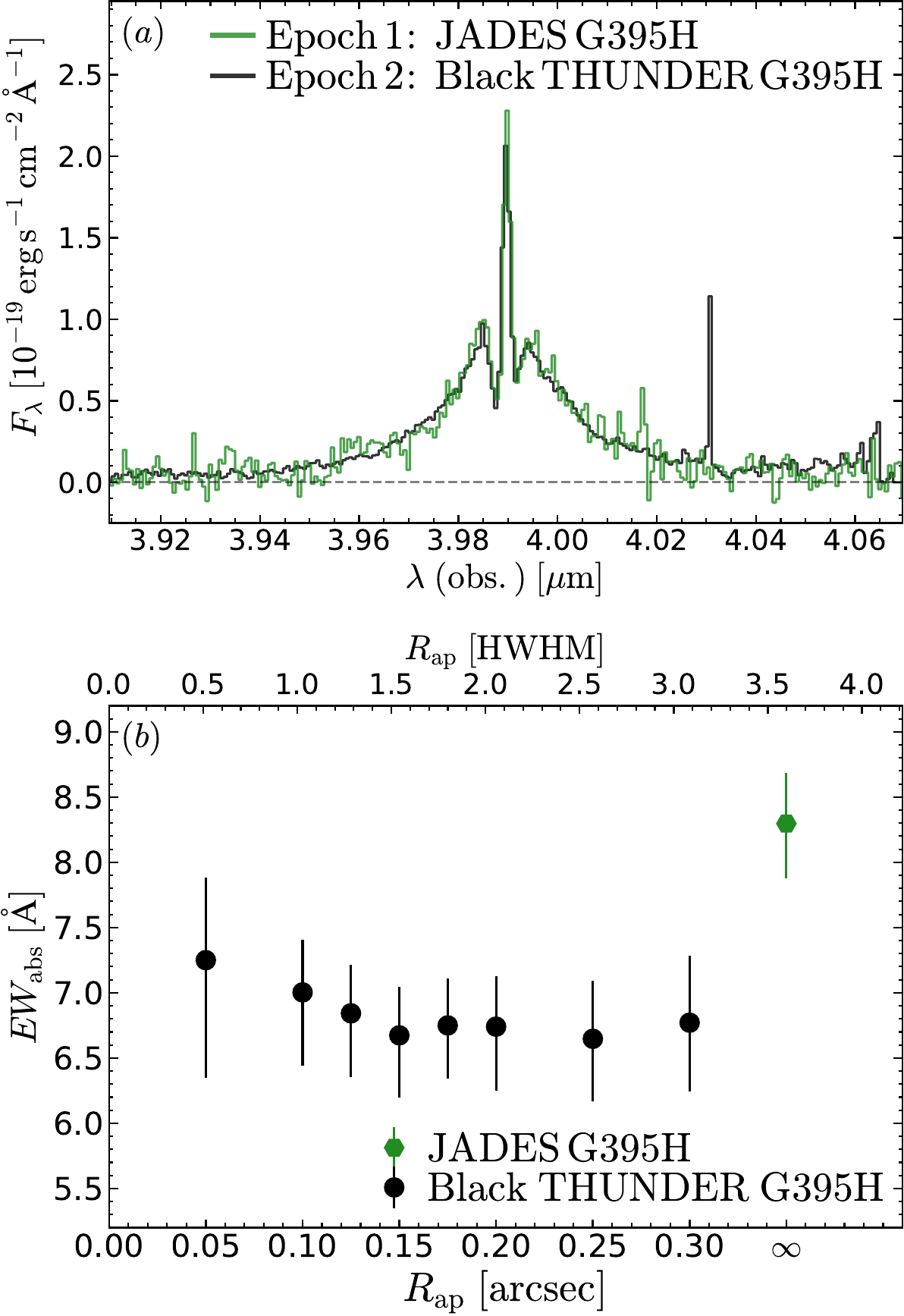}
\caption{Panel~\subref{f.ewvar.a} compares the \Halpha emission from JADES (Epoch 1, December 2023, green) and \blackthunder (Epoch 2, black; December 2024). Panel~\subref{f.ewvar.b} shows the EW of the \Halpha absorption (JADES, green hexagon; \blackthunder, black circles); the black circles show EW as a function of the aperture radius. The top axis indicates the extent of the radial aperture in units of the
half-width at half maximum (HWHM) of the NIRSpec/IFS PSF. The fiducial measurements differ by 2.6~\textsigma.}\label{f.ewvar}
\end{figure}

Intriguingly, the EW of the \Halpha absorption is different between the JADES and \blackthunder G395H observations. In Fig.~\ref{f.ewvar.a} we compare directly the two emission lines. Note that there may be small flux-calibration systematics affecting the normalization. In panel~\subref{f.ewvar.b}, we compare the EW measured from JADES (hexagon) to the
EWs measured from \blackthunder. To check for
systematics, for \blackthunder we 
extended the measurements of Section~\ref{s.an.ss.g395h.bt} to a set of apertures of
increasing radius. It is clear that the value from
\blackthunder is systematically lower, with 2.6-\textsigma significance (P-value$<0.005$).
This result is unchanged if we repeat the
\blackthunder measurements but modelling only \Halpha, without including broad \Hbeta, hence the
systematic difference is not due, e.g., to tying
the absorber kinematics and covering factor between
the two Balmer lines. Moreover, we can rule 
out that the observed difference is due to errors in the flux calibration, because EW measurements are completely
insensitive to flux calibration. In principle, aperture
losses could play a role, since our EW measurements
combine flux emitted on different spatial scales;
however, the \blackthunder data also rules out this
possibility, because the EW stays constant (within
the uncertainties) as we change the aperture radius.

Under the hypothesis that the absorber is in front of both
the broad lines and the continuum, the EW of a non-saturated
absorber would not change with the underlying continuum.
This would suggest that it may be the absorber itself that is
varying on a rest-frame timescale of two months.
The picture is complicated by the fact that in the JADES
observations, the absorber is nearing saturation
($\tau_0(\Halpha) \sim 3$; Table~\ref{t.pars}). New observations
are needed to both confirm the time variability, and to
establish if it originates in the absorber itself or rather in the underlying continuum and BLR.

\section{Nearby sources}\label{s.sats}

\subsection{Aperture spectra}\label{s.sats.ss.apspec}

\begin{figure}
    \centering
    {\phantomsubcaption\label{f.satap.a}
     \phantomsubcaption\label{f.satap.b}
     \phantomsubcaption\label{f.satap.c}
     \phantomsubcaption\label{f.satap.d}
     \phantomsubcaption\label{f.satap.e}
     \phantomsubcaption\label{f.satap.f}
     \phantomsubcaption\label{f.satap.g}
     \phantomsubcaption\label{f.satap.h}
     \phantomsubcaption\label{f.satap.i}
     \phantomsubcaption\label{f.satap.j}
     \phantomsubcaption\label{f.satap.k}
     \phantomsubcaption\label{f.satap.l}}
    \includegraphics[width=\columnwidth]{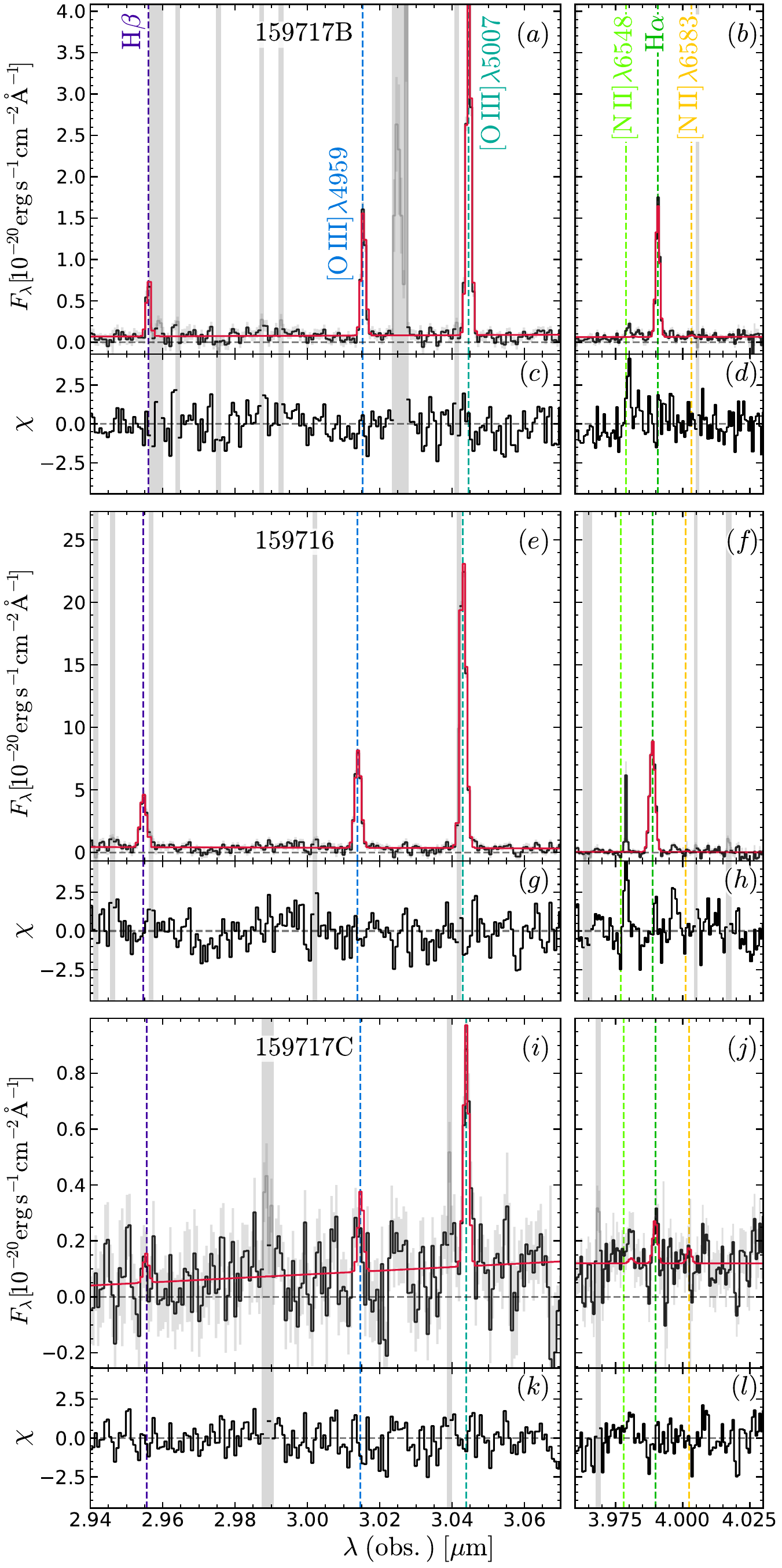}
    \caption{\blackthunder G395H aperture spectra for the three nearby sources,
    showing \Hbeta and \OIIIall (left) and \Halpha (right). The three sources show a
    range of line widths and ionization properties.}\label{f.satap}
\end{figure}

The aperture spectra of three associated sources are shown in Fig.~\ref{f.satap}.
\satlarge and \satsmall are undoubtedly detected and associated with \target
(panels~\subref{f.satap.a}--\subref{f.satap.f}).
To the south west, we find evidence for a spatially detached emission-line source,
detected only in \OIIIL, which we call \sattiny (panels~\subref{f.satap.i}
and~\subref{f.satap.j}). Below we report the emission-line
properties of these three sources, using a simplified version of the model from
Section~\ref{s.an.ss.g395h.bt}, where we removed both the broad-line \Hbeta and \Halpha,
and the gas absorber.

\satsmall (Fig.~\ref{f.satap.a}--\subref{f.satap.d}) is spatially unresolved (Section~\ref{s.an.ss.size})
and has Gaussian line profiles with $\sigma_\mathrm{n}=41\pm1$~\kms, narrower than for \target. The continuum in this location is dominated by the bright spiral arm of
\interlop, which prevents us from robustly detecting \satsmall's own continuum.
Using the Balmer lines, and under the same
assumptions as for \target (Section~\ref{s.an.ss.g395h.bt}), we infer
$A_V=0.1\pm0.3$~mag, consistent with no dust. We estimate the star-formation rate
(SFR) using the \Halpha-to-SFR scaling of \citet{shapley+2023}. After applying an
aperture correction of 1.5, we obtain $\log(\mathrm{SFR}[\Msun~\peryr]) = -0.5\pm0.1$.
Lacking an accurate continuum deblending, we do not attempt to measure a stellar mass for this target, but we note that
if it lay on the star-forming sequence it would have a stellar mass
$\log(\mstar/\Msun) = 7.6\text{--}8.1$, where the lowest value is from
\citet[][their $z=4.5\text{--}5$ redshift bin]{cole+2023}, while the 
largest value is from the bias-corrected theoretical work of 
\citet{mcclymont+2025} and the corresponding empirical measurements of 
\citet{simmonds+2025}. Of course, this estimate is very uncertain,
not just due to the large scatter about the main sequence
\cite[0.3~dex;][]{cole+2023}, but primarily due to the possibility that the
galaxy may not lie on the main sequence \citep[e.g., due to
environment-driven quenching in close environments;][]{alberts+2024,deugenio+2025d}, and to the possibility that the emission
lines are not powered by star-formation photoionization. In fact, we cannot rule
out that \satsmall may be externally photo-ionized (see also Section~\ref{s.sats.ss.photoion}).

\satlarge (Fig.~\ref{f.satap.e}--\subref{f.satap.h}) is much more extended.
This source is clearly detected in the continuum, both in NIRSpec and NIRCam, hence it
is clearly a satellite galaxy with stellar content (Section~\ref{s.sats.ss.mstar}).
Its emission lines are also Gaussian, but have
velocity dispersion $\sigma_\mathrm{n} = 67\pm2$~\kms, broader than both \satsmall
and \target.

The last source, \sattiny, is a tentative detection (Fig.~\ref{f.satap.i}--\subref{f.satap.l}). As for \satsmall, there is substantial contamination from \interlop
in the foreground, hence the continuum level is unreliable.
This system has a diffuse morphology. While the SNR is low, three arguments support the detection
of this system. In addition to \OIIIL in the aperture spectrum (8-\textsigma detection), we also
detect this source in the emission-line maps (Section~\ref{s.sats.ss.specres}). Moreover, there is
a marginal detection of \Halpha at the same redshift as \OIIIL (4-\textsigma significance). 
Finally, we have some indications that weak \OIIIL emission may be present in the NIRCam 
F200W-F277W image (Section~\ref{s.sats.ss.specres}), at a location that matches very well the
NIRSpec findings. This source has an intrinsic velocity dispersion $\sigma_\mathrm{n}=70\pm10$~\kms,
which is large for such a faint source. We also find a $\OIIIL/\Hbeta>10$.

Unfortunately, none of these sources has useful constraints on the dynamical mass.
For \satsmall and \sattiny, we lack an accurate measurement of their size and morphology.
Using the same formula as in Section~\ref{s.phys.ss.host}, and adopting conservatively
$\re<0.1$~pc, $q=1$ and $n=1$, we obtain $\log(\mdyn/\Msun)<9.1$ and $<9.6$ for
\satsmall and \sattiny, respectively.
For \satlarge, the use of the virial theorem seems inappropriate, since this galaxy may
be out of stationary equilibrium, as discussed in Section~\ref{s.sats.ss.specres}.

\subsection{Stellar mass measurement for \satlarge}\label{s.sats.ss.mstar}

\begin{figure}
    \centering
    {\phantomsubcaption\label{f.mstar.a}
     \phantomsubcaption\label{f.mstar.b}
     \phantomsubcaption\label{f.mstar.c}
     \phantomsubcaption\label{f.mstar.d}}
    \includegraphics[width=\columnwidth]{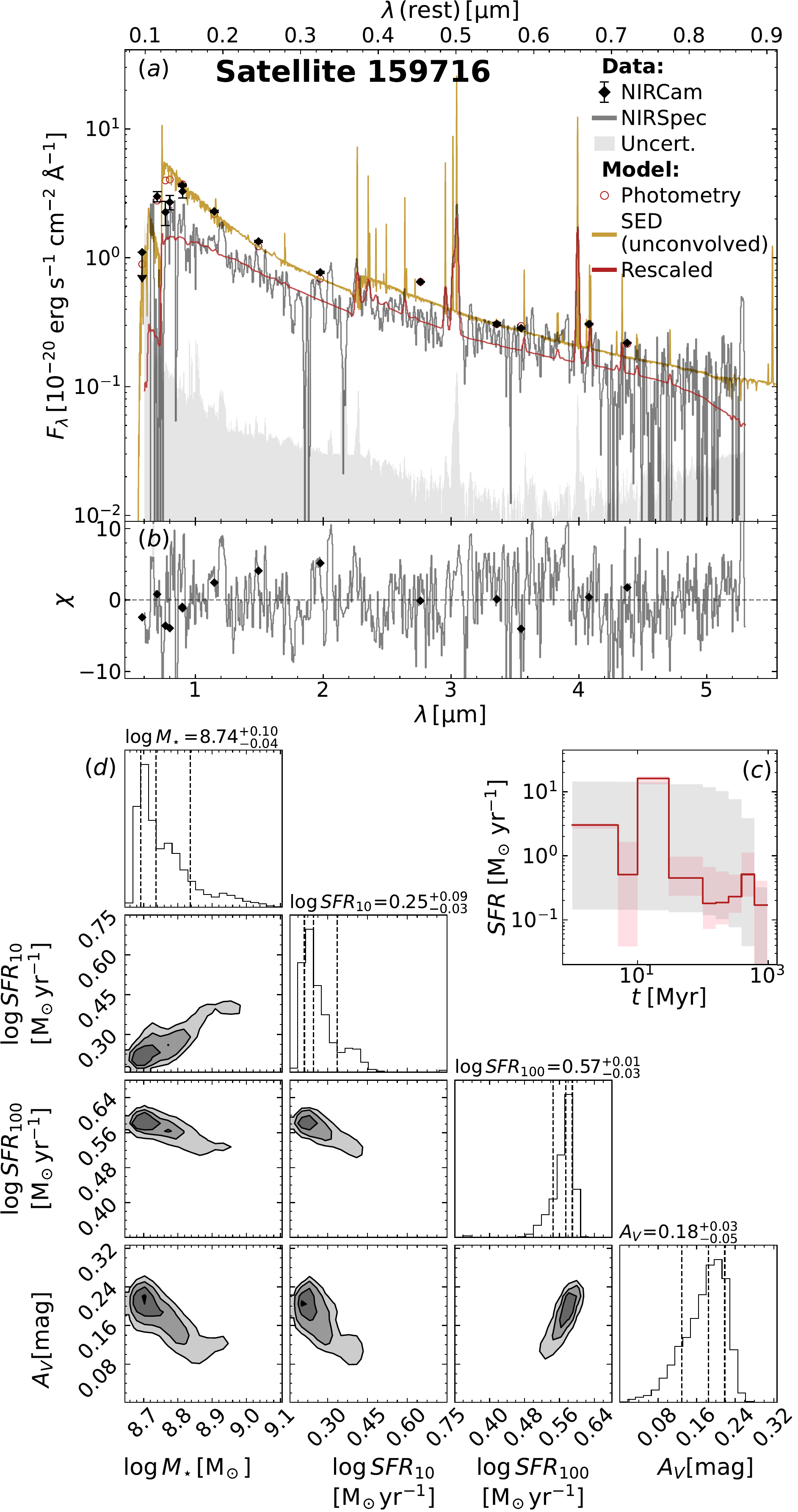}
    \caption{Summary of the \prospector inference of the physical properties of \satlarge.
    The data and best-fit model are shown in panel~\subref{f.mstar.a}; a weak Balmer
    break is evidence in both the NIRSpec data (grey line) and in the fiducial model
    (the sand and red lines are the model, and the model rescaled to match the shape
    and normalization of the spectrum). The residuals are shown in panel~\subref{f.mstar.b}.
    Panel~\subref{f.mstar.c} shows the
    SFH, with the prior in grey. The corner diagram (panel~\subref{f.mstar.d}) displays
    a subset of the model parameters.
    }\label{f.mstar}
\end{figure}

We measure the stellar mass of only one galaxy, \satlarge. Our method of 
choice is the Bayesian inference framework \prospector 
\citep{johnson+2021}. The SED model is derived from 
\citet{tacchella+2022}, while the comparison with observations follows the
setup of \citet{deugenio+2025b}, with improvements described in 
\citet{baker+2025,carniani+2025}. Here we summarize the most important
aspects. We use stellar-population template spectra derived from the
\textsc{mist} isochrones \citep{choi+2016} and \textsc{C3K} model
atmospheres \citep{conroy+2019}, synthesized using \textsc{fsps} \citep{conroy+2009,conroy_gunn_2010}. The nebular continuum uses
pre-computed \textsc{cloudy} models \citep{byler+2017}. The dust 
attenuation law follows \citet{noll+2009} and the implementation of
\citet{kriek+conroy2013}. A dual dust screen captures attenuation due
to diffuse dust, and the additional birth-cloud attenuation toward star-forming regions \citep{charlot+fall2000}, following the parametrization
of \citet{tacchella+2022}. We adopt a non-parametric star-formation history
(SFH) with 9 time bins between $z=20$ and the epoch of observation $t=0$.
The first four bins are manually set to $t=5$, 10, 30, and 100~Myr, while
the remaining five bins are logarithmically spaced between 100~Myr and
$z=20$. We use a `rising' probability prior \citep{turner+2025}, where the 
SFH follows the increasing mass accretion rate on dark-matter haloes 
\citep{tacchella+2018}. We also use a 2\textsuperscript{nd}-order Chebyshev
polynomial to rescale the best-fit model to the level and shape of the 
observed spectrum; this means that our extensive quantities (\mstar, SFR) 
are driven by the photometry, while the high-frequency spectral 
information (e.g., emission-line ratios) is preserved. Overall, this model
has 22 free parameters, and we calculate the posterior probability
distribution using the Markov-Chain Monte-Carlo method.
The data are shown in Fig.~\ref{f.mstar.a}, with the \blackthunder aperture
spectrum in grey. There is a systematic flux mismatch between the NIRCam photometry
and NIRSpec spectrum, probably due to the photometry suffering from contamination while
the spectroscopy suffers from aperture losses.
The spectrum of this satellite galaxy displays a clear Balmer break. Unlike for LRDs,
the shape of the continuum is blue both bluewards and redwards of the break, as expected 
from a Balmer break due to an older stellar population.
This break therefore suggests that there was likely a decline in the SFR in the last few 
10's million years. The Balmer break
is also supported by NIRCam, since the flux in F335M and F356W (which does not have
strong line contamination) is considerably higher than the SED extrapolation from the UV.
The \prospector SFH shows a drop, which drives the observed Balmer break. We show only a
subset of the model parameters (Fig.~\ref{f.mstar.d}), highlighting the most relevant
physical properties. In particular, we infer $\log(\mstar/\Msun) = 8.74_{-0.04}^{+0.10}$.

\subsection{Spatially resolved gas kinematics}\label{s.sats.ss.specres}

\begin{figure*}
    \centering
    {\phantomsubcaption\label{f.sats.a}
     \phantomsubcaption\label{f.sats.b}
     \phantomsubcaption\label{f.sats.c}
     \phantomsubcaption\label{f.sats.d}
     \phantomsubcaption\label{f.sats.e}}
    \includegraphics[width=\textwidth]{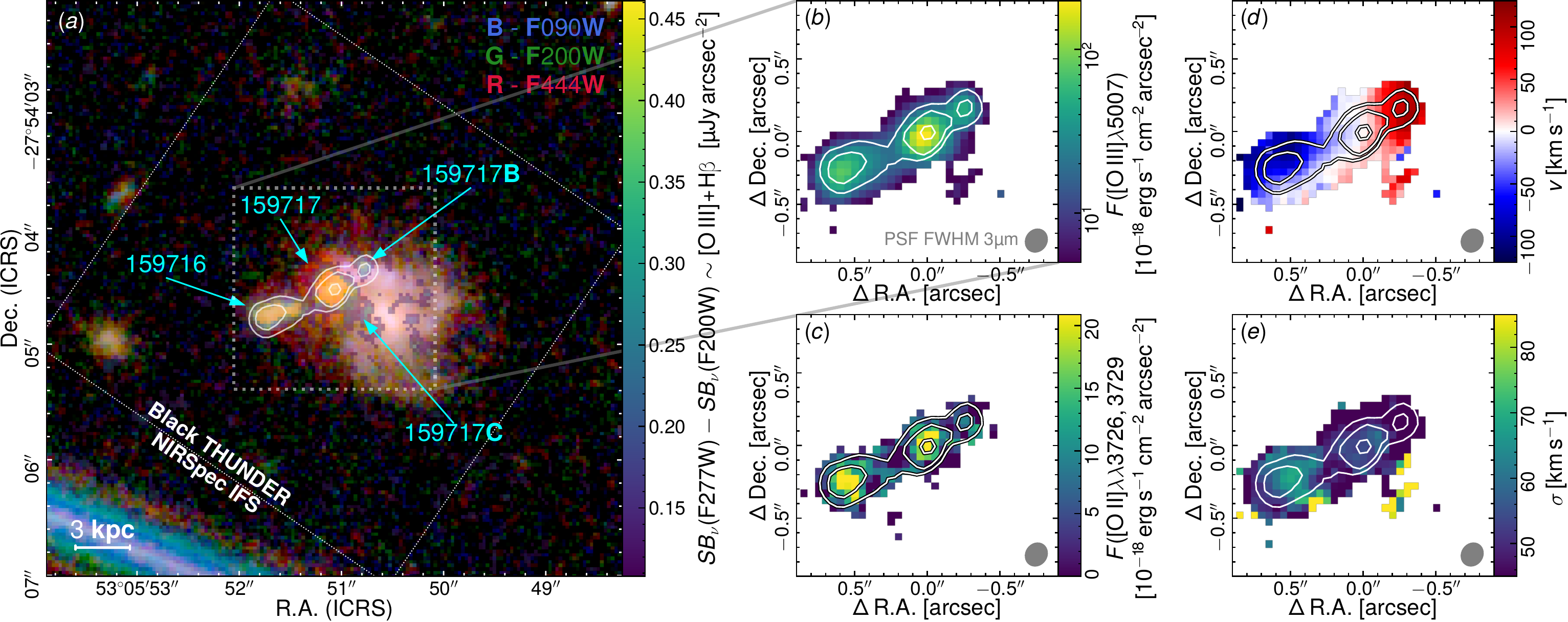}
    \caption{Same as Fig.~\ref{f.ifsimg}, but also highlighting \OIIIall emission (white contours), estimated using
    the photometric excess between NIRCam F277W and
    F200W. The map identifies another possible satellite to the north-west of \target, \satsmall, which is
    also confirmed by \blackthunder in \OIIIall
    and \Halpha.
    The bright source at the centre of \jadesgs{\interlop} is the galaxy's bulge, which is
    prominent in this map due to its relatively red F200W-F277W colour.
    }\label{f.sats}
\end{figure*}

To fit the spatially extended emission, we use \qubespec \citep{scholtz+2025}.
The narrow lines are modelled as single-component Gaussians, and broad Gaussians
are added to model the BLR \Halpha. \OIIIall can also use a broader component,
but a single component is favoured everywhere, which is in agreement with the
expectations, since we barely detect a second, broader \OIIIall component even
in the aperture-integrated spectrum, which has much higher SNR than individual
spaxels (Section~\ref{s.an.ss.outflows}). For the same reason, no model of the
BLR \Hbeta is used.

The SNR map of \OIIIL is overlaid as white contours on the NIRCam RGB image
in Fig.~\ref{f.sats.a}, highlighting values of 15, 30 and 60. The colour bar
measures the emission-line map derived from NIRCam F277W-F200W
(Section~\ref{s.an.ss.size}), also overlaid in transparency. The two
independent emission-line maps match reasonably well.
NIRSpec detects an extended ionized-gas bridge connecting all three of \satlarge,
\target and \satsmall, with a noticeable gap separating \sattiny (Fig.~\ref{f.sats.b}).
The \OIIall emission-line map from the prism observations (Fig.~\ref{f.sats.c})
shows significant detections both in \target and in \satlarge, indicating the presence of some
lower-ionization gas too, except for \sattiny.

The velocity map (Fig.~\ref{f.sats.d}) displays a smooth gradient across the
entire gas distribution, with an end-to-end amplitude of 200~\kms, while the
velocity dispersion map is relatively low at around 50~\kms (Fig.~\ref{f.sats.e}).
There is a 70-\kms $\sigma$ peak about one spaxel east of the centre of \satlarge, and a second,
weaker peak between \target and \satsmall. The fact that $\sigma$ is overall narrow,
with the highest values found between the surface-brightness maxima suggests a
merger scenario, where gas is being removed from the galaxies and has formed an
extended, common reservoir.
Alternatively, the superposition along the line of sight of multiple systemic
kinematic components could also broaden the velocity dispersion. However, in this
case the observed broadening (50--70~\kms) requires such a small
difference in systemic velocity, as to make interaction a more plausible scenario.
Significant disturbance of stationary equilibrium
can be inferred from the kinematics of \satlarge; for this galaxy, the
photometric major axis is aligned with the kinematic axis, implying prolate
rotation which is dynamically unstable for such an elongated object
\citep{fridman+poliachenko1984,merrit+hernquist1991}.
The disturbed nature of \satlarge suggests an advanced interaction stage.
This hypothesis resonates with the presence of the fainter \satsmall on the far
side with respect to \target, possibly due to tidally stripped material during
the first passage. In this case, \target would be interpreted naturally as the
most massive of the three systems.
Overall, these observations imply that \target, far from being isolated, is
located at the centre of a relatively dense region, where gas is collapsing
, leading to both star formation and SMBH accretion.

\sattiny instead represents a separate case; it has velocity dispersion
$\sigma\sim100$~\kms, which is small, but noticeably higher than the rest of the
system -- a fact that is even more striking when we recall that this is the
faintest of the three sources. This larger dispersion is confirmed by the aperture
spectra, although the measurement uncertainties are large, due to the low
SNR. This source could represent a tidally disrupted system or a tidal tail. Its
location, orthogonal to the axis connecting \satsmall to \satlarge, disfavours an
origin from either of \target, \satsmall or \satlarge, but we cannot fully rule out
\sattiny being a smaller system involved in the merger. Still, the relatively
large dispersion,
together with the non detection in both \Hbeta and \Halpha (the latter line has
SNR = 4), could point to \sattiny being a ionized gas outflow, or,
alternatively, a gas cloud photoionized by the AGN. It is located at right angles to both the
elongated gas distribution and to the spatially resolved component of \target
(Section~\ref{s.an.ss.size}; Table~\ref{t.size}), which also favours the interpretation of an
outflow or photoionized cloud. If we accept the hypothesis that \target is the
dominant galaxy in the system, and given the clear presence of a SMBH,
it is natural to attribute the origins of this outflow to \target itself.

\subsection{Gas ionization sources}\label{s.sats.ss.photoion}

\begin{figure}
  {\phantomsubcaption\label{f.bpt.a}
   \phantomsubcaption\label{f.bpt.b}
   \phantomsubcaption\label{f.bpt.c}
   \phantomsubcaption\label{f.bpt.d}}
  \includegraphics[width=\columnwidth]{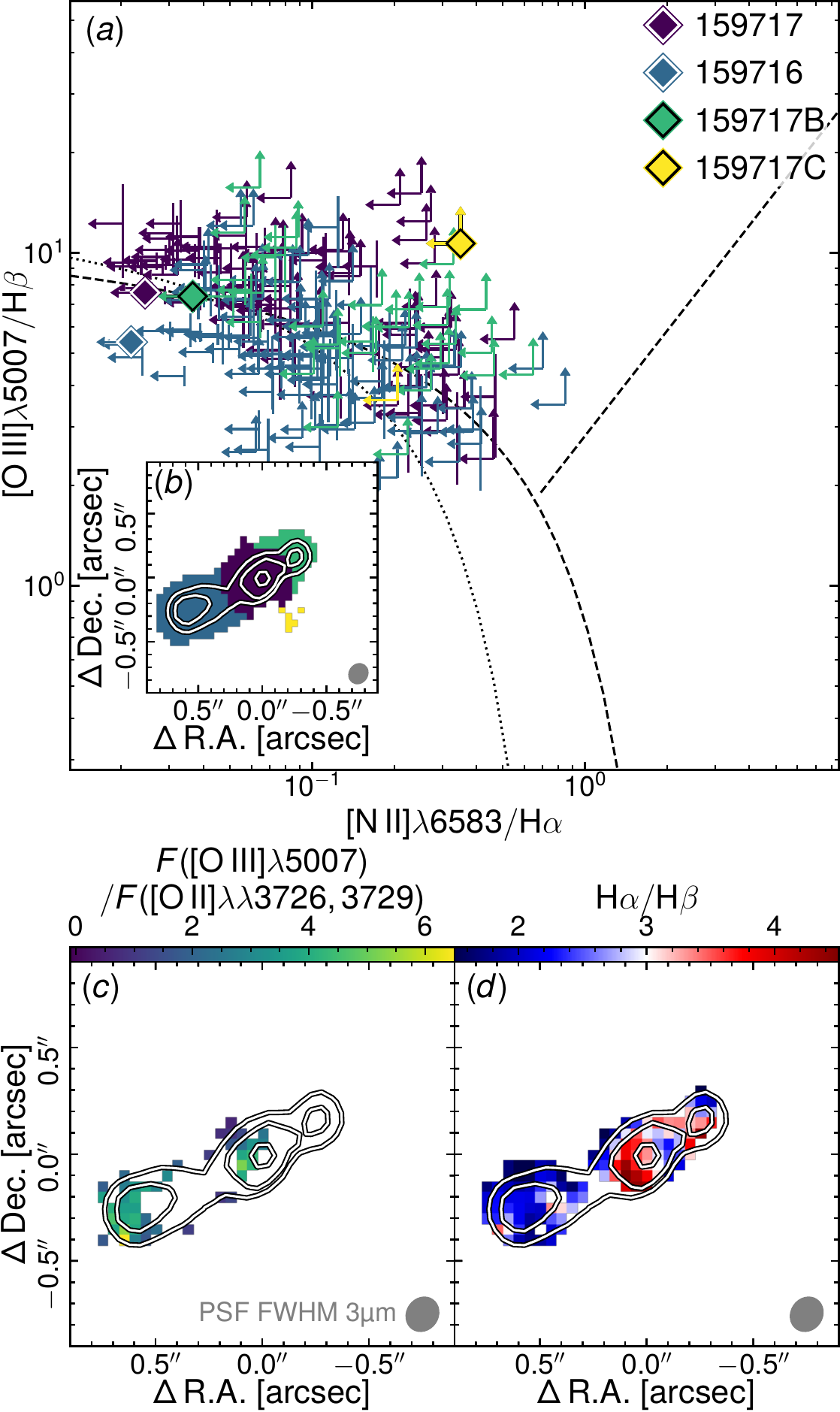}
  \caption{Panel~\subref{f.bpt.a}; BPT diagram, colour coded as indicated in the
  segmentation map of panel~\subref{f.bpt.b} \citetext{the dashed and dotted
  demarcation lines separate star-forming, AGN and low-ionization emission at $z\sim0$, and are
  taken from \citealp{kewley+2001}, \citealp{kauffmann+2003c}, \citealp{kewley+2006},
  and \citealp{schawinsky+2007}}.
  The BPT diagram indicates that three out of four systems are located in a
  similar region of the BPT (large diamonds), while \sattiny (large yellow diamonds) has
  noticeably larger \OIIIL/\Hbeta. The individual spaxels (colour coded as in
  panel~\ref{f.bpt.b}) show a broad range of \OIIIL/\Hbeta, while \NIIL is
  not detected.
  Panel \subref{f.bpt.c}; the ionization diagnostic \OIIIL/\OIIall;
  we do not apply a dust attenuation correction, hence the intrinsic ratio should
  be lower in \target than in \satlarge, given that the observed ratios are
  comparable, but \target is more dusty than \satlarge. Panel~\subref{f.bpt.d}; the Balmer
  decrement from the narrow lines is largest in \target, indicating that it has
  the strongest dust attenuation among the four systems.
  }\label{f.bpt}
\end{figure}

We do not detect \NIIL or other low-ionization lines useful for the BPT
\citep{baldwin+1981} and VO diagrams \citep{veilleux+osterbrock1987},
but thanks to the depth of these observations, we can provide stringent
limits on the \NIIL/\Halpha. In Fig.~\ref{f.bpt.a} we highlight
the four components of the system around \target as large diamonds, which
are colour coded to match the segmentation map in panel~\ref{f.bpt.b}.
The largest three systems, \target, \satlarge and \sattiny  display
\OIIIL/\Hbeta ratios (or lower limits) consistent with star-forming galaxies
\citep{cameron+2023} and type-1 AGN \citep{juodzbalis+2025} at the same epoch.
These three sources have very similar locations on the BPT diagram, despite
\target being seemingly dominated by AGN emission. Nevertheless, their location
cannot discriminate between AGN-driven and star-formation driven photoionization \citep{kocevski+2023,ubler+2023}, since the demarcation lines derived for local
star-forming and AGN-host galaxies (dotted and dashed lines in Fig.~\ref{f.bpt.a}) do not apply at $z\gtrsim 5$.
In Fig.~\ref{f.bpt.a} we also show individual spaxels, with the same
colour coding. When detected, \OIIIL/\Hbeta displays a relatively broad
range  of values, from 3 to 12, including high-SNR detections. This variation
suggests that the ionization conditions vary across the system. This is
confirmed by the \OIIIL/\OIIall map (Fig.~\ref{f.bpt.c}), derived from fitting
the \blackthunder prism data. 
The Balmer decrement map is shown in Fig.~\ref{f.bpt.d}; in \target the
decrement is larger than
the Case-B value of 2.86, while \satlarge shows no appreciable deviation,
indicating little dust content.
Overall, these findings are consistent with the spatially integrated view of
star-forming galaxies and LRDs at $z\sim5$ \citep{shapley+2023,mcclymont+2024,
sandles+2024,taylor+2024,juodzbalis+2025}.

Probably the most interesting source is \sattiny. While the lower limit
on \OIIIL/\Hbeta is not constraining, in Fig.~\ref{f.bpt.a} we show instead the ratio
$2.86\cdot \OIIIL/\Halpha$, where we treat the 4-\textsigma measurement of
\Halpha as a detection, and we derive \Hbeta from Case-B recombination and
no dust. This lower limit is larger than the integrated value for the other three
galaxies, suggesting a different source of photoionization. The harder ionizing spectrum
implied by the large inferred \OIIIL/\Hbeta supports the hypothesis that \sattiny is not
a star-forming clump or tidal tail, but either an outflow or a AGN-illuminated cloud
(Section~\ref{s.sats.ss.specres}).

\section{Discussion}\label{s.disc}

\subsection{Overmassive black hole}\label{s.disc.ss.overmassive}

\begin{figure*}
  \includegraphics[width=\textwidth]{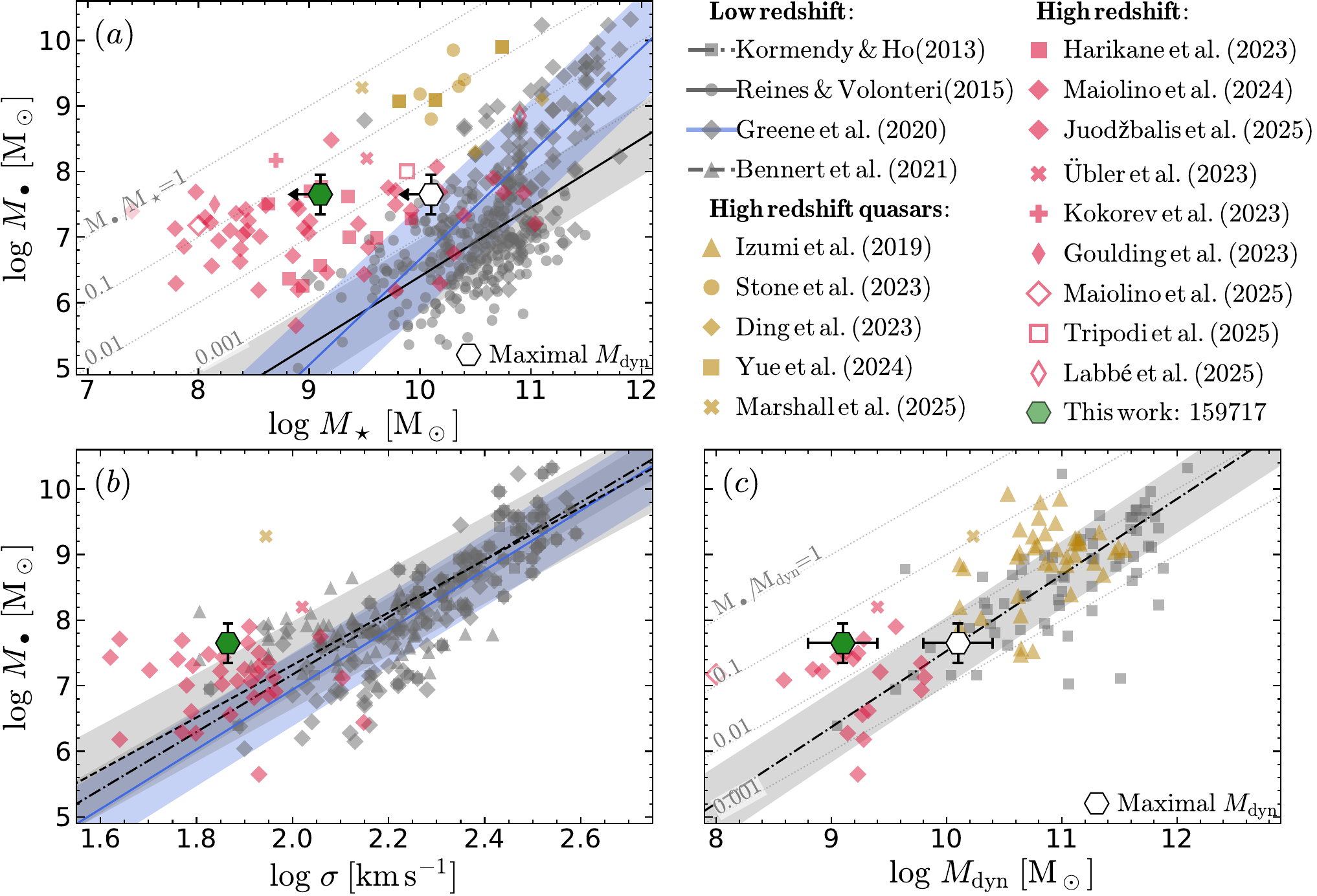}
  {\phantomsubcaption\label{f.mbhgal.a}
   \phantomsubcaption\label{f.mbhgal.b}
   \phantomsubcaption\label{f.mbhgal.c}}
  \caption{When compared to local scaling relations between \mbh and \mstar
  (panel~\subref{f.mbhgal.a}), the SMBH in \target (green hexagon) is over-massive,
  similar to the low-luminosity, high-redshift AGN samples from earlier works \citep[e.g.,][]{harikane+2023,maiolino+2024}. Even our most conservative estimate (white hexagon, Section~\ref{s.phys.ss.host}) lies above the local relation from \citet[][but is consistent with \citet{greene+2020}]{reines+volonteri2015}. Similarly to
  \citet{maiolino+2025} and \citet{ji+2025}, we do not have an actual \mstar measurement, but we use our largest
  estimates of \mdyn as upper limits on \mstar.
  In contrast to relations with \mstar, and in agreement with \citet{maiolino+2024},
  \target seems to follow the \mbh--$\sigma$ relation (panel~\subref{f.mbhgal.b}), while
  agreement with the \mbh--\mdyn relation depends on the adopted \mdyn value (panel~\subref{f.mbhgal.c}). The low-redshift data and fitted relations are from
  \citet{kormendy+ho2013}, \citet{reines+volonteri2015},
  \citet{greene+2020}, and \citet{bennert+2021}; for the relations,
  we show the 16\textsuperscript{th}--84\textsuperscript{th}
  prediction interval as a shaded region.
  High-luminosity AGNs
  (quasars, golden) are from \citet{izumi+2019}, \citet{ding+2023}, \citet{stone+2023}, \citet{yue+2024b}, and \citet{marshall+2025}. High-redshift,
  low-luminosity AGNs (red) are from \citet{harikane+2023}, \citet{maiolino+2024}, \citet{juodzbalis+2025}, \citet{ubler+2023}, \citet{kokorev+2023}, \citet{goulding+2023}, \citet{labbe+2025}, and \citet{maiolino+2025}.
  This figure is available as a \href{https://github.com/fdeugenio/black_hole_scaling_relations}{github repository}.
  }\label{f.mbhgal}
\end{figure*}

The stellar mass of the host galaxy of \target is difficult to determine, due to a
combination of intrinsic and accidental difficulties (sections~\ref{s.an.ss.dust}
and~\ref{s.phys.ss.host}).
Nevertheless, maximal stellar models for this kind of source consistently exceed
the physical limit imposed by dynamical mass \citetext{\citealp{wang+2024a};
\citetalias{juodzbalis+2024b}; \citealp{ma+2024}; \citealp{ji+2025}; \citealp{deugenio+2025c}; \citealp{akins+2025}}.
For this reason, projection effects alone cannot explain why $\mstar>\mdyn$ is very common among
broad-line AGN with `v'-shaped SEDs. Unless one assumes a preferential viewing angle to the host
galaxy, the maximal mass scenario seems ruled out.
In addition, the finding of time variability in the rest-frame optical continuum in Abell~2744-QSO1 \citep{ji+2025}
disfavours a stellar-dominated continuum in \target too, since the two objects share at least four similarities:
the `v'-shaped SED, a relatively high \mbh/\mdyn ratio (discussed below), a broad \Halpha line that is best fit by a
double Gaussian, and a strong \Halpha absorber in the rest frame \citep{deugenio+2025c}. 
In Fig~\ref{f.mbhgal.a} we show \target on the \mbh--\mstar plane, together with sources from the literature. Since
we cannot measure a reliable \mstar, we use \mdyn as an upper limit; the conclusion of an over-massive black hole
(relative to local scaling relations) seems thus warranted, in agreement with previous findings in the redshift range
$z=5\text{--}7$ \citep{harikane+2023,maiolino+2024}. While recent works have challenged the inference of over-massive
black holes \citep{sun+2024,li+2025}, both these works focus on lower redshifts $z\sim3$ and more massive galaxies.

It is important to note that, in addition to the uncertainties on their \mstar,
LRDs also have uncertain \mbh. While virial SMBH mass estimates from single-epoch spectra are widely used, they are subject to systematic uncertainties from assumptions about the geometry, kinematics, and ionization of the BLR.
Calibrations can vary by up to 0.5 dex depending on the line used
\citep[e.g., \Halpha \textit{vs} \CIVall;][]{beteremes+2024} and the adopted
virial factor or line width \citep[e.g.,][]{dallabonta+2025}.
These systematics must be kept in mind, especially when comparing across redshifts, and particularly for a seemingly new class of AGN, such as LRDs. While a few
direct \mbh measurements do exist beyond the local Universe \citep{abuter+2024}, these 
do not extend to LRDs.

More recently, \citet{rusakov+2025} proposed that broad-line profiles that cannot 
be modelled by a single Gaussian are best modelled with an exponential profile, 
which represents the effect of electron scattering by ionised gas. With their 
approach, the intrinsic FWHM is much narrower than what
inferred from a single-Gaussian fit, leading to much smaller SMBH masses.
However, since we use two Gaussians, and given the inferred flux ratio between the two components,
the decrease in FWHM between the fiducial fit (Table~\ref{t.pars}) and the exponential fit
(Appendix~\ref{a.exponential}) is only a factor of two, leading to four times smaller  SMBH mass
of $\log(M_\bullet/\Msun) = 7.0_{-0.3}^{+0.2}$. Even with this reduced \mbh value, \target
remains above the local \mbh--\mstar relation of \citet{reines+volonteri2015}, indicating
that it is overmassive under the scenario of \citet{rusakov+2025}. This demonstrates that overmassive black holes can exist even against the most conservative modelling assumptions.

Our results are consistent with the redshift evolution between a regime of large
\mbh--\mstar scatter at  early epochs and the settling of the local scaling relation around $z\sim3$, as proposed by \citet{juodzbalis+2025}.
In contrast to stellar mass, our target is fully consistent with the \mbh--\mdyn relation (Fig.~\ref{f.mbhgal.b}), in agreement with
the sample of \citet{maiolino+2024}.

It is interesting to note that both the stellar mass and the upper limit on the
dynamical mass of \satlarge are of the same order as the dynamical mass of \target.
If true, this would imply that we are witnessing a major merger,
which intuitively seems at odds with the disrupted nature of \satlarge. Admittedly,
this different behaviour could also be explained by an encounter between galaxies of
similar mass but significantly different size, such that the more diffuse nature of
\satlarge relative to \target could have made the former more susceptible to external
perturbations, unlike \target.

\subsection{Gas metallicity}\label{s.disc.ss.metallicity}

For the metallicity, we find a gas-phase value of $Z = 0.1~\Zsun$, but these measurements are
subject to large uncertainties, due to insufficient information to explore the
possibility of a multi-zone ISM. The key unknown is the possible presence
of a high-density zone ($\nelec\gtrsim10^5~\pcm$) in the narrow-line region
\citep[e.g.,][]{binette+2024}. If present, this zone would significantly affect the
\OIIIL[4363]/\OIIIL ratio, yielding a higher $\Telec(\mathrm{O}^{++})$, a higher
emissivity, and, therefore, lower metallicity.
The auroral-to-strong \permittedEL[O][iii] ratio may be enhanced in the densest zone
of the narrow-line region due to collisional suppression of \OIIIL. This would explain
both the high \nelec inferred from high-ionization species in local
narrow-line AGN \citep[$\nelec\sim10^4~\pcm$;][]{binette+2024} and the anti-correlation
between the ratio \OIIIL to narrow \Hbeta and \OIIIL[4363]/\OIIIL in type 1 AGN (where,
according to the unified model of AGNs, we should be seeing the deepest
zone of the narrow-line region). As a reference, from the observed \OIIIL/\Hbeta and
assuming $\Telec=15,000$~K and $\nelec=10^5~\pcm$, \pyneb gives
$12+\log\mathrm{(O^{++}/H)}=7.9$~dex, an O$^{++}$ abundance that is three times higher
than the fiducial value
reported in Table~\ref{t.pars}. Nevertheless, the scatter about the metallicity relations
at these redshifts is large, hence our value is still plausible for a galaxy with the
dynamical mass of \target.
We do not attempt a multi-line approach due to the relatively low number of emission
lines, with key density diagnostics undetected (\CIIIall, \SIIall), or not adequately
resolved (\OIIall).

We find substantial dust attenuation through the narrow-line ratios, in agreement with
previous works \citep{deugenio+2025c}. This result is somewhat in tension with two findings.
On one hand, our dust attenuation is larger than for typical star-forming galaxies at $z=5$ 
\citep{shapley+2023,sandles+2024}. This could be alleviated by attributing some of the
narrow-line emission to AGN photoionization.
In addition, LRDs seem to require low dust attenuation, based on constraints from
mid- and far-infrared observations \citep{casey+,axins+20,setton+2025,casey+2025}.

\subsection{The broader picture}\label{s.d.ss.sats}

The fact that \target is found at the centre of an extended distribution of ionized gas
seems significant. Indeed, many LRDs seem accompanied by fainter blue dots
\citep{rinaldi+2024,chen+2025}, suggesting
a physical association. The presence of dense gas clouds in \target could be due to gas
being accreted towards the central regions, aided by gravitational torques during the
ongoing merger with \satlarge. LRDs are certainly capable of driving outflows, which are
generally observed through nuclear hydrogen absorption
\citetext{\citealp{matthee+2024}; \citetalias{juodzbalis+2024b}; \citealp{rusakov+2025}},
or through broad, centrally concentrated \OIIIL emission \citepalias{juodzbalis+2024b}.
In \target, for the first time, we see tentative evidence of a spatially detached 
ionized gas outflow (Section~\ref{s.sats}). If confirmed, this detection could indicate
past feedback.
The gap between this cloud and \target may indicate that feedback was interrupted,
perhaps in relation to the recent merger with \satlarge. For instance, if the rate of
gas accretion towards the central regions was too high, it could have overwhelmed the
outflows driving an accumulation of gas.

This scenario would be in agreement with recent works, which have highlighted the
complex morphology of many LRDs \citep{rinaldi+2024,torralba+2025}. Although there are notable
examples of LRDs without prominent companions \citep[e.g.,][]{furtak+2023,furtak+2024,
naidu+2025,carranza-escudero+2025}, several others are known to have neighbours \citep{tanaka+2025,merida+2025}.

\subsection{Gas absorber}

The observed \Halpha absorption is undoubtedly originating in a gas cloud between the BLR and the observer. A stellar origin is excluded because when measuring the EW of \Halpha with respect to the continuum (by subtracting the un-absorbed model BLR from the data), 
we find $EW(\Halpha) = 100~\AA$, much higher than in any stellar-atmosphere model.
Similarly, if the absorber was not positioned in front of the BLR, the absorption depth would reach unphysical negative values.
However, even with high-quality, high-resolution data, it is impossible to provide a tight constraint on the total column density, due to the uncertain conversion between the $n=2$ hydrogen population and the other levels, and between the bound and ionized hydrogen populations. This uncertainty is reflected in the broad range of column and volume densities that can reproduce our observations (Fig.~\ref{f.cloudy}). Tighter constraints are available when \Hbeta absorption is also strongly detected \citep{juodzbalis+2024b}, but this would require deeper spectra -- particularly for sources with large Balmer decrement, such as LRDs \citep{brooks+2024,furtak+2024,deugenio+2025g,torralba+2025b,nikopoulos+2025}. In the future, deep \jwst programs using high-resolution G235H and G395H gratings may be able to better characterize Balmer absorption in AGNs at redshifts $z=2.5\text{--}7$. The range $z=2.5\text{--}3.9$ would additionally cover \HeIL[1.08\mum], useful to characterize the ionization state of the absorbing gas and thereby the distance of the absorber to the black hole accretion disk \citep{juodzbalis+2024b}. From current data, we can constrain the total column density of hydrogen to be $\log(\mathrm{N_H}/\pcm[2]) = 20.5\text{--}24$, with the upper range reaching the Compton-thick threshold; such high values
have been proposed to explain the low X-ray to bolometric luminosity ratio of low-luminosity AGNs at $z\sim 5$ \citetext{\citealp{maiolino+2024x,yue+2024}; \citetalias{juodzbalis+2024b}}, and would be consistent with the non-detection of \target in X-rays, despite lying in one of the deepest zones of the 7-Ms \textit{Chandra} mosaic of GOODS-S \citep{luo+2017}.

Just like for \qsoone \citep{furtak+2023,furtak+2024}, also in \target we can rule out
completely a stellar origin of the absorber, based on the same two arguments as used in \citet{deugenio+2025c}. First, the absorber
EW is too large to absorb only the continuum. As noted also by \citetalias{juodzbalis+2024b},
the depth of the absorption is so large that if we were to remove the BLR line, we would be left
with negative flux. Additionally, thanks to the superior resolution of G395H, we can infer that the absorber
turbulence is $v_\mathrm{turb} \equiv \sqrt{2} \cdot \sigabs \approx 140\text{--}150$~\kms.
This value is similar to what has been found independently in \qsoone by \citet{ji+2025}
and \citet[][using respectively the shape of the Balmer break and the shape of \Halpha absorption]{deugenio+2025c}. Such a large turbulence parameter is an order of magnitude higher than what is
found in the stellar atmospheres of luminous stars \citep{smith+1998}, hence constituting a
second, independent argument ruling out stellar absorption.
To these two compelling arguments, we can add EW variability (Section~\ref{s.phys.ss.absorb}).
While admittedly tentative, time variability is completely inconsistent with stellar absorption,
while being in perfect agreement with a scenario where the absorbing clouds are located relatively 
near to the SMBH.

For a typical gas temperature of $T\sim10^4$~K, the speed of sound is of order of 10--15~\kms,
implying a Mach number around 10 for the absorbing gas. Although it remains unclear how such
high turbulence is sustained, the consistency with previous LRD studies reporting
$v_\mathrm{turb}\sim100~\kms$ supports the measurement \citep{ji+2025,deugenio+2025c,naidu+2025}.
Similar turbulent velocities are also required to explain \permittedEL[Fe][ii] emission
regularly seen in the UV spectra of local AGN and QSOs, yet the physical origin of this turbulence 
remains unresolved \citep{baldwin+2004}.

As an alternative to absorption, scenarios such as Balmer-line resonant scattering have
been considered \citep{naidu+2025}. In this case, the intrinsic Balmer lines would be
much narrower than the observed profiles, with the broadening primarily due to scattering.
A key advantage of this scenario is that it explains why the absorber appears near
rest-frame wavelengths. However, this mechanism would likely produce different line
profiles for \Halpha and \Hbeta. While our data cannot rule out such differences, deep
observations of bright targets \citetext{e.g., \citetalias{juodzbalis+2024b};
\citealp{ma+2025}} will be able to test this possibility.

\subsection{Towards atmospheric models?}\label{s.disc.ss.atmo}

An intriguing finding is that the optical depth of \Hbeta and \Halpha are inconsistent with
the absorption cross sections of these two lines, which are set by the oscillator strengths.
In our model, this ratio should be exactly equal to the value dictated by atomic physics, 
which is 0.139. While of course this mismatch could be only apparent, due to
the low SNR of our observations, this has now been confirmed in other objects too
\citetext{e.g., \citealp{deugenio+2025c,deugenio+2025g}}. At face value, the only possible
explanation then is that a simple passive absorber modelled as a gas slab may be inadequate, much
like in the atmosphere of A-type stars the absorption strength of \Hbeta and higher-order
Balmer lines is larger than for \Halpha -- contrary to the expectations from the relative
cross sections of these lines.
This explanation could imply that \Halpha and \Hbeta arise from different depths in
the absorbing gas, mimicking what happens in stellar atmospheres. For reference, the
density where \Halpha arises in A-type stars is $~10^{12}~\pcm$, larger but not too different from the findings of \citet{naidu+2025}.

\subsection{Absorber kinematics}

Gas absorbers along the line-of-sight of AGNs are interpreted as inflows/outflows
\citetext{\citealp{matthee+2024}; \citetalias{juodzbalis+2024b}; \citealp{wang+2024a}}.
In our case, the absorbing cloud is consistent with no velocity component along the
line of sight (relative to the systemic velocity determined from the narrow lines).
In principle, this low projected velocity could be due to orbital motion of a
cloud bound to the SMBH. In the local Universe, this configuration would require both
a high-inclination orbit, line-of-sight alignment and the right orbital phase, as found
in local AGNs \citep[NGC~1365;][]{maiolino+2010}.
However, the combination of all these occurrences is unlikely, making NGC~1365 one
of very few cases where the inclination is suitable for observing transiting clouds.
The particular geometric configuration seen in NGC~1365 should not be common at
$z=5$ either. Moreover, in this case, there is no direct evidence of a dusty torus,
with several claims of LRDs being deficient in mid-IR emission \citep[e.g.,][]{wang+2024a,
setton+2025,degraaff+2025}, and the highest-redshift MIR detection of an LRD unambiguously
attributed to a dusty torus being at $z=2.26$ \citepalias{juodzbalis+2024b}.
A rotating configuration is therefore possible in our case, although other absorbers have been measured with
line-of-sight velocities between $-340$ and $+50~\kms$
\citetext{\citealp{matthee+2024}; \citetalias{juodzbalis+2024b}; \citealp{rusakov+2025}},
which are clearly inconsistent with rotation and favour non-equilibrium configurations.

While the remarkably low outflow velocity could be due to projection effects
(with the inclination of the outflow axis close to $i \approx 90$\textdegree),
this solution seems disfavoured too. For example, observing only $\vabs =
-13~\kms$ from a typical deprojected outflow velocity of 350~\kms \citep{carniani+2024},
 we would need $\vabs/\cos i \equiv \vout \leq -350~\kms$ which requires
$i\geq88$ \textdegree. For a randomly oriented galaxy, the corresponding probability
is only $P<0.04$. While this is not a small value, \target is already the third
case of an absorbing cloud very near to rest frame \citep{deugenio+2025c,naidu+2025},
out of $\sim10$ LRDs with clearly detected absorption. This suggests that the low velocity
is not due to projection, but reflects low intrinsic velocity.

This scenario would indicate therefore a long-lived cloud lingering near the
SMBH, or even a `stalling' outflow, caused by insufficient AGN power or by inefficient
coupling \citep{arakawa+2022,fabian+2008}.
Insufficient energy seems at odds with our relatively high Eddington ratio
($\lambda_\mathrm{E}=0.45$). The other possibility, that feedback is inefficient,
e.g. due to poor coupling with the gas, seems also unlikely. Such a scenario could be
plausible in low-metallicity gas, if we assume that most of the opacity is driven by
metals and dust \citep{fabian+2008}. However, in our case, the galaxy appears decidedly
dusty, based on the narrow-line ratios ($\Avhatn=1-1.6$~mag) and the metallicity of
the host galaxy seems fairly high too ($>0.06~\Zsun$; Table~\ref{t.pars}).
Furthermore, our metallicity values could be underestimated, since any 
contribution from high-density regions to the \OIIIL[4363] emission would
decrease $T_\mathrm{e}$ and hence further increase metallicity
(Section~\ref{s.disc.ss.metallicity}).

\citet{kido+2025} argue that typical LRDs occupy a region of the \mbh--\mstar plane
where strong outflows are expected. The general absence of such outflows in sources like
\target therefore suggests an alternative scenario: the dense gas envelope model proposed
by \citet{naidu+2025}, \citet{degraaff+2025}, and \citet{kido+2025}. In the case of
\target, not only is the combination of its large \mbh/\mstar\ ratio and lack of outflows
notable, but its relatively high $\lambda_\mathrm{Edd}$ and rest-frame Balmer absorption
are also consistent with the predictions of the envelope model.

\subsection{Breathing-mode accretion?}\label{s.disc.ss.breathing}

Hydrodynamic simulations of cold-gas accretion onto SMBHs which include radiative
transfer suggest a self-regulated behaviour, with highly turbulent gas flows near the
ionization front \citep{park+2017,inayoshi+2022}, but even in this case the turbulence is ten times lower than what inferred from our measurements.

\citet{park+2017} also find that during quiescent periods between bursts of accretion,
the ionized bubble around the SMBH
shrinks in size, leading to an increase in the density of the gas near and before the
ionization front. Moreover, the dense gas is not dissipated, but following the
accretion rate on the central SMBH, gives rise to an oscillatory behaviour of
expansion and contraction.
This `breathing mode' of the gas could explain the observation of \Halpha absorption
with a range of inflow, outflow, and rest velocities. Depending on the distance
travelled, the gas would spend different amounts of time near the inversion points.
The formation of a shell of neutral gas around accreting black holes is also consistent
with the models of \citet[][for $\mbh \sim 100~\Msun$]{milosavljevic+2009} and
\citet[][for $\mbh = 100\text{--}10^5~\Msun$]{park+ricotti2012}, thus further supporting
the presence of gas capable of Balmer absorption.
If this breathing-mode picture is true, then the relative abundance of absorbers with
different velocities could hold valuable information about the interface between the accreting
SMBH and the host galaxy. As we have seen, the incidence of Balmer absorbers may be
underestimated by current medium-resolution observations (Section~\ref{s.resol}).
This is particularly true at high redshifts, where most LRD spectra use either the
NIRSpec medium-resolution gratings or the prism (but to some extent it is a problem
also in the local Universe, where large surveys such as SDSS
and LAMOST have $R\lesssim 2,000$, in the relevant wavelength region). Clearly,
current \jwst/NIRSpec observations underestimate the total incidence of absorbers.
The problem must be relatively more severe for rest-frame absorbers, since symmetric
narrow-line infill would be more effective at removing the absorber, and would still
result in a symmetric line profile. This suggests that the fraction of red- and
blue-shifted absorbers relative to rest-frame ones could also be over-estimated by
current data.

Still, key issues with the model of \citet{park+2017} remain, most notably the large
discrepancy in density between their simulations and those inferred for LRD absorbers 
\citetext{$10^9\text{--}10^{11}$
\pcm[3]; \citetalias{juodzbalis+2024b}; \citealp{ji+2025,naidu+2025}}. The models
of \citet{park+2017} reach maximum densities of only $10,000~\pcm[3]$.
This discrepancy may be due to the \textit{ad-hoc} assumptions of the
\citet{park+2017} simulations, which may not consider the appropriate cosmological context
for LRDs -- particularly if LRDs occupy somehow atypical dark-matter haloes compared to
the general galaxy population \citep[e.g.,][]{pacucci+2025}. For instance, theory supports
the possibility of rapid gas inflow towards the central region of galaxies
\citep[e.g.,][]{tacchella+2016}. While these results apply to dark matter haloes that
are more massive than for LRDs \citep{pizzati+2025}, recent zoom-in
simulations such as THESAN-Zoom \citep{kannan+2025,mcclymont+2025} suggest that
the accretion rate onto the central regions of lower-mass, high-redshift galaxies can be 
substantial too, potentially supporting
the accumulation of high-density gas near the SMBHs powering LRDs.
However, while THESAN-Zoom preserves the cosmological context \citep{kannan+2025}, it still lacks the spatial resolution of
high-resolution simulations like those of \citet{park+2017} and thus it cannot resolve the
immediate environment around SMBHs. A promising path would be to use constraints from zoom-in simulations like THESAN-Zoom to supply the boundary conditions for even higher-resolution
simulations targeting galactic nuclei.

\subsection{Soft ionizing spectrum}\label{s.disc.ss.soft}

The absence of high-ionization lines in LRDs, including metallicity independent lines, such as \HeII, could be due to an intrinsically red SED \citep{lambrides+2024,wang+2024b},
However, there are convincing reports of high-ionization UV lines in a few LRDs
\citep{tripodi+2025a,tang+2025,akins+2024,labbe+2025}, while preliminary analysis suggests 
an incidence rate of $\sim12$~percent across the population \citep{tang+2025}, albeit with
large uncertainties.
A possibility to reconcile these differences is that ionizing photons from an intrinsically
hard SED are being absorbed by hydrogen \citep{inayoshi+maiolino2025,ji+2025,naidu+2025}.
This would be in agreement with the weakness of some bright LRDs at mid- and far-IR wavelengths
\citep{setton+2025}. This hydrogen absorption should drive damped \Lyalpha absorption (DLA) and
possibly a Balmer break, which are not seen in this and in many other LRDs \citep[e.g.,][]{
greene+2024,rusakov+2025, juodzbalis+2025}.
Arguably, the Balmer break could be hidden by dust reddening, which in this and in other
objects is inferred independently from the Balmer decrement \citetext{e.g.,
\citealp{killi+2024}; \citetalias{juodzbalis+2024b}; \citealp{deugenio+2025c}}, in combination with outshining from the host galaxy \citep{naidu+2025}. Either way, if Balmer
breaks were common to all LRDs (whether hidden or not), we should always see Balmer-line
absorption. While this is seemingly contradicted by current observations, we showed that
this conclusion may be driven by inadequate spectral resolution (Section~\ref{s.resol}).
Future surveys using high-resolution spectroscopy would be an easy way to check the
incidence of (possibly hidden) Balmer breaks.

\section{Conclusions}\label{s.conc}

We presented the broad-line AGN \target at $z\approx5.077$, observed by the 
JADES and \blackthunder programmes using \jwst/NIRCam, NIRSpec/MSA in multiple
dispersers, and NIRSpec/IFS. While a full analysis of this system is complicated
by the foreground system, extensive \jwst data and deep, multi-epoch high-resolution
observations provide new insights into the nature of low-luminosity, `Little Red Dot'
AGN.
\begin{itemize}
\item The prism spectroscopy shows a `v'-shaped continuum spectrum (Fig.~\ref{f.data}),
characterizing this source as a LRD. In the medium- and high-resolution spectrum, the
\Halpha line has multiple components, including two broad Gaussians, with an overall
broad-component width $\fwhm = 1510\pm70~\kms$. \OIIIL has a narrow
$\sigma_\mathrm{n}=47\text{--}54~\kms$ ($\fwhm = 127~\kms$). Combined, this is evidence for a broad-line region.
\item The resulting SMBH mass of $\log \mbh/\Msun = 7.5$~dex places \target very near the
$\mbh-\sigma$ scaling relation of nearby galaxies. From the dynamical mass, used as an upper limit on \mstar,
we infer that the SMBH is overmassive relative to the local scaling relations (Fig.~\ref{f.mbhgal}).
\item The overmassive nature of \target remains true (but less severe) even if we model
the broad line as a single Gaussian plus electron scattering \citep{rusakov+2025}.
\item The metallicity of this system from the direct method is 0.06~\Zsun; while \target
is metal poor, it is far from pristine, also in agreement with the large Balmer decrement
of the narrow Balmer lines ($\Avhatn = 1.0\pm0.2$~mag).
\item Far from being isolated, \target is found in the process of interacting with a
companion, \satlarge, in what is a possible major merger. Interestingly, \satlarge
exhibits a Balmer break likely driven by a recent period of lowered star formation.
Another source, \satsmall,
may be material stripped from \satlarge, while the whole triple is immersed in a pool
of \OIIIL emitting gas. The merger may be helping in funnelling gas towards the central
regions of \target, ultimately feeding the SMBH.
\item Despite their differences in luminosity, morphology, and AGN presence, the three
sources have similar ionization properties, such as a similar location on the BPT diagram.
\item One more source, \sattiny, is spatially detached. Its higher ionization and dispersion
suggest a possible outflow origin. If confirmed, this would be the first evidence of
kpc-scale AGN feedback in a LRD.
\item \Halpha presents clear absorption with velocity and velocity dispersion of
$\vabs = -13~\kms$ and $\sigabs = 120~\kms$. We interpret the low projected velocity and
large turbulence as a `stalling' outflow, or lingering gas cloud.
\item Intriguingly, there is tentative evidence (2.6~\textsigma) of time variability
in the EW of the \Halpha absorber. If confirmed, this would indicate a highly dynamic
environment.
\item This LRD, like others, lacks high-ionization lines, which could be due to hydrogen
bound-free absorption. We do not observe a Balmer break, but this could be due to dust
attenuation, since dust is seen in the narrow lines too. If Balmer breaks were widespread
in LRDs (including many hidden by dust), this should still give rise to Balmer line
absorption.
\item We highlight that the absorber in \target, despite being clearly evident and even
dominant in the G395H observations, is not detected in the medium-resolution G395M data,
implying that current incidence of absorbers may be severely underestimated.
\end{itemize}

These findings resonate with earlier results pointing to LRDs being `more than
just a dot' \citep{rinaldi+2024}. The presence of rest-frame Balmer absorption,
determined here with unprecedented precision, highlights how dense gas near
LRDs may display properties that are not common in massive SMBHs, as highlighted
by several studies \citep{inayoshi+maiolino2025,ji+2025,deugenio+2025c,rusakov+2025,
naidu+2025,ma+2025}.
Overall, our findings highlight the exciting opportunities offered by multi-epoch,
high-resolution observations of LRDs, to investigate both their local environment,
SMBHs, and time variability. The use of high-resolution NIRSpec spectroscopy is
essential for a complete census of gas absorbers in broad-line AGN, because estimates
based on NIRSpec medium-resolution spectroscopy may severely under-estimate the
incidence of absorbers (Fig.~\ref{f.abs}).

\section*{Acknowledgements}

We thank Debora Sijacki, Sergio Martin-Alvarez, William McClymont, Gabriele Pezzulli,
and Harley Katz for useful discussions.
FDE, RM, XJ, JS, IJ and GCJ acknowledge support by the Science and Technology Facilities Council (STFC), by the ERC through Advanced Grant 695671 ``QUENCH'', and by the
UKRI Frontier Research grant RISEandFALL. RM also acknowledges funding from a research professorship from the Royal Society. IJ also acknowledges support by the Huo Family Foundation through a P.C. Ho PhD Studentship.
MP, SA and BRP acknowledge grant PID2021-127718NB-I00 funded by the Spanish Ministry of Science and Innovation/State Agency of Research (MICIN/AEI/ 10.13039/501100011033). MP also acknowledges the grant RYC2023-044853-I, funded by  MICIU/AEI/10.13039/501100011033 and European Social Fund Plus (FSE+).
GM and H\"U acknowledge funding by the European Union (ERC APEX, 101164796). Views and opinions expressed are however those of the authors only and do not necessarily reflect those of the European Union or the European Research Council Executive Agency. Neither the European Union nor the granting authority can be held responsible for them.
SC and GV acknowledge support by European Union's HE ERC Starting Grant No. 101040227 - WINGS.
AJB acknowledges funding from the ``FirstGalaxies'' Advanced Grant from the European Research Council (ERC) under the European Union's Horizon 2020 research and innovation program (Grant agreement No. 789056).
ECL acknowledges support of an STFC Webb Fellowship (ST/W001438/1).
KI acknowledges support from the National Natural Science Foundation of China (12073003, 11721303, 11991052), and the China Manned Space Project (CMS-CSST-2021-A04 and CMS-CSST-
2021-A06).
YI is supported by JSPS KAKENHI Grant No. 24KJ0202.
ZJ, BDJ, BER and CNAW acknowledge support from the NIRCam Science Team contract to the University of Arizona, NAS5-02015. BER also acknowledges support from JWST Program 3215.
ST acknowledges support by the Royal Society Research Grant G125142.
The research of CCW is supported by NOIRLab, which is managed by the Association of Universities for Research in Astronomy (AURA) under a cooperative agreement with the National Science Foundation.
JW gratefully acknowledges support from the Cosmic Dawn Center through the DAWN Fellowship. The Cosmic Dawn Center (DAWN) is funded by the Danish National Research Foundation under grant No. 140.
The authors acknowledge use of the lux supercomputer at UC Santa Cruz, funded by NSF MRI grant AST 1828315.

This work is based on observations made with the NASA/ESA/CSA James Webb Space Telescope. The data were obtained from the Mikulski Archive for Space Telescopes at the Space Telescope Sci-
ence Institute, which is operated by the Association of Universities for Research in Astronomy, Inc., under NASA contract NAS 5-03127 for JWST. These observations are associated with program \#1286.

This work made extensive use of the freely available \href{http://www.debian.org}{Debian GNU/Linux} operating system.
We used the \href{http://www.python.org}{Python} programming language \citep{vanrossum1995}, maintained and distributed by the Python Software Foundation. We made direct use of Python packages
{\sc \href{https://pypi.org/project/astropy/}{astropy}} \citep{astropy+2013,astropy+2018},
{\sc \href{https://pypi.org/project/corner/}{corner}} \citep{foreman-mackey2016},
{\sc \href{https://pypi.org/project/emcee/}{emcee}} \citep{foreman-mackey+2013},
{\sc \href{https://pypi.org/project/jwst/}{jwst}} \citep{alvesdeoliveira+2018},
{\sc \href{https://pypi.org/project/matplotlib/}{matplotlib}} \citep{hunter2007},
{\sc \href{https://pypi.org/project/numpy/}{numpy}} \citep{harris+2020},
%{\sc \href{https://pypi.org/project/ppxf/}{ppxf}} \citep{cappellari+emsellem2004, cappellari2017, cappellari2022},
{\sc \href{https://pypi.org/project/astro-prospector/}{prospector}} \citep{johnson+2021} \href{https://github.com/bd-j/prospector}{v2.0},
{\sc \href{https://pypi.org/project/PyNeb/}{pyneb}} \citep{luridiana+2015},
{\sc \href{https://pypi.org/project/python-fsps/}{python-fsps}} \citep{johnson_pyfsps_2023},
{\sc \href{https://pypi.org/project/pysersic/}{pysersic}} \citep{pasha+miller2023},
{\sc \href{https://github.com/honzascholtz/qubespec/}{qubespec}} \citep{scholtz+2025},
and {\sc \href{https://pypi.org/project/scipy/}{scipy}} \citep{jones+2001}.
We also used the software packages {\sc \href{https://github.com/cconroy20/fsps}{fsps}} \citep{conroy+2009,conroy_gunn_2010}, {\sc \href{https://www.star.bris.ac.uk/~mbt/topcat/}{topcat}}, \citep{taylor2005}, {\sc \href{https://github.com/ryanhausen/fitsmap}{fitsmap}} \citep{hausen+robertson2022} and {\sc \href{https://sites.google.com/cfa.harvard.edu/saoimageds9}{ds9}} \citep{joye+mandel2003}.

%%%%%%%%%%%%%%%%%%%%%%%%%%%%%%%%%%%%%%%%%%%%%%%%%%
\section*{Data Availability}

The data used in this article are all available through the MAST archive. The JADES data are publicly available also on the \href{https://jades-survey.github.io/}{JADES Collaboration website}. The specific data-reduction 
version used in this work has been published in \citet[][\href{http://dx.doi.org/10.17909/8tdj-8n28}{10.17909/8tdj-8n28}]{deugenio+2025a}. The most recent updates to the reduction have been presented in \citet{scoltz+2025b}.

\section*{Affiliations}
\noindent
{\it
\hypertarget{aff11}{$^{11}$} INAF - Osservatorio Astrofisico di Arcetri, largo E. Fermi 5, 50127 Firenze, Italy\\
\hypertarget{aff12}{$^{12}$}Centre for Astrophysics Research, Department of Physics, Astronomy and Mathematics, University of Hertfordshire, Hatfield AL10 9AB, UK\\
\hypertarget{aff13}{$^{13}$}Steward Observatory, University of Arizona, 933 N. Cherry Ave., Tucson, AZ, 85721, USA\\
\hypertarget{aff14}{$^{14}$}Kavli Institute for Astronomy and Astrophysics, Peking University, Beijing 100871, China\\
\hypertarget{aff15}{$^15$}Center for Astrophysics $|$ Harvard \& Smithsonian, 60 Garden St., Cambridge MA 02138 USA \\
\hypertarget{aff16}{$^{16}$}Department for Astrophysical and Planetary Science, University of Colorado, Boulder, CO 80309, USA\\
\hypertarget{aff17}{$^{17}$}Department of Astronomy and Astrophysics University of California, Santa Cruz, 1156 High Street, Santa Cruz CA 96054, USA\\
\hypertarget{aff18}{$^{18}$}Institut d'Astrophysique de Paris, Paris, 98 bis Boulevard Arago, 75014 Paris, France\\
\hypertarget{aff19}{$^{19}$}NSF National Optical-Infrared Astronomy Research Laboratory, 950 North Cherry Avenue, Tucson, AZ 85719, USA\\
\hypertarget{aff20}{$^{20}$}NRC Herzberg, 5071 West Saanich Rd, Victoria, BC V9E 2E7, Canada\\
\hypertarget{aff21}{$^{21}$}Cosmic Dawn Center (DAWN), Copenhagen, Denmark\\
\hypertarget{aff22}{$^{22}$}Niels Bohr Institute, University of Copenhagen, Jagtvej 128, DK-2200, Copenhagen, Denmark
}



\bibliography{bibliography} % if your bibtex file is called example.bib
\bibliographystyle{config/mnras}

%%%%%%%%%%%%%%%%%%%%%%%%%%%%%%%%%%%%%%%%%%%%%%%%%%

%%%%%%%%%%%%%%%%% APPENDICES %%%%%%%%%%%%%%%%%%%%%

\appendix

\section{Interloping galaxy.}\label{a.interl}

Emission from \target is plagued by a foreground spiral galaxy, JADES~ID 
159715 (Fig.~\ref{f.data.a}). By inspecting \hst imaging at wavelengths 
shorter than 7,000~\AA (the \Lyalpha drop in \target), we can readily 
identify a star-forming region near to the line of sight to \target; 
this region is detected in \OIIIall and \Halpha in the 2-d and 1-d NIRSpec 
spectra (see Fig.~\ref{f.data}). The redshift of these emission lines is 
consistent with what found from the \blackthunder high-resolution spectrum,
where we find \Paalpha at 3.75~\mum, giving a redshift of $z_\mathrm{spec}=1.00115\pm0.00001$. This is independently confirmed by \citet{deugenio+2025h}
using deeper grating observations in G235M and G395M.

\begin{figure}
\includegraphics[width=\columnwidth]{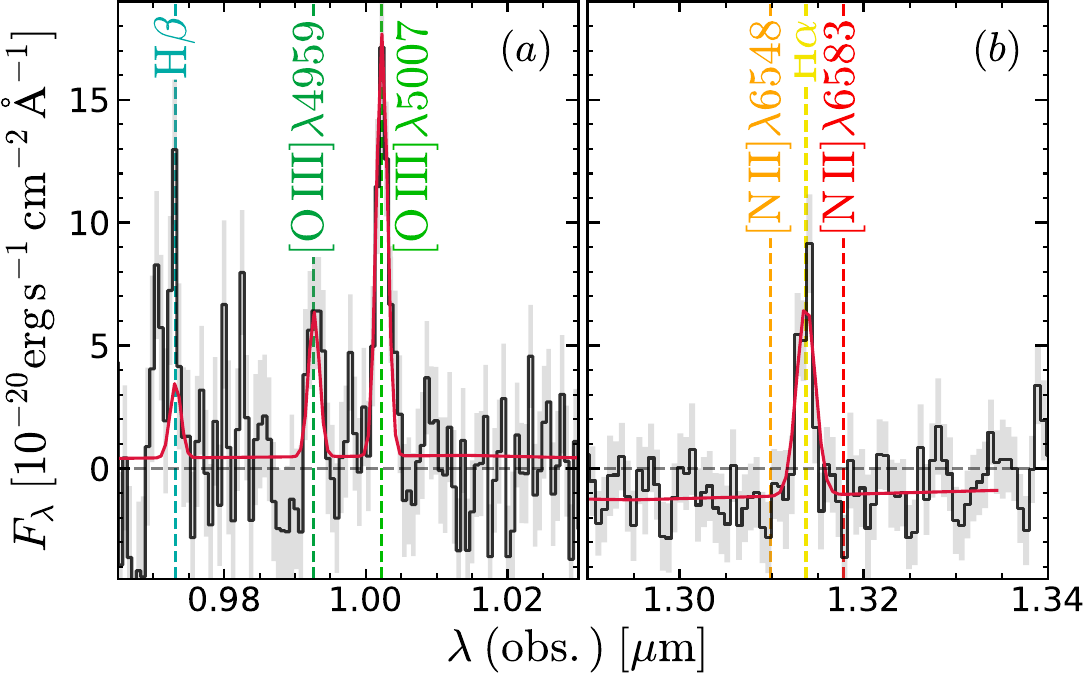}
  {\phantomsubcaption\label{f.inter.a}
   \phantomsubcaption\label{f.inter.b}}
  \caption{Medium-resolution G140M grating data of the interloping star-forming regions from galaxy 159715. The model (red) assumes a physically motivated \Halpha/\Hbeta ratio, due to a noise spike at the location of \Hbeta. Background over-subtraction is present throughout the spectrum, due to the spatial footprint of the main target \target and of the interloper itself on the NIRSpec shutters (Fig~\ref{f.data.a}).}\label{f.inter}
\end{figure}

To analyse this lower-redshift interloper, we start by correcting the 
G140M/F070LP spectrum for foreground Milky-Way attenuation, using the 
\citet{cardelli+1989} extinction law with $E(B-V)=0.0067$ from 
\citep{schlafly+2011}. This has minimal impact (0.02~mag) on the recovered 
$A_V$. We extract the 1-d spectrum from three spaxels of the JADES 2-d
spectrum, identified from the 2-d SNR map as displaying clear evidence of
\OIIIL emission in the prism spectrum (cf. Fig.~\ref{f.data.b}). The
resulting spectrum is shown in Fig.~\ref{f.inter}, with clear evidence of 
both \OIIIall (panel~\subref{f.inter.a}) and \Halpha 
(panel~\subref{f.inter.b}), while \Hbeta is superimposed to a noise feature.
We model this spectrum using two local backgrounds around \Hbeta--\OIIIall 
and \Hbeta, and four Gaussians, the latter having the same redshift and 
(instrument-convolved) velocity dispersion. The \Halpha flux is a free 
parameter, while the \Hbeta flux is derived from \Halpha assuming the 
intrinsic Case-B recombination ratio of 2.86 for 
$T_\mathrm{e}=10,000$~K and $n_\mathrm{e}=100$~cm$^{-3}$, and a
\citetalias{gordon+2003} dust law with free $A_V$. We also limit the 
observed \OIIIL/\Hbeta ratio to be less than 20, which effectively limits 
$A_V$ to realistic values of $A_V<3$~mag.
The model is fit to the data using the Bayesian approach described in 
sections~\ref{s.an.ss.fitting} and~\ref{s.an.ss.g395h.bt}.
We find a redshift $z_\mathrm{spec}=1.0012\pm0.0002$, higher than the
fiducial value from \blackthunder, but still compatible, given the
known wavelength calibration discrepancies between different NIRSpec
dispersers \citep[e.g.,][]{deugenio+2025a}. The resulting dust attenuation
is $A_V=0.4^{+0.7}_{-0.3}$~mag, consistent with no attenuation. A consistent result is obtained by fitting jointly \Paalpha together with
\Hbeta, \OIIIall and \Halpha, where we assume an intrinsic \Paalpha/\Hbeta=0.339, consistent with the assumptions for the intrinsic \Halpha/\Hbeta. This fit is performed on the prism spectrum, because \Paalpha is not detected in the gratings. This
model gives $A_V = 0.4^{+0.5}_{-0.3}$~mag.

The most stringent constraints come from the \blackthunder data; we extract prism spectra from two apertures, one centred on the target AGN, and one capturing the brightest star-forming region to the north west of the AGN. Modelling these spectra  yields respectively  $A_V = 0.35\pm0.15$ and $A_V = 0.55\pm0.10$~mag. Of these two values, the most relevant one for the foreground screen is the lowest, given that it is closer to the line of sight of \target.

In any case, \Hbeta and \Halpha light from \target passing through the interloper had observed-frame wavelengths of 1.48 and 1.99~\mum. At these 
wavelengths, with the fiducial $A_V=0.35$ of the interloper, the 
attenuation would be dominated by dust within \target, with a foreground 
attenuation of only 6 and 4~per cent for \Hbeta and \Halpha, respectively.

\begin{figure}
  {\phantomsubcaption\label{f.nolya.a}
   \phantomsubcaption\label{f.nolya.b}}
\includegraphics[width=\columnwidth]{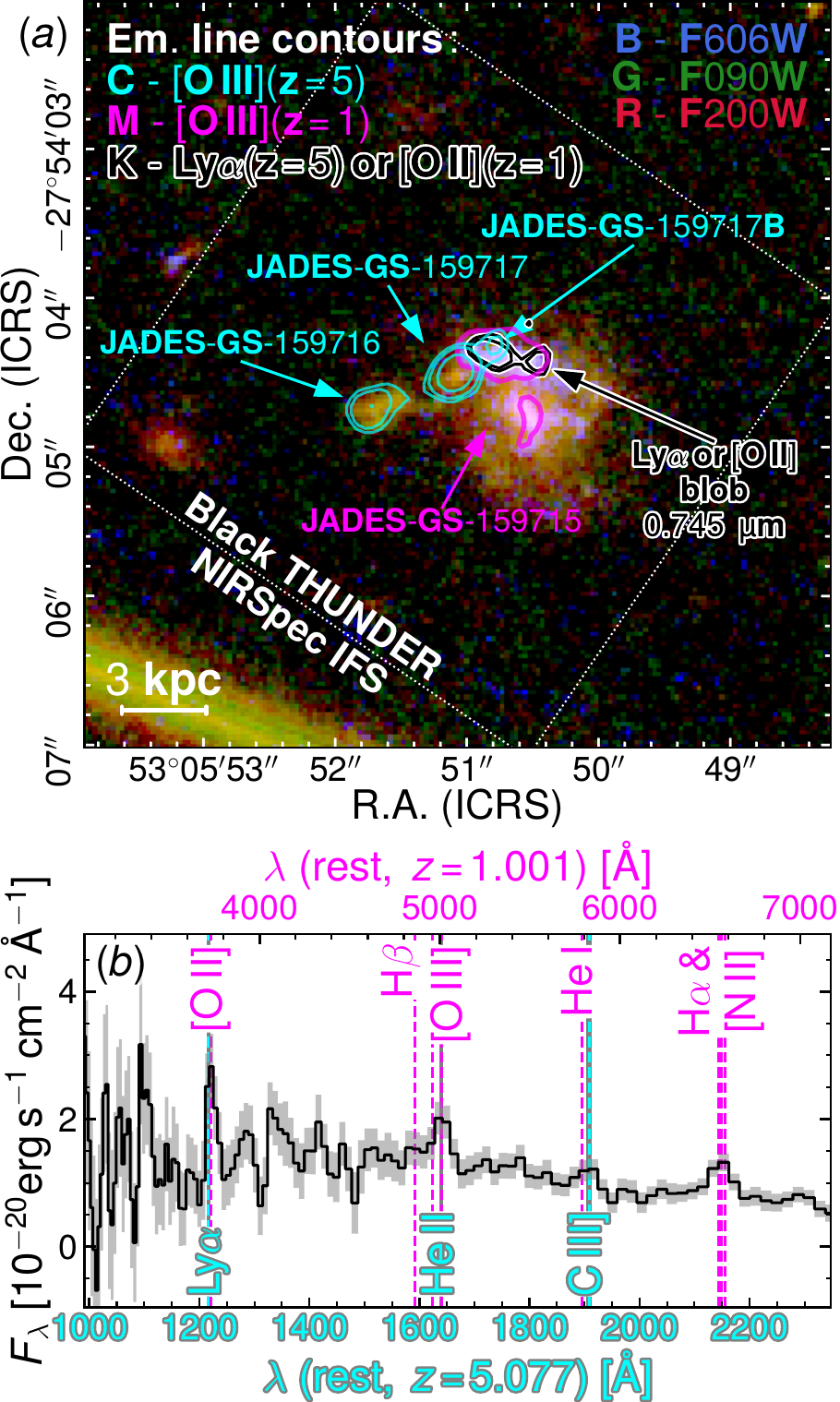}
  \caption{Panel~\subref{f.nolya.a}: a morphological analysis favours interpreting the 0.745-\mum emission
  line (black contours) as \OIIall from the $z=1$ interloper, instead of \Lyalpha from
  \target. The false-colour
  RGB image highlights in blue the bulge of JADES~ID 159715 and a star-forming
  spiral arm just north of the bulge. The contours show the location of various
  emission lines from the \blackthunder prism observations. The 0.745-\mum line
  emission coincides spatially and morphologically with the star-forming spiral arm
  of JADES~ID 159715, which is also traced by the \OIIIL emission line at $z=1$
  (magenta). In contrast, \OIIIL at
  $z=5.0775$ (cyan) is either spatially offset (\target and \satlarge), or
  morphologically different (\satsmall).
  Panel~\subref{f.nolya.b} reports the prism spectrum, annotating notable emission lines at the two redshift solutions.
  .}\label{f.nolya}
\end{figure}

To assess the \Lyalpha emission, we compare the spatial distribution of secure
emission lines in the main \target and in the interloper. In Fig.~\ref{f.nolya.a},
we use the \blackthunder prism data to create emission-line maps at the wavelengths
of \OIIIL at $z=5.078$ (cyan), of \OIIIL at $z=1$ (magenta), and at the coincident
wavelengths of \Lyalpha (at $z=5.078$) and \OIIall (at $z=1$; black contours).
The cyan contours are clearly misaligned with respect to the black contours, but
this in itself is not sufficient reason to discard the \Lyalpha interpretation,
since \Lyalpha can have substantially different morphology than optically thin
emission lines \citep[e.g.,][]{torralba+2025}. To further test the
\Lyalpha or \OIIall origin of the black contours, all emission-line contours are
overlaid on a RGB image that shows the rest-frame UV and
optical from the interloper in \hst/ACS F606W and \jwst/NIRCam F090W, respectively.
UV-bright regions are seen in pink hues, and coincide with the core of the foreground
galaxy and with its northern spiral arm. The latter also coincides perfectly with
both the northern magenta contour (\OIIIL at $z=1$) and with the black contours,
suggesting the latter should be conservatively identified as \OIIall arising from
the foreground. The prism spectrum is shown in Fig.~\ref{f.nolya.b}, where the top
and bottom axes report the rest-frame wavelengths for \interlop and \target,
respectively. We annotate notable emission lines for the two redshift solutions.
Note that \OIIIall and \Halpha are independently confirmed by the G140M detection
(Fig.~\ref{f.inter}).

\section{Gravitational lens}\label{s.lens}

Finding a bright LRD within only 0.6-arcsec separation from a foreground galaxy 
raises the question of possible gravitational-lens magnification \mulens. To 
evaluate this possibility and to infer \mulens, we need to estimate the mass of the 
lens galaxy inside the circle of radius $\theta = 0.6$-arcsec, corresponding to
4.94~kpc at $z=1.00115$. We start from the stellar mass measurement $\mstar = 10^{8.9}~\Msun$ from the CANDELS catalogue \citep{grogin+2011,koekemoer+2011}, and we use the stellar-to-halo mass 
relation of \citep{behroozi+2013} to derive a dark-matter halo mass of
$M_\mathrm{DM}=10^{11.8}~\Msun$, with a scatter of 0.2--0.3~dex \citep{behroozi+2013,moster+2013}.
For this halo, the virial radius is $R_\mathrm{vir} = 90$~kpc \citep{bullock+2001}.
Assuming a NFW profile \citep{navarro+1997}, and 
a concentration value $c=7\text{--}10$ \citep{dutton+maccio2014}, we can estimate both the scale density
and scale radius of the NFW profile, from which we derive enclosed masses of $M_\mathrm{DM}(R<4.94~\mathrm{kpc}) = 3.3\text{--}2.7\times10^{10}~\Msun$. This value is still much
larger than \mstar, without even considering that some of the stellar mass lies beyond the separation radius $\theta = 0.6$ arcsec, because the half-light radius of 
the lens is $\re = 0.43\pm0.05$ arcsec \citep{vanderwel+2012}.
We derive a magnification factor $\mulens < 1.1$, which is below the typical 
uncertainties on both \mstar, \re and \mbh.
From similar considerations, applied to the source \sattiny (which lies closer to
the centre of \interlop, about 0.1-arcsec away), we still derive $\mulens<1.1$.

\section{Point-source continuum in \target}\label{a.compcont}

In Section~\ref{s.an.ss.size}, we study the morphology of \target and its neighbours
using NIRCam and NIRSpec/IFS. Aiming to isolate the sources at $z\sim5$ from the
foreground, we target \OIIIall emission, which is done in NIRCam by taking the difference
between F277W and F200W, and in NIRSpec by creating an emission-line map centred on \OIIIL.
The two approaches  have different strengths, but they yield different morphological
properties for \target. Here we show that \target has a strong point-source continuum,
which NIRCam F277W-F200W cannot separate from the line emission, unlike NIRSpec/IFS.
Such emission could easily explain the different inferences between NIRCam and NIRSpec,
as reported in Section~\ref{s.an.ss.size}.

\begin{figure}
\centering
\includegraphics[width=0.9\columnwidth]{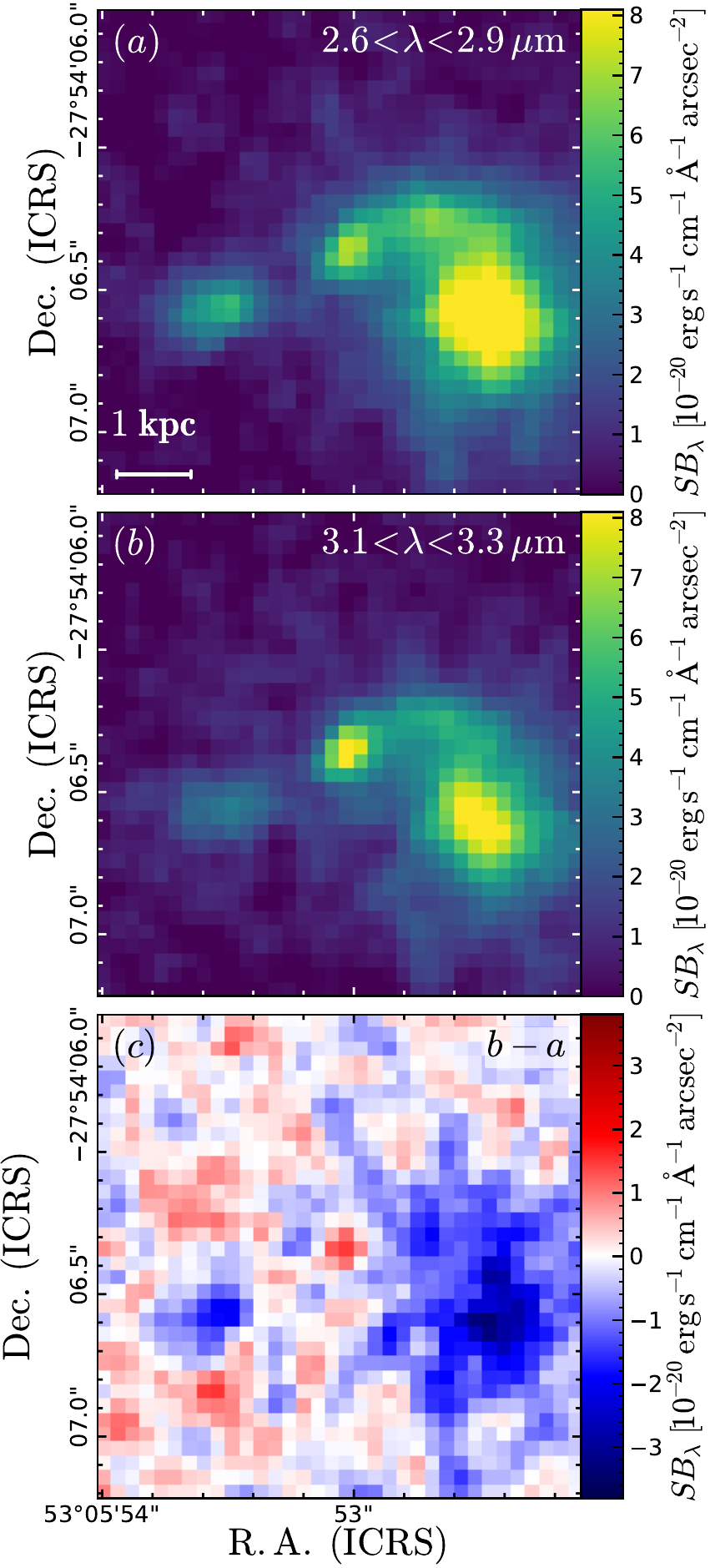}
  {\phantomsubcaption\label{f.nscont.a}
   \phantomsubcaption\label{f.nscont.b}
   \phantomsubcaption\label{f.nscont.c}}
	\caption{Synthetic images from NIRSpec/IFS prism, in two narrow wavelength intervals blueward and redward of the \Hbeta--\OIIIall complex (panels~\subref{f.nscont.a} and~\subref{f.nscont.b}).
Panel~\subref{f.nscont.c} shows the flux difference between the two images, acting as a pseudo-colour (but in linear space). The negative regions underscore the blue continuum colour of \satlarge and of the foreground galaxy \interlop, while \target appears in red, indicating the presence of a spectrally red, point-source continuum.}\label{f.nscont}
\end{figure}

In Fig.~\ref{f.nscont}, we show synthetic images derived from the NIRSpec/IFS prism observations,
targeting the continuum blueward and redward of the \Hbeta--\OIIIall emission-line group,
$2.6<\lambda<2.9~\mum$ and $3.1<\lambda<3.3~\mum$, respectively.
\target is clearly detected at both wavelengths -- as are other sources in the field of view
(panels~\ref{f.nscont.a} and~\ref{f.nscont.b}). However, the difference image $b-a$
(Fig.~\ref{f.nscont.c}) shows that \target is clearly red, which would then leave continuum
flux in the F277W-F200W image (Fig.~\ref{f.size.a}), while the narrow wavelength range of
the NIRSpec/IFS line map removes the continuum completely (Fig.~\ref{f.size.e}).

\section{NIRCam/WFSS fits}\label{a.nircamwfss}

\begin{figure}
  \includegraphics[width=\columnwidth]{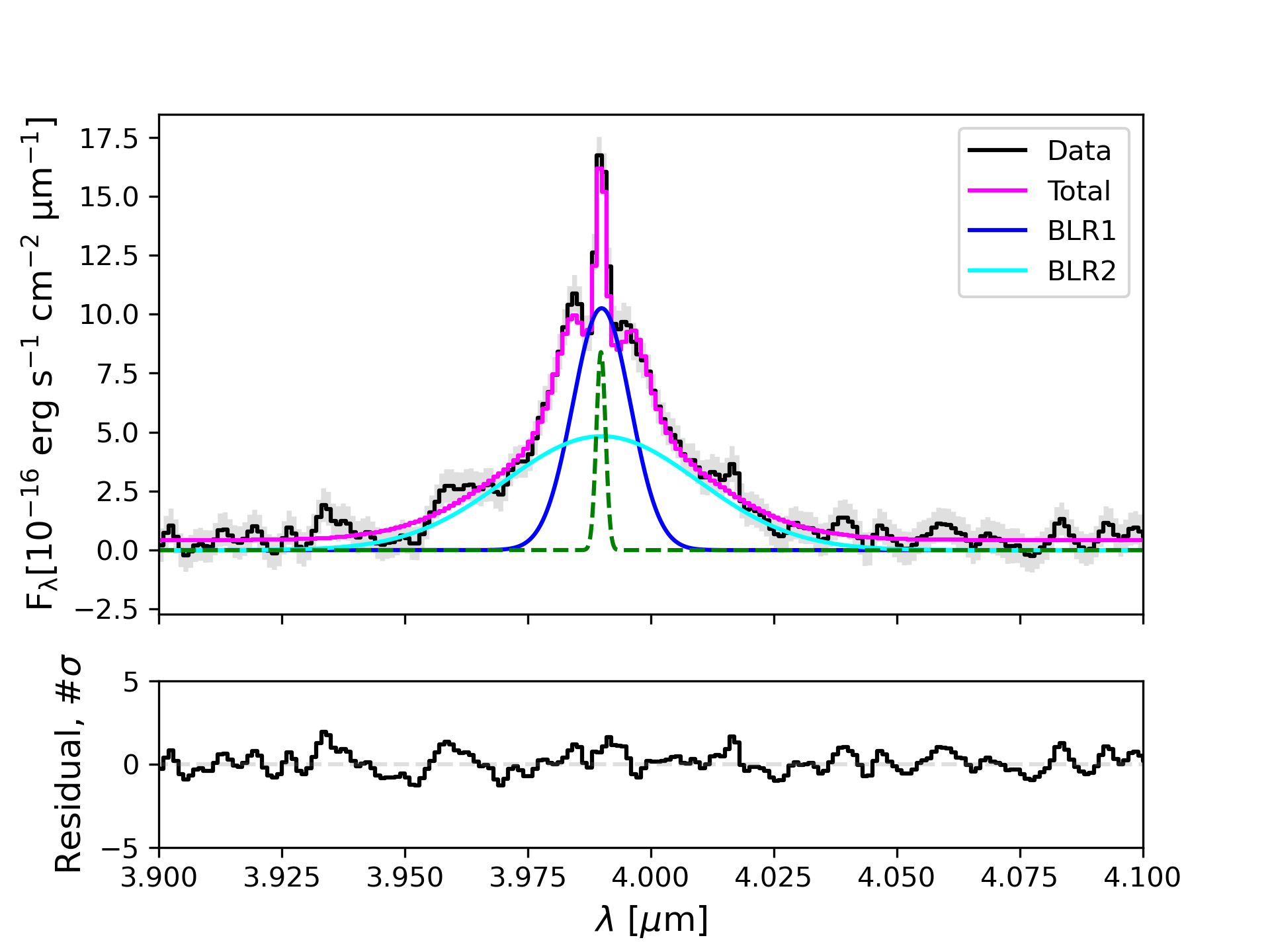}
  \caption{Mock observations of broad \Halpha in \target, using the
  spectral resolution of NIRCam/WFSS. While $R\sim1,600$ can 
  capture the presence of the rest-frame absorber (unlike the NIRSpec 
  $R\sim1,000$ gratings, Fig.~\subref{f.abs.a}), in practice the
  typical SNR of NIRCam/WFSS is significantly lower, due primarily to the 
  brighter background. This makes slitless spectroscopy excellent for
  identifying absorbers, but only in bright sources. The lines are the
  same as Fig.~\ref{f.abs}.}\label{f.nircamwfss}
\end{figure}

To probe the ability of NIRCam/WFSS to identify rest-frame absorbers, we
mock a NIRCam spectrum by convolving the G395H JADES data to match the
resolution of NIRCam F444W \citep{greene+2017} (Fig.~\ref{f.nircamwfss}).
We then fit this spectrum with the same setup as Fig.~\ref{f.abs}, with
the resulting absorber parameters shown in the corner diagram of
Fig.~\ref{f.trian} (purple). With resolution $R\sim 1,600$, NIRCam/WFSS
can clearly identify the absorber, although we note that the actual
SNR achieved here would require significantly longer than the 2~hours
used for NIRSpec. Nevertheless, NIRCam/WFSS seems very well suited to
assess the incidence of rest-frame (and other) absorbers in bright
sources. As for the inferred parameters, NIRCam/WFSS performs significantly
better than the medium-resolution NIRSpec gratings (Fig.~\ref{f.trian}),
but the posterior probabilities are both significantly broader and biased
compared to the results from the high-resolution NIRSpec grating.
\section{Line broadening due to electron scattering}\label{a.exponential}

The broad-line profiles of some AGN display a distinctively exponential profile \citep{laor2006}.
This has also been found in LRD-like AGN \citep{rusakov+2025}. In this section, we repeat the
fit on the high-SNR aperture, using the scattering assumption. We use the same framework from
Section~\ref{s.an.ss.g395h.bt}, but the broad line is modelled as a single emitted Gaussian, which
we convolve with a symmetric exponential kernel in wavelength space, $K(\lambda) = 1/(2 W)
\exp(-|\lambda/W|)$, where $W$ is a free parameter. The emitted and transmitted Gaussian are then
rescaled by factors of $\exp(-\tau)$ and $1-\exp(-\tau)$, respectively, where the optical depth
$\tau$ is also a free parameter.

\begin{figure}
\includegraphics[width=\columnwidth]{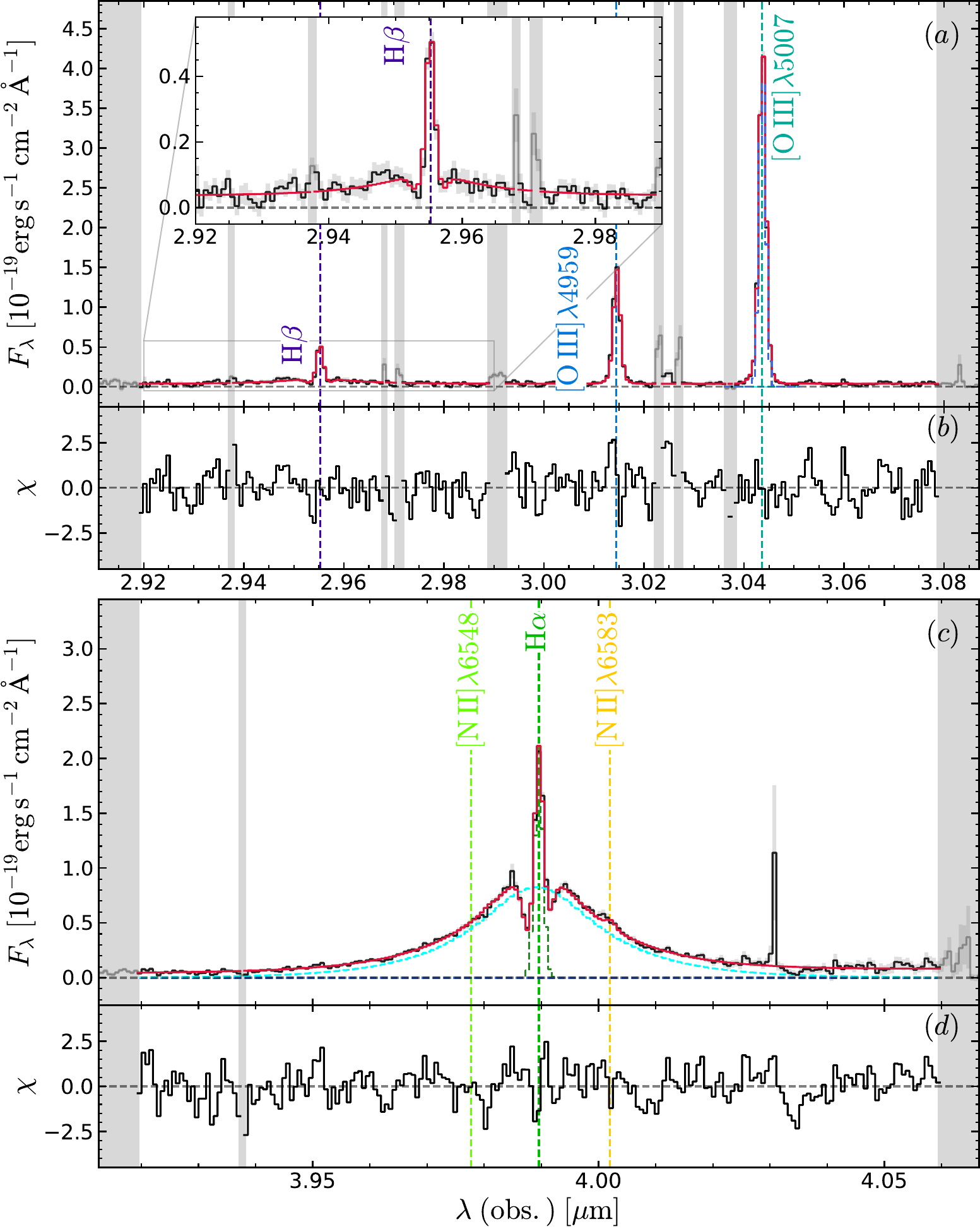}
  \caption{Modelling the broad lines with a single Gaussian and electron scattering,
  following the approach of \citet{laor2006} and \citet{rusakov+2025}. This model finds a four
  times smaller SMBH mass, but our overall conclusions would remain unchanged.
  }\label{f.exp}
\end{figure}

The resulting model is illustrated in Fig.~\ref{f.exp}. We find $\tau = 3.7$, implying that
over 98~per cent of the intrinsic broad line has been scattered. $W$ \citetext{$\sigma$, in the
terminology of \citealp{laor2006}} instead takes a value of
870~\kms, while the broad-line FWHM is $800\pm200$~\kms, two times narrower than the
fiducial value. With these numbers, and using the calibration from \citet{reines+volonteri2015},
we obtain a SMBH mass $\log (M_\mathrm{\bullet,ism}/\Msun) = 7.0_{-0.3}^{+0.2}$, four times smaller than the fiducial value, implying a super-Eddington ratio of order 1.6.

The best-fit model is illustrated in Fig.~\ref{f.exp}, with the same meaning as the fiducial
fit in the main text (Fig.~\ref{f.g395h.bt}). Both models reproduce the data, but there is a possible
inconsistency between the large optical depth $\tau$ and the relatively narrow scatter parameter $W$.
Using the relation from \citet{rusakov+2025}, we can in fact estimate $\tau$ directly from
$W$, with $\tau(W)$ scaling as $W = (428\,\tau(W) + 370) * \sqrt{T_\mathrm{e}/10^4~\mathrm{K}}$.
With the narrow value of $W$ inferred from the model, we need $T_\mathrm{e}\sim2,000$~K
to have $\tau = \tau(W)$, i.e., the large optical depth inferred from the line shapes does not
match the width of the exponential. This is also confirmed if we repeat the inference procedure
with a model where $W$ is not a free parameter, but is inferred using the relation from
\citet{rusakov+2025}, with $\tau$ and $T_\mathrm{e}$ as free parameters.

%%%%%%%%%%%%%%%%%%%%%%%%%%%%%%%%%%%%%%%%%%%%%%%%%%

% Don't change these lines
\bsp	% typesetting comment
\label{lastpage}
\end{document}